\documentclass[fleqn,usenatbib]{mnras}

\usepackage{newtxtext,newtxmath}
 
\usepackage[T1]{fontenc}

\DeclareRobustCommand{\VAN}[3]{#2}
\let\VANthebibliography\thebibliography
\def\thebibliography{\DeclareRobustCommand{\VAN}[3]{##3}\VANthebibliography}

\usepackage{graphicx}	
\usepackage{amsmath}	
\usepackage{amssymb}	
\usepackage{pdflscape}	

\usepackage{pgfplotstable}
\usepackage{csvsimple}
\usepackage{booktabs} 

\newcommand{\Msun}{{$M_{\odot}$}}

\title[Dust in SMC SNRs]{{\it Spitzer} and {\it Herschel} studies of dust in supernova remnants in the Small Magellanic Cloud}

\author[M. Matsuura et al.]{
Mikako Matsuura,$^{1}$\thanks{E-mail: matsuuram@cardiff.ac.uk (MM)}
Victoria Ayley,$^{1}$
Hannah Chawner,$^{1,2}$
M.D. Filipovi\'c,$^{3}$ \newauthor
Warren Reid,$^{3,4,5}$
F.D. Priestley,$^{1}$
Andy Rigby,$^{1}$
M.J. Barlow,$^{6}$
Haley E. Gomez,$^{1}$
\\
$^{1}$School of Physics and Astronomy, Cardiff University,  The Parade, Cardiff CF24 3AA, UK\\
$^{2}$School of Chemistry, University of Bristol, Queens Road, Bristol, BS8 1QU, UK \\
$^{3}$Western Sydney University, Locked Bag 1797, Penrith South DC, NSW 2751, Australia\\
$^{4}$Department of Physics, Macquarie University, Sydney, NSW 2109, Australia \\
$^{5}$Observatory Hill Waitoki, 130 Dormer Rd, RD2 Helensville, 0875, New Zealand\\
$^{6}$Department of Physics and Astronomy, University College London, Gower Street, London WC1E 6BT, UK\\
}

\date{Accepted XXX. Received YYY; in original form ZZZ}

\pubyear{2021}

\begin{document}
\label{firstpage}
\pagerange{\pageref{firstpage}--\pageref{lastpage}}
\maketitle

\begin{abstract}
%
With the entire Small Magellanic Cloud (SMC) mapped by the {\it Spitzer Space Telescope} and {\it Herschel Space Observatory}, we were able to search 8--250\,$\mu$m images in order to identify infrared (IR) emission associated with SMC supernova remnants (SNRs). 
A valid detection had to correspond with known X-ray, H$\alpha$ and radio emission from the SNRs.
From the 24 known SNRs, we made 5 positive detections with another 5 possible detections.
Two detections are associated with pulsars or pulsar wind nebula, and another three detections are part of the extended nebulous emission from the SNRs.
We modelled dust emission where fast moving electrons are predicted to collide and heat dust grains which then radiate in IR.
With known distance (62.44 $\pm$ 0.47kpc), measured SNR sizes, electron densities, temperatures from X-ray emission as well as hydrogen densities, the modelling of SMC SNRs is straightforward.

If the higher range of hydrogen and electron densities were to be accepted, we would expect almost all SMC SNRs to be detected in the IR, at least at 24\,$\mu$m, but the actual detection rate is only 25\,\%.
One possible and common explanation for this discrepancy is that small grains have been destroyed by the SNRs shockwave.
However,  within the uncertainties of hydrogen and electron densities,  we find that infrared dust emission can be explained reasonably well, without invoking dust destruction.
There is no conclusive evidence that SNRs destroy swept-up ISM dust.
\end{abstract}




\section{Introduction}

Dust grains are small particles made of minerals, and are typically composed of silicon or carbon, together with other metals, such as magnesium, oxygen and iron.
The detection of a vast amount of dust grains ($>10^8$ \Msun) in the interstellar medium (ISM) of galaxies within the Local Universe and out to high redshifts
\cite[e.g.][]{Bertoldi:2003p15519, Watson:2015in, Laporte:2017eu}
triggered a debate over how these dust grains could have been acquired in cosmic time.
In the Local Universe, asymptotic giant branch (AGB) stars, the late phase of low- and intermediate-mass stars may be able to account for a substantial amount of this dust mass  \citep{1989IAUS..135..445G, Dwek:1998p3931, Matsuura:2009p29906, Matsuura:2013js}, with a possible contribution from ISM grain growth  \citep{2010pcim.book.....T, Galliano:2018ev}.
However, the evolutionary timescale for low- and intermediate-mass stars \citep{Morgan:2003p15480} and the ISM grain growth \citep{Ferrara:2021uc} might be too slow to achieve the large amounts of dust seen in high redshift galaxies.
Supernovae (SNe) have been suggested as an alternative and additional source of dust in the ISM of galaxies \citep[e.g.][]{Morgan:2003p15480, Maiolino:2004p10092, Matsuura:2009p29906, Cherchneff:2010p28841, Hjorth:2013be}.

Although SNe eject newly synthesised heavy elements into the ISM, questions remain over the proportion of the metals that condense into dust grains, and how much of the SN and ambient ISM dust can survive SN shocks \cite[e.g.][]{Nozawa:2003p29406, Schneider:2004cn, Cherchneff:2010p28841, Sarangi:2015he}.
Chemical modelling predicts that a SN may produce approximately 0.1--1\,\Msun\, of dust \citep{Dwek:2011gp, Sarangi:2015he}, however, their survival rate after encountering reverse shocks is currently predicted as anywhere between 1\% and 90\% \citep[e.g.][]{Micelotta:2016jk, Micelotta:2018gl, Kirchschlager:2019cq}, a large uncertainty.
Moreover, while SNRs expand at speeds ranging between a few hundred and a few thousand km\,s$^{-1}$, the forward shock can destroy existing ambient ISM dust \citep{McKee:1980jz, Jones:1996bia, Bocchio:2014es, Micelotta:2018gl}.
Even though dust grains may be produced by AGB stars and SN ejecta, the destruction of dust by forward shocks can be so efficient as to lower or eliminate the total ISM dust budget.  This has  led some to conclude that the ISM of galaxies might not  be able to maintain a large amount of dust grains \citep[e.g.][]{Jones:1996bia, Michaowski:2010p29825}.

 While modelling predicts both production and destruction of dust by SNe and SNRs, recent observations reveal that indeed SNe can produce a substantial mass of dust in their ejecta.
 Amongst remnants of core-collapsed SNe, a large amount ($\sim$0.1--1\,\Msun) of dust has been detected, including SN\,1987A \citep{Matsuura:2011ij, Matsuura:2015kn, Wesson:2014gs}, Cassiopeia\,A \citep{Rho:2008p414, Barlow:2010p29287, DeLooze:2017cz, NiculescuDuvaz:2021wv}, Crab Nebula \citep{Gomez:2012fm, temim:2013eb, DeLooze:2019jg}.
 Additionally, there are reports  of significant quantities of dust in  SMC 1E\,0102$-$7219 (SNR J0104.0$-$7202) \citep{Stanimirovic:2005p12420, Sandstrom:2009jxa}.
 However, challenges remain regarding how to measure the amount of dust that can  survive passage through the reverse shock, its wake and the follow-up radiation from the SN.
 These SNRs with dust measurements are relatively young, estimated at $\sim$30--1000\,years only.
  In order to evaluate the true impact of SN shocks on dust destruction, the ageing effects on dust should be examined using older SNRs.
Somehow among older SNRs, those with pulsar wind nebulae (PWNe), whose ejecta are heated by radiation from the neutron stars \citep{Chevalier:1977ef, Chevalier:ba, Hester:2008kx}
 have been frequently reported to have a substantial quantity of dust, including
 G54.1+0.3 \citep{temim:2017ev, Rho:2018ee}, G11.2$-$0.3, G21.5$-$0.9, G29.7$-$0.3 \citep{Chawner:2019dn},
SNR 0540$-$69.3 \citep{Lundqvist:do}.
It is still yet unknown why  the detection rate of ejecta dust in PWNe appears to be higher than in general SNRs. It might be because the presence of a heating source, pulsars, warm the ejecta, thereby allowing the PWN dust to stand out against the cooler ISM dust \citep{Chawner:2019dn}. 

The dust destruction rate by forward shocks is even more difficult to estimate.
\citet{Lakicevic:2015iwa} estimated infrared emission towards Large Magellanic Cloud (LMC) SNRs and surrounding ISM, suggesting ISM dust density is lower towards SNRs.
However, the reduced ISM dust density towards the SNRs could be artificially caused by hotter SNR dust emission that can hide a larger amount of colder ISM dust emission along the line of sight \citep{Micelotta:2018gl}. 
The method used by \citet{Lakicevic:2015iwa} gives inconclusive results for the SNR dust destruction rate.
\citet{temim:2015bs} estimated the dust destruction rate in the LMC, however, the actual destruction rate still relies on the theoretical dust destruction rate.
Therefore, measuring the dust destruction rate is very challenging.
One of the better methods to estimate the dust destruction rate is to consider the energy balance between collisional heated dust grains (by electrons) and radiation \citep{Dwek:1992da}.
Since heating by electron collision and infrared radiation depends on the dust properties, including dust mass and grain size, this method can potentially inform the dust destruction rate, if mass and size of ISM dust grains prior to the shocks can be estimated or assumed.
\citet{Williams:2006cn} suggested that infrared dust emissions observed in three SNRs are consistent with collisionally heated dust, and that the estimated dust mass is smaller than the expected SNR swept-up dust mass, hence, dust destruction is occurring.
Unfortunately, there was little description of the uncertainties  in the estimated dust mass, and it is unclear how accurate the dust destruction rate is.
A more recent survey of Galactic and LMC SNRs found that infrared dust emission is overwhelmingly fainter than  what the collisionally heated dust model predicts \citep{Seok:2013fz, Seok:2015jo, Koo:2016hba}.
This discrepancy might either be due to a lack of small dust grains, which have already been destroyed \citep{Dwek:2008p28793}, or due to dust grains that are heated by radiation, rather than by collisions \citep{Koo:2016hba}.

To date, studies of dust in SNRs have been limited by small sample sizes.
In order to increase the sample size and improve the evaluation of SNR dust emission,  we conducted this search in SMC,
a dwarf irregular galaxy at a distance of approximately 60\,kpc \citep{2007glg..book.....V, Graczyk.2020}.
Due to our viewing angle within the Galactic plane, the evaluation of infrared emission from SNRs often suffers from severe confusion with general ISM emission \citep{Chawner:2019dn, Chawner2020}, but viewed away from the Galactic plane, the SMC is much less prone to the ISM confusion problem.
A second advantage of studying SMC SNRs is that together with the general known distance, the SMC has an estimated depth of only 4--10 kpc \citep{1993MNRAS.261..873H, Subramanian.2009.AA, Graczyk.2020}, translating to distance uncertainties of only about 15--20\% for the SNRs.
This is in contrast to Galactic SNR distances, where uncertainties often amount to a few factors.
Distance uncertainties propagate into the sizes and inferred dust masses of SNRs, so the distance uncertainties are an important factor in dust analysis.
Thirdly, the local star-formation history has been analysed, so that the ages of SNRs may be more estimated \citep{Lopez:2014fo, Anonymous:c6obe9b8}.
Fourthly, there is a complete X-ray image of the SMC, which can inform the electron temperature and electron density measurements \citep{Anonymous:c6obe9b8, Maggi:2019wq}.
The electron temperature and density are crucial parameters for a collisionally heated dust model.

The SMC has the disadvantage of being more distant than the Milky Way, so that SNRs tend to be fainter, and 
have fewer expansion velocity measurements available. 
 Above all, there are only about 25 SNRs or SNR candidates known in the SMC \citep{Maggi:2019wq}, compared with over 250 SNRs and candidates in the Milky Way \citep{Green:2009wh}.
Nevertheless, with electron density and temperature measurements, important in estimating the dust destruction rate, plus infrared images, we were able to conduct the following analysis of SMC SNRs.

Note that the SMC is known to have a much lower metallicity than the Milky Way or the LMC, and it has been estimated that the dust-to-gas ratio in the SMC ISM is much lower \citep{Gordon:2014bya}.


\section{Source catalogue}

We assembled a list of known SNRs in the SMC from several references, mainly radio and X-ray surveys.

The optical H$\alpha$ and [N {\small II}] emission line surveys were originally used to find large nebulous objects in the SMC \citep{Davies:vz}, but their classification as SNRs had to await the application of early radio detections. 
Radio surveys are effective at detecting the synchrotron radiation produced by SNRs, clearly separating them from other large nebulous objects such as H{\small II} regions. 
Since the year 2000, a new generation of radio surveys at the Australia Telescope Compact Array (ATCA) has led to the identification of nearly 30 SNRs and SNR candidates \citep[e.g.][]{Filipovic:2002kx, Filipovic:2008fv, Anonymous:TowVvoI8}. 
In each case the radio brightness and spectral index were used to determine the synchrotron component.

Most SNRs stand out in the X-ray wavelength \citep{Borkowski:2006dh, Bozzetto:2017cv}, providing an extra means of identification. 
One of the first X-ray surveys of the SMC resulted in the {\it Chandra} Catalogue of Magellanic Cloud Supernova Remnants which listed 14 in the SMC \footnote{https://hea-www.harvard.edu/ChandraSNR/snrcat$_-$lmc.html}.  
A further catalogue from \citet{2010MNRAS.407.1301B} listed 23 remnants with X-ray and radio band detections using existing catalogues such as  \citet{2005MNRAS.364..217F}. 
Most of the remnants listed by the {\it Chandra} Catalogue appeared in \citet{2010MNRAS.407.1301B}, apart from SNR B0044$-$73.4, N19, HFPK-285 and SXP-1062. 
The {\it XMM-Newton}  study by \citet{Haberl:to} included the remnants catalogued by  \citet{2010MNRAS.407.1301B}, while adding SXP-1062, and three new candidates, bringing the total to 30 SMC remnants. 
Finally,  \citet{Anonymous:c6obe9b8} derived elemental abundances and SN types, describing whether SNRs are the result of core collapse or not. 
Based on elemental abundances, they eliminated some SNR candidates, leading  to a total of 25 SNRs and SNR candidates.
It was important that the catalogue contained celestial coordinates and sizes, so the location of the infrared-counterpart could be localised. 
The catalogue also contained X-ray brightness, as the plasma detected in X-rays is thought to be responsible for infrared dust emission from SNRs with shock heated gas \citep{Dwek:2008p28793, Koo:2016hba}.

More recently, a catalogue of SMC SNRs was used to investigate SNR populations, the star-formation history, and dust destruction rates \citep{2010MNRAS.407.1301B, temim:2015bs}, as well as SNR morphologies and the impact they make to their environments \citep{Lopez:2014fo}.
 These catalogues laid the foundations for multi-wavelength studies of SMC SNRs spanning X-ray to radio wavelengths, leading to measurements of metallicities and estimates of progenitor star masses \citep{Maggi:2019wq}. \citeauthor{Maggi:2019wq}. listed {\it XMM}, optical and radio images of SMC SNRs, which we include as reference images to compare with {\it Spitzer} and {\it Herschel} images.
 



\clearpage

\begin{landscape}
\begin{center}
\begin{table}
\caption{List of SMC SNRs and SNR candidates \label{detectionlevel}}
\begin{filecontents*}{list3.csv}
SNR,SNR,,SNR,RA,December,Size,Size reference wavelengths,Detected wavelengths,Type,Ref,,IR Det.,Age,Progenitor mass,Ref,Auchettl Mass,Rad Vel,N,O,n_0 from Temim,TeV Maggi,TeV uncertainties,TeV uncertainties,EM,Uncertainties,,n_H Maggi,F24 maggi,F70 Maggie,F24 Temim,F70 temim
SNR J0040.9$-$7337,B0039$-$73.53,DEM S5,,00h40m55s,$-$73d36m55s,2.5,R,RX,,ABFHM,,Detection,,,,8--12.5 (100\%),108$\pm$94,7.4,8.2,0.51,0.65,+0.09,-0.12,4.01,+0.96,-0.39,0.04,0.04,0.7,0.4,7
SNR J0046.6$-$7307,B0044$-$73.4,DEM S32,,00h46m41s,$-$73d07m55s,3.5,X,RX,,CM,,Possible detection,,,,$>$21.5 (92\%); 8--12.5 (8\%),345$\pm$45,,,1.67,0.60,+0.11,-0.11,7.71,+3.45,-1.64,,,,,
=SNR J0046.6$-$7309,,,,00h46m39s,$-$73d08m39s,2.3,R,,CC,ABHM,,,,,,,,,,,,,,,,,,,,,
SNR J0047.2$-$7308,B0045$-$73.4,,IKT 2,00h47m12s,$-$73d08m26s,1.5,X,RX,CC,ABCFHM,,Unlikely detection,,,,$>$21.5 (92\%); 8--12.5 (8\%),180$\pm$35--326$\pm$73,6.8--7.4,7.6--8.5,1.76,0.60,+0.04,-0.05,13.7,+4.87,-0.45,0.03,0.9,4.4,7.2,34
SNR J0047.5$-$7306,B0045$-$73.3,,,00h47m29s,$-$73d06m01s,2.7,R,RX,,AGFHM,In N19 / HII,Non detection,,15--22.5,K,$>$21.5 (92\%); 8--12.5 (8\%),232$\pm$52--338$\pm$40,6.8--7.1,7.6--8.2,1.71,0.63,0.25,-0.13,5.64,+4.87,-4.78,,,,,
=SNR J0047.5$-$7307,,NS19?,HFPK 419?,00h47m32s,$-$73d07m45s,4.5,X,X,,C,,,,,,,,,,,,,,,,,,,,,
SNR J0048.4$-$7319,B0046$-$73.5,DEM S42E,IKT 4,00h48m25s,$-$73d19m24s,2.0,X,RX,Ia,ABCFHM,,Unlikely detection,,,,$>$21.5 (59\%); 12.5--21.5 (17\%); 8--12.5 (24\%),216$\pm$23--283$\pm$57,6.8--7.1,7.6--8.5,1.89,1.07,0.76,-0.18,1.03,+0.65,-0.58,,,,,
SNR J0049.0$-$7306,,,,00h49m00s,$-$73d06m17s,1,X,X,CC,H,SNR cand.,Unlikely detection,,,,,,,,,,,,,,,,,,,
SNR J0049.1$-$7314,B0047$-$73.5,DEM S49,IKT 5,00h49m07s,$-$73d14m35s,2.7,R,RX,Disputed,ABCFHM,,Unlikely detection,,,,$>$21.5 (59\%); 12.5--21.5 (17\%); 8--12.5 (24\%),,7.1--7.7,7.9--8.5,1.48,0.52,+0.15,-0.14,3.16,+3.63,-1.64,,,,,
SNR J0051.1$-$7321,B0049$-$73.6,DEM S53,IKT 6,00h51m07s,$-$73d21m29s,3.0,X,RX,CC,AGCFHM,,Unlikely detection,14000,13--15 or 15--22.5,HK,12.5--21.5 (54\%); 8--12.5 (46\%),139$\pm$37--289$\pm$160,6.8--7.1,7.9--8.2,1.13,0.21,+0.01,-0.01,171,+22.4,-33.6,0.05,0.2,4.0,0.8,16
SNR J0052.6$-$7238,B0050$-$72.8,DEM S68SE,,00h52m33s,$-$72d37m35s,5.4,X,RX,,ABFHM,,Detection,,,,8--12.5 (100\%),185$\pm$40,,,0.79,0.38,+0.66,-0.16,27.4,+308,-24.6,,,,,
=SNR J0052.9$-$7237,,HFPK 285,,00h53m01s,$-$72d36m52s,5.0,X,X,,Cf,,,,,,,,,,,,,,,,,,,,,
SNR J0056.5$-$7208,,,,00h56m30s,$-$72d08m12s,2.0,X,X,,HM,SNR cand.,Non detection,,,,,85$\pm$50,,,,,,,,,,,,,,
SNR J0057.7$-$7213,,,,00h57m46s,$-$72d13m04s,3.0,X,X,,HM,SNR cand.,Non detection,,,,,,,,,,,,,,,,,,,
SNR J0058.3$-$7218,B0056$-$72.5,,IKT 16,00h58m16s,$-$72d18m05s,4.5,X,RX,CC,HFM,,Unlikely detection,,,,,,6.8,7.9,0.92,0.37,+0.26,-0.09,28.5,+60.5,-22.1,,,,,
SNR J0059.4$-$7210,B0057$-$72.2,DEM S103,IKT 18,00h59m25s,$-$72d10m10s,2.3,X,RX,CC,ABCFHM,,Possible detection,200000,,N,,,8.0,9.1,0.64,0.66,+0.03,-0.03,25.9,+2.37,-2.15,0.04,0.8,4.4,3.0,17
SNR J0100.3$-$7134,B0058$-$71.8,DEM S108,,01h00m21s,$-$71d33m40s,3.5,R,RX,,ABFHM,,Non detection,,,,$>$21.5 (100\%),,,,0.50,0.49,+0.14,-0.16,3.28,+4.96,-0.99,,,,,
SNR J0103.2$-$7209,B0101$-$72.6,,IKT 21,01h03m17s,$-$72d09m42s,1.5,R,RX,CC,ABFHM,,Unlikely detection,,,,$>$21.5 (92\%); 8--12.5 (8\%),221$\pm$34--188$\pm$19,6.8--7.4,7.6--7.8,0.89,1.0,+0.35,-0.20,1.97,+0.45,-0.40,0.03,0.3,1.5,2.6,15
SNR J0103.5$-$7247,,HFPK 334,,01h03m30s,$-$72d47m20s,1.5,R,RX,,ABCFM,,Possible detection,,,,Ia,,,,0.70,1.3,,,0.93,+0.45,-0.35,,,,,
SNR J0103.6$-$7201,,,,01h03m36.5,$-$72d01m35s,1.5,,Opt,,M,,Possible detection,,,,,,,,,,,,,,,,,,,
SNR J0104.0$-$7202,B0102$-$72.3,DEM S124,IKT 22,01h04m02s,$-$72d01m48s,1.0,X,RXOpt,IIb,ABCFHM,,Detection,2050,$<$15 or 25--35,BKX,$>$21.5 (49\%); 12.5--21.5 (20\%); 8--12.5 (31\%),,,,0.85,,,,,,,,,,,
SNR J0105.1$-$7223,B0103$-$72.6,DEM S125,IKT 23,01h05m04s,$-$72d22m56s,3.5,X,RX,CC,ABCFHM,,Non detection,18000,$\sim$18 or 15--22.5,KP,8--12.5 (100\%),,,,0.67,0.19,+0.01,-0.01,448,+51.7,-51.7,0.06,0.4,6.5,0.7,12
SNR J0105.6$-$7209,,DEM S128,,01h05m27s,$-$72d10m38s,3.2,R,RX,Ia,ABFHM,,Possible detection,,,,$>$21.5 (76\%); 12.5--21.5 (15\%); 8--12.5 (9\%),,,,0.91,0.68,+0.20,-0.11,4.44,+1.72,-0.78,0.05,0.02,0.6,0.5,12
SNR J0106.2$-$7205,B0104$-$72.3,,IKT 25,01h06m20s,$-$72d05m18s,2.5,R,RX,Disputed,ABCFHM,,Detection,17600,18--25 or 25$^{a}$,L,$>$21.5 (76\%); 12.5--21.5 (15\%); 8--12.5 (9\%),216$\pm$25--297$\pm$43,6.8,7.9,0.90,0.72,+0.07,-0.08,9.95,+1.59,-1.51,0.04,0.2,1.8,2,18
SNR J0106.5$-$7242,,,,01h06m32.1s,$-$72d42m17.0,2.58,R,,,M,SNR cand.,Non detection,,,,,,,,,,,,,,,,,,,
SNR J0109.7$-$7318,,,,01d09m43.6s,$-$73d18m46s,1.75,,Opt,,M,SNR cand.,Non detection,,,,,,,,,,,,,,,,,,,
SNR J0127.7$-$7332,,SXP 1062,,01h27m46s,$-$73d32m56s,3.5,Opt,Opt,Disputed,CHM,Pulsar or HXB?,Detection,,,,,,,,,0.19,+0.02,-0.01,25.8,+19.1,-12.7,,,,,
,,,,,,,,,,,,,,,,,,,,,,,,,,,,,,,
Non-SNRs?,,,,,,,,,,,,,,,,,,,,,,,,,,,,,,,
SNR J0047.8$-$7317,,NS 21,,00h47m48s,$-$73d17m27s,2.0,R,R,,ABFH,SNR cand. / HII region,Possible detection,,,,,,,,,,,,,,,,,,,
SNR J0051.9$-$7310,,,IKT 7,00h51m54s,$-$73d10m24s,1.6,X,X,CC,ABHM,XB,Unlikely detection,,,,,,,,,,,,,,,,,,,
SNR J0114.0$-$7317,,N83C,,01h14m00s,$-$73d17m08s,0.5,R,R,,ABHM,HII region?,Non detection,,,,$>$21.5 (100\%),,,,,,,,,,,,,,,
\end{filecontents*}
\csvreader[tabular= l l c c cc ll l c l l l c l,
				table head=\hline Name & & &&  RA & Dec & D  & D. Wav. & S. Wav. & Type & Ref & Comment \\ \hline\hline, 
				table foot=\hline ] 
				{list3.csv}
				{} 
				{\csvcoli & \csvcolii & \csvcoliii & \csvcoliv & \csvcolv & \csvcolvi & \csvcolvii & \csvcolviii  & \csvcolix & \csvcolx & \csvcolxi &  \csvcolxii  } 
\\
Some SNRs (J0046.6$-$7307, J0047.5$-$7307 and J0052.9$-$7237) are listed with slightly different coordinates arising from observations made at different wavelengths, so both of them are listed after `='.
D: Diameter is in arcmin. D.Wav.:  Wavelength used for measuring the diameter (R: radio and X: X-ray). S. Wav.: Studied wavelengths in the literature.
References:
A: \citet{Anonymous:c6obe9b8}, B: \citet{2010MNRAS.407.1301B}, C: Chandra Catalogue, F: \citet{2005MNRAS.364..217F}, f: \citet{Filipovic:2008fv}, H: \citet{Haberl:to}, M: \citet{Maggi:2019wq} and N: \citet{Nota:dk}.
XB: X-ray binary. HXB: high-mass X-ray binary
\end{table}
\end{center}
\end{landscape}


\begin{landscape}
\begin{table}
\begin{center}
\caption{ Infrared detection level with properties of SNRs: abundance, age and progenitor mass, if the SNR is of core-collapse origin. \label{table-detection}}
\csvreader[tabular= l l  lll cc  rr  ll ll l c l l l c l,
				table head=\hline Name &  IR Det & kT  & EM & $n_{\rm H}$ & $n_{\rm H, ISM}$ & $F_{24}$ & $F_{70}$ & $F'_{24}$ & $F'_{70}$ & Age & Mass (\Msun) & Ref & Mass possibilities (\%)\\ \hline\hline, 
				table foot=\hline ] 
				{list3.csv}
				{} 
				{\csvcoli &  \csvcolxiii & \csvcolxxii$^{\csvcolxxiii}_{\csvcolxxiv}$  & \csvcolxxv$^{\csvcolxxvi}_{\csvcolxxvii}$   & \csvcolxxviii &  \csvcolxxi  &  \csvcolxxix & \csvcolxxx & \csvcolxxxi &  \csvcolxxxii   & \csvcolxiv   & \csvcolxv & \csvcolxvi & \csvcolxvii } 
\\
\end{center}
Infrared (IR)  detection (either {\it Spitzer} or {\it Herschel}, or both).
kT (keV): electron temperature estimated from X-ray spectra \citep{Maggi:2019wq},
EM: emission measure   \citep{Maggi:2019wq},
$n_{\rm H}$: hydrogen density estimated from EM.
$n_{\rm H, ISM}$: average hydrogen density in the line of sight, estimated from H{\small I} line \citep{temim:2015bs}.
$F_{24}$ and $F_{70}$: predicted surface brightnesses (MJy\,sterad$^{-1}$) of a SNR at 24 and 70\,$\mu$m, using $n_{\rm H}$ (Sect.\,\ref{section-model}).
$F'_{24}$ and $F'_{70}$: the same as $F_{24}$ and $F_{70}$, but using $n_{\rm H, ISM}$.
Age: estimated age in years; Mass: estimated progenitor mass in \Msun; Ref: reference; Mass possibilities: estimated progenitor mass (\Msun) from star-formation history at the SNR location and the probabilities (\%) of the given mass \citep{Anonymous:c6obe9b8}.
$^a$: There is a dispute as to whether SNR J0106.2$-$7205 is of core collapse or type Ia origin. The estimated mass here is based on the core collapse case.
References:
B: \citet{Blair:iw}
H: \citet{Hendrick:2005bk},
L: \citet{Lopez:2014fo},
K: \citet{Katsuda:2018gs}, 
P: \citet{Park:jf},
X: \cite{Xi:fj},
\end{table}
\end{landscape}


\section{Spitzer, Herschel and Ancillary data}

The {\it Spitzer Space Telescope} \citep{Werner:2004jt} and {\it Herschel Space Observatory} \citep{Pilbratt:2010p29312} surveyed the SMC at mid- and far-infrared wavelengths.
The Spitzer imaging survey of the SMC project,  SAGE-SMC  \citep{Gordon:2011jq},
used in total 7 bands, IRAC 3.6, 4.5, 5.8 and 8.0\,$\mu$m \citep{Fazio:2004p8080} and MIPS 24, 70 and 160\,$\mu$m \citep{Rieke:2004ku}. 
The HERITAGE is the {\it Herschel} Magellanic Cloud survey 
\citep{Meixner:2010kj, Meixner:2013kr} at 100, 160\,$\mu$m from PACS  \citep{Poglitsch:2010p28964}, and 250, 350 and 500\,$\mu$m from SPIRE \citep{Griffin:2010p29303}.
The HERITAGE surveyed the Magellanic Clouds with PACS and SPIRE Parallel mode, which has a choice of only two PACS filters (100, 160\,$\mu$m) out of three available filters, and whose fast scan speed limits its PACS spatial sampling rate and sensitivities.
For some SMC SNRs, dedicated {\it Herschel} pointed observations are available (Proposal IDs: OT1$_-$jsimon01$_-$1 and OT2$_-$ksandstr$_-$3), and these observations have better spatial sampling and better sensitivities.
HERITAGE included these pointed observations in the data reduction. 
In the case of SNR J0104.0$-$7202 (or 1E102.2$-$7219) from Proposal ID of OT2$_-$ksandstr$_-$3, an additional PACS 70\,$\mu$m image had been taken \citep{2018AAS...23124108L}. 
The  PACS  70\,$\mu$m image has higher angular resolution (FHWM of the point spread function is 5.46"$\times$5.76" at medium scan rate according to the PACS Observer Manual)  than the  {\it Spitzer} MIPS 18" resolution at 70\,$\mu$m, so that we used PACS 70\,$\mu$m images.
Filter bands and sensitivities are summarised in Table\,\ref{sensitivities}.

From all available Spitzer/IRAC bands, we give more weight to IRAC 8.0\,$\mu$m, as SNRs tend to stand out most at 8.0\,$\mu$m compared to other IRAC bands \citep{Reach:2006cl}.
IRAC 8.0\,$\mu$m represents [Ar {\small II}], H$_2$ lines and synchrotron emission, with a possible contribution of dust emission.
The SNR emission at MIPS 24\,$\mu$m band is usually dominated by dust emission \citep{PinheiroGoncalves:2011ff}, however, lines such as 
26\,$\mu$m [Fe {\small II}]  and [O {\small IV}] \citep{Bouchet:2006p2168, Rho:2008p414, Sandstrom:2009jxa}  might contribute to the brightness.

Although both filters are called `70\,$\mu$m band', the MIPS 70\,$\mu$m band covers 61 to 80\,$\mu$m, while the PACS 70\,$\mu$m band covers a slightly different 60--85\,$\mu$m.
The PACS 100\,$\mu$m band ranges from 85 to 125\,$\mu$m.
Flux from SNRs at these bands are dominated by dust emission in general, as found in grey body profiles in the far-infrared regime \citep{Gomez:2012fm,Chawner:2019dn}.
Contaminations of lines (63\,$\mu$m [O {\small I}] and 88\,$\mu$m [O\,{\small III}]) are mostly negligible:
for example, the fraction of line contaminations are estimated at 5\,\% and 6\,\%--13\,\% at PACS 70 and 100\,$\mu$m bands respectively for the Crab Nebula \citep{Gomez:2012fm}, and only 
6\,\% and 2\,\% for Galactic SNR Kes 75 \citep{temim:ve}.

X-ray images are taken from the Chandra Catalogue of Magellanic Cloud Supernova Remnants, if available, otherwise
the {\it XMM-Newton} survey of the Magellanic Clouds \citep{Haberl:to} is used. 
Unless otherwise stated in Figure captions, X-ray images shown are from {\it XMM-Newton} 0.2--1.0\,keV, a band that indicates the thermal spectrum SNRs, with some contributions from lines, such as oxygen and Fe \citep{Maggi:2019wq}.
We use these {\it XMM-Newton} images of SNRs, having previously been studied by \citet{Haberl:to} and \citet{Maggi:2019wq}, to guide our astrometry for related emission in infrared wavelengths.


Optical H$\alpha$,  [O\,{\small III}]\,5007\,\r{A} and [S{\small II}]\,6716 and 6731\,\r{A} images from the Magellanic Cloud Emission Line Survey (MCELS) \citep{Smith:tza} are also used to identify the structure of SNRs when combined with the infrared images.
 MCELS surveyed the Magellanic Clouds, with sampling of 2.4 arcsec pixels and 3--4 arcsec resolution \citep{Smith:tza} and has already been used to identity emission from SNRs, planetary nebula, and H{\small II} regions \citep[e.g.][]{Reid:tf}.
The  MCELS images have also been used to identify optical counterparts of SNRs detected in X-ray and radio bands \citep{Maggi:2019wq}. Importantly, these images can also reveal the locations of H{\small II} regions, thereby helping to distinguish SNRs from H{\small II} regions in our infrared images.

Radio continuum images of the SMC that we used are from the Australian Square Kilometre Array Pathfinder (ASKAP) \citep{Joseph2019}, that covered at two bands at 960\,MHz and 1320\,MHz. The bean sizes are 30"$\times$30" and 16.3"$\times$15.1", and the sensitivities are 186 and 165\,$\mu$Jy per beam and respectively.
We used a 1320\,MHz image to trace the synchrotron radiation from SNRs.


\begin{table*}
\begin{center}
\caption{The SMC surveys used in this study \label{sensitivities} }
\begin{filecontents*}{filters.csv}
Project,Band,Wav,D Wav,Res,Sensitivity,Reference
MCLES,H$\alpha$,0.6563,0.03,3--4,3$\times10^{-17}$~erg~cm$^{-2}$~s$^{-1}$~arcsec$^{-2}$,S
,[S {\small II}],0.6724,0.05,3--4,,S
,[O {\small III}],0.5007,0.04,3--4,,S
Spitzer,3.6 $\mu$m,,,2,0.04 MJy sr$^{-1}$ (5$\sigma$),B
,4.5 $\mu$m,,,2,0.05 MJy sr$^{-1}$ (5$\sigma$),B
,5.8 $\mu$m,,,2,0.2 MJy sr$^{-1}$ (5$\sigma$),B
,8.0 $\mu$m,,,2,$\sim$0.1 MJy sr$^{-1}$ (5$\sigma$),G
,24 $\mu$m,,,6,0.3 MJy sr$^{-1}$ (5$\sigma$),G
,70 $\mu$m,,,18,2.5 MJy sr$^{-1}$ (5$\sigma$),G
Herschel,70 $\mu$m,,,5.2--5.8$^{1}$,$^{2}$,
,100 $\mu$m,,,6.7$\times$6.9,85--90 MJy sr$^{-1}$ (10 $\sigma$)$^{3}$,M
,160 $\mu$m,,,10.7$\times$12.1,32--50 MJy sr$^{-1}$ (10 $\sigma$)$^{3}$,M
,250 $\mu$m,,,18.2,6 MJy sr$^{-1}$ (10 $\sigma$),M
,350 $\mu$m,,,24.9,3 MJy sr$^{-1}$ (10 $\sigma$),M
,500 $\mu$m,,,36.3,2 MJy sr$^{-1}$ (10 $\sigma$),M
ASKAP,960 MHz (32 cm),,,30$\times$30,186 $\mu$Jy beam$^{-1}$ (RMS),J
,1320 MHz (23cm),,,16.3$\times$15.1,165 $\mu$Jy beam$^{-1}$ (RMS),J
\end{filecontents*}
\csvreader[tabular= l l c c cc ll l c l l l c l,
				table head=\hline Project & Band &  Res (") & Sensitivity & Ref \\ \hline\hline, 
				table foot=\hline ] 
				{filters.csv}
				{} 
				{\csvcoli & \csvcolii  & \csvcolv & \csvcolvi & \csvcolvii    } 
\\
\end{center}
Note:
$^{1}$: PACS angular resolutions depend on mapping scan speed, which varies.
$^{2}$: PACS 70\,$\mu$m was not used by the HERITAGE project, but used for localised imaging for some dedicated area. Sensitivity varies on the projects.
$^{3}$: HERITAGE sensitivities. Dedicated local mapping might have better sensitivities.
References:
B: \cite{Bolatto:2007hha} \cite{Smith:tza}
G: \cite{Gordon:2011jq}
J: \cite{Joseph2019}	
M: \cite{Meixner:2013kr}	
S: \cite{Smith:tza}	
\end{table*}


\section{Detections of infrared emission from supernova remnants}

Following the classification of detection levels from 
\citet{Reach:2006cl}, \citet{PinheiroGoncalves:2011ff} and \citet{Chawner:2019dn, 2020MNRAS.493.2706C},
we classified Infrared (IR)  detection (either {\it Spitzer}  or both {\it Spitzer} and {\it Herschel}) from SNRs into four levels, described in Table.\,\ref{category}.
The detection levels of individual SNRs are summarised in Table\,\ref{detectionlevel}. This should refer to Table 2.


\begin{table}
	\centering
	\caption{The description of detection levels.}
	\label{category}
	\begin{tabular}{ll} 
		\hline
		Level & Description \\
		\hline
		Detection & IR emission is clearly correlated with radio  \\
		   & or X-ray structure of a SNR and can be  \\
		   & distinguished from ISM emission \\
		Possible detection & IR emission in the region of the \\ 
		   & SNR, potentially related to radio or X-ray  \\
		   & structure but confused with ISM emission \\
		Unlikely detection & Detection of IR emission which is \\
		   & probably unrelated to the SNR \\
		No detection \\
				\hline
	\end{tabular}
\end{table}


\subsection{Descriptions of Individual SNRs}

 We present {\it Spitzer} and {\it Herschel} images of SNRs and their surroundings, with comparison X-ray, optical and radio images.
The default X-ray images are {\it XMM Newton} 0.2--1.0\,keV images, but occasionally other bands (1.0--2.0 or 2.0--4.5\,keV) or {\it Chandra} images are used.
The optical images are mostly  H$\alpha$ but an [O\,{\small III}] image is used for SNR J0104.0$-$7202, as this oxygen-rich SNR allows more detail to be seen in [O\,{\small III}].
The default radio images are 1320\,MHz but some figures display at 960\,MHz.
All alternations of the selected bands are mentioned in the figure captions.

 In this Section, only detections and possible detections (Table\,\ref{table-detection}) are included, while unlikely, non-detections and SNR candidates/non-SNRs are included in on-line material (Sections \ref{sect-non-detection} and \ref{sect-non-SNR}).


\subsubsection{SNR J00409$-$7337, B0039$-$72.53, DEM S5: detection}

\begin{figure*}
  \begin{minipage}[c]{0.70\textwidth}
    \includegraphics[width=\textwidth]{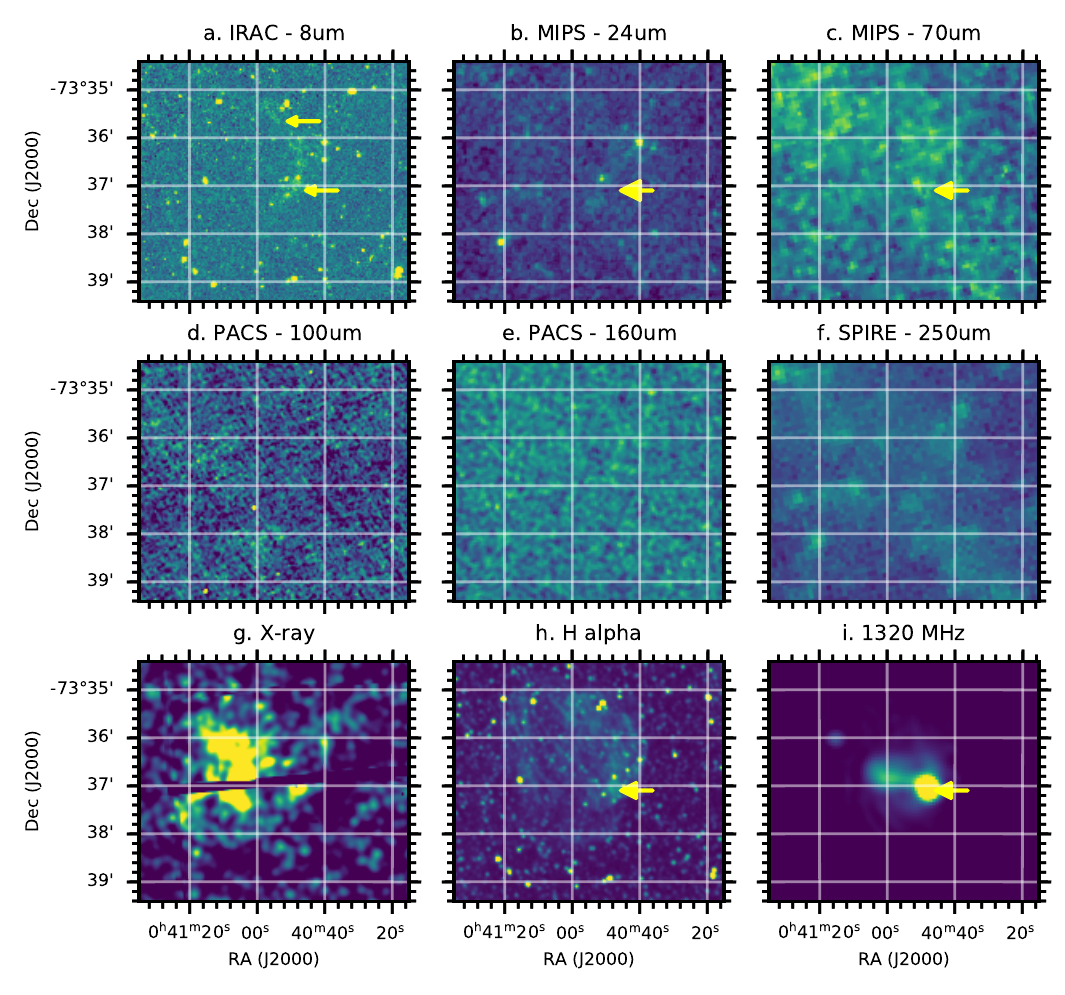}
  \end{minipage}\hfill
  \begin{minipage}[c]{0.30\textwidth}
    \caption{SNR J0040.9$-$7337 (B0039$-$73.53, DEM S5): detection. 
    The pulsar powered bow shock is clearly seen in the 1320\,MHz radio image (i), and that part of the SNR is brightest in the H$\alpha$ image. 
    The corresponding emission is found at 8\,$\mu$m. 
    There is a discrete source detected across 8, 24, and 70\,$\mu$m, as well as H$\alpha$, with a very close correlation to a strong X-ray source. Whether this is a part of the pulsar wind nebula or the pulsar itself requires further confirmation.
    The filaments extend from this bow shock to the north, and part of the filaments is detected at 8\,$\mu$m, as if it is an arc, with a hint at 70\,$\mu$m.
     } \label{fig-00409}
  \end{minipage}
\end{figure*}

 A pulsar-powered bow shock was discovered in SNR J00409$-$7337 \citep{Alsaberi:2019gl}, using the Australia Telescope Compact Array (ATAC) at 2048\,MHz and IRAC 8\,$\mu$m. This pulsar is a bright point source in the 1320\,MHz image (Fig.\,\ref{fig-00409}.i).
The bow shock was detected as faint filaments both north and south of the pulsar in H$\alpha$ and IRAC 8\,$\mu$m images, as reported by \citet{Alsaberi:2019gl}.
Based on the IRAC mid-IR emission diagnostic, this bow shock is thought to trace shock-excited molecular hydrogen lines
\citep{Alsaberi:2019gl}.


The brightest discrete region of the bow shock in H$\alpha$ (Fig.\,\ref{fig-00409}.h) is also the brightest region in the 8\,$\mu$m image, however, it is not particularly strong at 24 and 70\,$\mu$m. The detection of a strong X-ray point source at this same position and the presence of the pulsar or PWN was suggested from X-ray spectra \citet{Alsaberi:2019gl}. Presently the source of this IR emission is unclear, either pulsar/PWN or part of the bow shock. 


The most prominent feature of the SNR is the filamentary arc extending north/south and curving around at each end. This can clearly be seen to correlate between the H$\alpha$ and IRAC 8\,$\mu$m images.
Emission possibly associated with this arc is also found at 70\,$\mu$m but less so at 160\,$\mu$m.


\subsubsection{SNR J0046.6$-$7307, B0044-73.4, DEM S32: possible detection}

\begin{figure*}
  \begin{minipage}[c]{0.70\textwidth}
    \includegraphics[width=\textwidth]{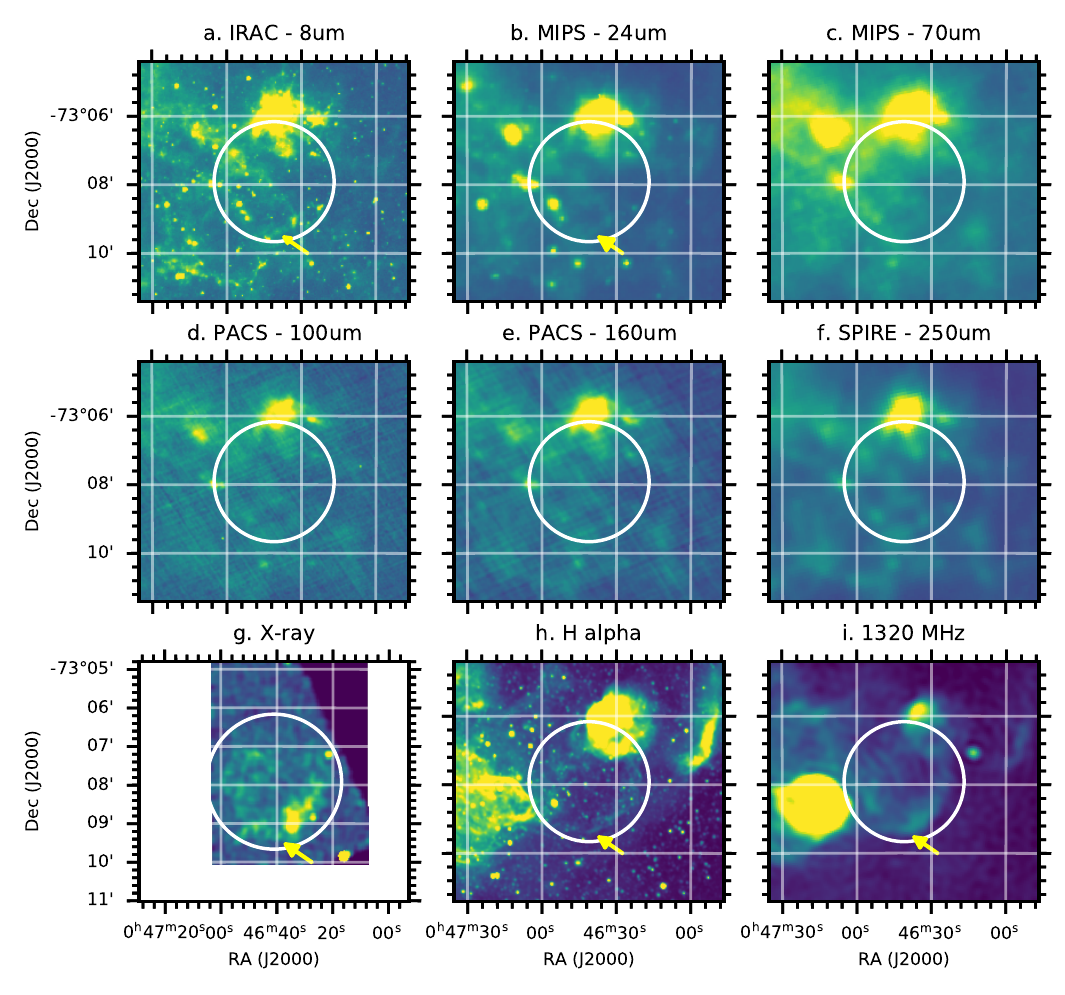}
  \end{minipage}\hfill
  \begin{minipage}[c]{0.30\textwidth}
    \caption{SNR J0046.6$-$7307 (B0044-73.4, DEM S32): possible detection. 
    H$\alpha$ and 1320\,MHz radio images clearly indicate shell structure, the south-west and potentially south-east part of the shell are detected in {\it Spitzer} 8 and 24\,$\mu$m images and possibly PACS and SPIRE 100--250\,$\mu$m. 
    The brightest clouds found in the north of the shell are molecular clouds. 
    The bright radio source just outside of the white circle in the south-east is SNR\,J0047.2$-$7308.
    A 0.3--10.0\,keV band from  {\it Chandra} is used for the X-ray image (g). 
         } \label{fig-00466}
  \end{minipage}
\end{figure*}

SNR\,J0046.6$-$7307 was initially found in an {\it XMM-Newton} image \citep{vanderHeyden:2004ee} and is a candidate core-collapse SNR \citep{vanderHeyden:2004ee}. 
Together with SNR\,J0047.2$-$7308 (IKT 2) and  SNR\,J0047.5$-$7306, it is one of the three SNRs located in a large complex H{\small II} region N19 \citep{Davies:vz, Maggi:2019wq}.
SNR\,J0046.6$-$7307 is a shell shaped SNR, as seen in H$\alpha$ and the 1320\,MHz radio image (Fig.\,\ref{fig-00466}) and is surrounded by at least eight molecular clouds \citep{Sano:2019jg}.
The bright emission found to the north of the SNR is the H{\small II} region NGC\,261.

\begin{figure}
	\includegraphics[trim={1.2cm 0.0cm 3.5cm 1cm},clip, width=1.0\columnwidth]{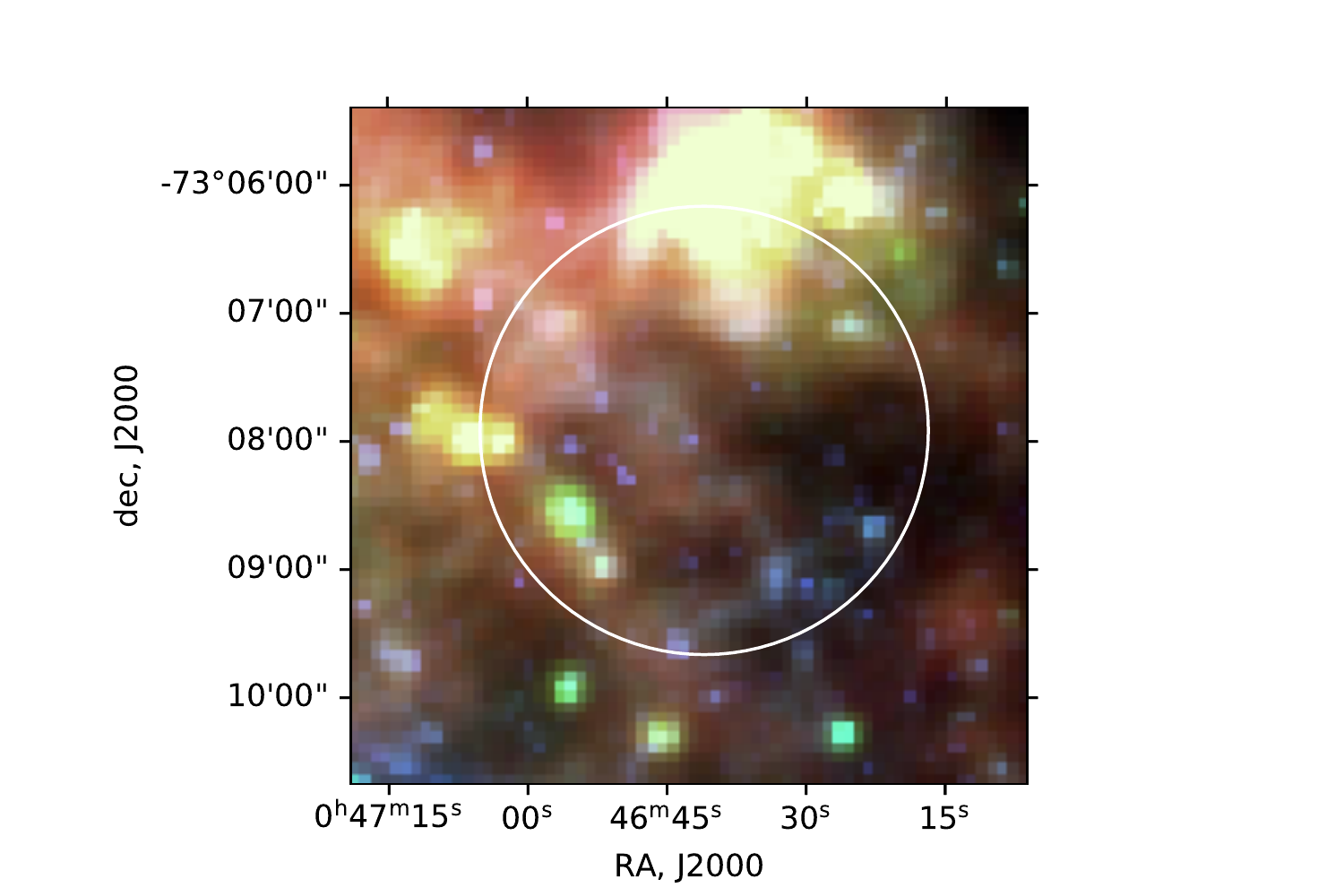}
    \caption{SNR\,J0046.6$-$7307 (B0044$-$73.4, DEM S32) three-colour image: 8\,$\mu$m (blue), 24\,$\mu$m (green), 70\,$\mu$m (red), indicating possible detection of the shell on south-east of the SNR, as traced by blue colour.}
    \label{fig-00466-3color}
\end{figure}

In {\it Spitzer} and {\it Herschel} images, part of the shell may be detected. 
The brightest part of the shell in the radio image is south-west, a position where faint emission can also be seen in the infrared images. 
At a minimum it is clearly detected at 8 and 24\,$\mu$m with associated emission probably detected at longer wavelengths. 
Since the majority of the arc comprising the shell in H$\alpha$ is slightly offset and a different shape to that found in 8\,$\mu$m, we prefer  to label this as a ``possible'' detection.
Fig.\,\ref{fig-00466-3color} shows an 8, 24, 70\,$\mu$m 3-color image of SNR\,J0046.6$-$7307, allowing the 8\,$\mu$m emission in south-west part of the shell to show prominently in blue, indicating relatively hotter dust than in the surrounding ISM. 
In Galactic studies, swept up dust in SNRs tends to be warmer than unrelated ISM dust \citep{Chawner:2019dn}, so it is still highly likely that this arc may be SNR related dust.
However, since the  potential of this feature being a be part of ISM filaments can not be totally excluded, it is classified as a ``possible detection".

Additionally, there is a diffuse cloud found crossing the centre of this SNR from north-east to south-west in all infrared images.
This  cloud is redder than the south-west shell (Fig.\,\ref{fig-00466-3color}), hence,
it is probably not associated with the SNR but rather the ISM. Indeed, no emission corresponding to this cloud is found in X-ray nor radio.

The south-east part of the shell contains point sources, and at least one of them is a star detected in the H$\alpha$ image. Since it is much brighter at 24\,$\mu$m than 8\,$\mu$m, compared with other stars within the field, it may be either a red star or a dust-enshrouded star.
If these point sources are part of molecular cloud D, according to \citet{Sano:2019jg}, which surrounds the SNR,
they could be recently formed stars, triggered by the compression of the gas from the SNR.



\subsubsection{SNR J0049.1$-$7314, B0047-73.5, DEM S49, IKT 5: unlikely detection but bow shock detected within a field}

Although we list this SNR as an unlikely detection, a notable detection of a bow shock allows it to be included in this section.

\begin{figure*}
  \begin{minipage}[c]{0.70\textwidth}
    \includegraphics[width=\textwidth]{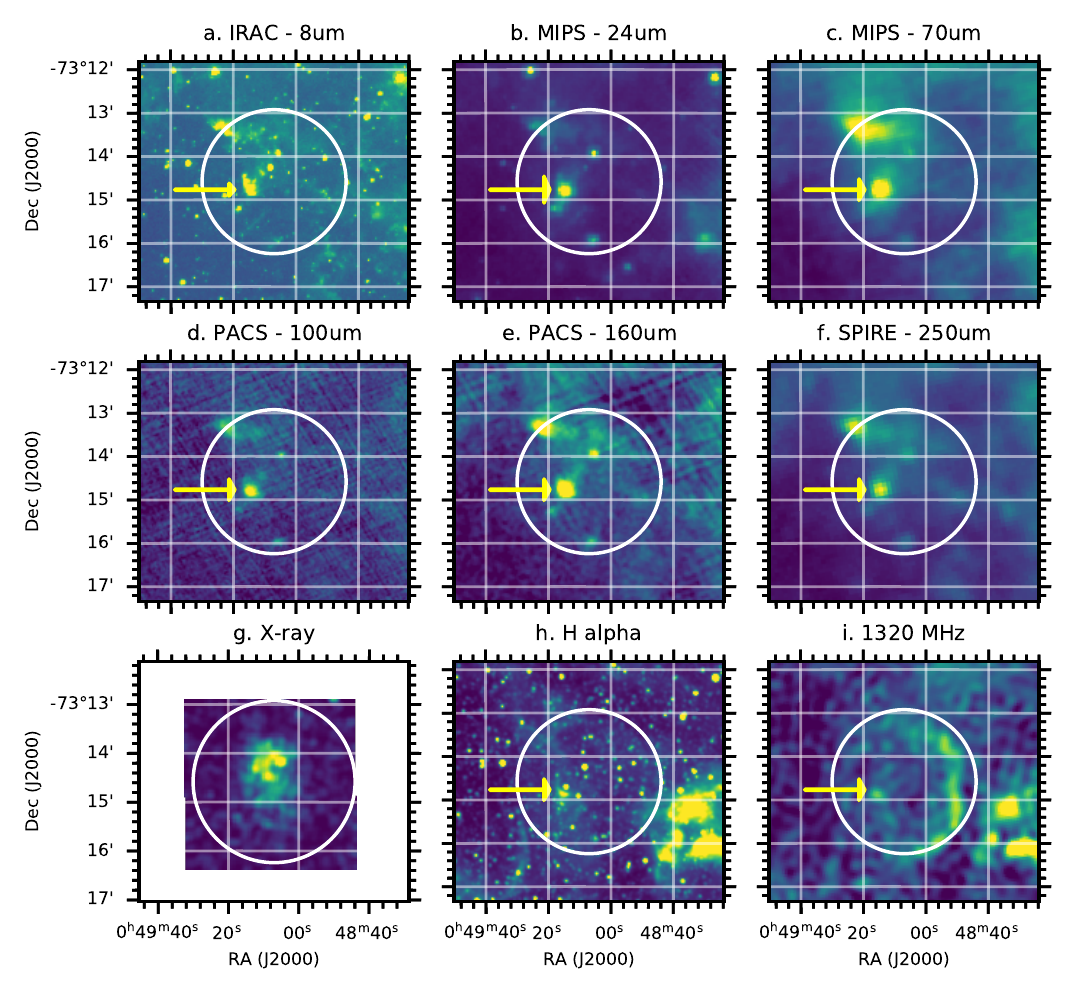}
  \end{minipage}\hfill
  \begin{minipage}[c]{0.30\textwidth}
    \caption{SNR J0049.1$-$7314, B0047-73.5, DEM S49, IKT 5: unlikely detection of the SNR itself, but it has a bow shock from an unknown source in the field. 
    Panel (g) is  a {\it Chandra} image in the 0.3--2.1 keV band.
       } \label{fig-00491}
  \end{minipage}
\end{figure*}

This SNR shows its shell type structure in the radio image, but emission from iron-rich ejecta is found in the centre in the X-ray image, suggesting type Ia origin \citep{Maggi:2019wq}.
There is no detection of this shell in the infrared images (Fig.\ref{fig-00491}).

\begin{figure}
	\includegraphics[trim={1.2cm 0.2cm 3.5cm 1.0cm},clip, width=\columnwidth]{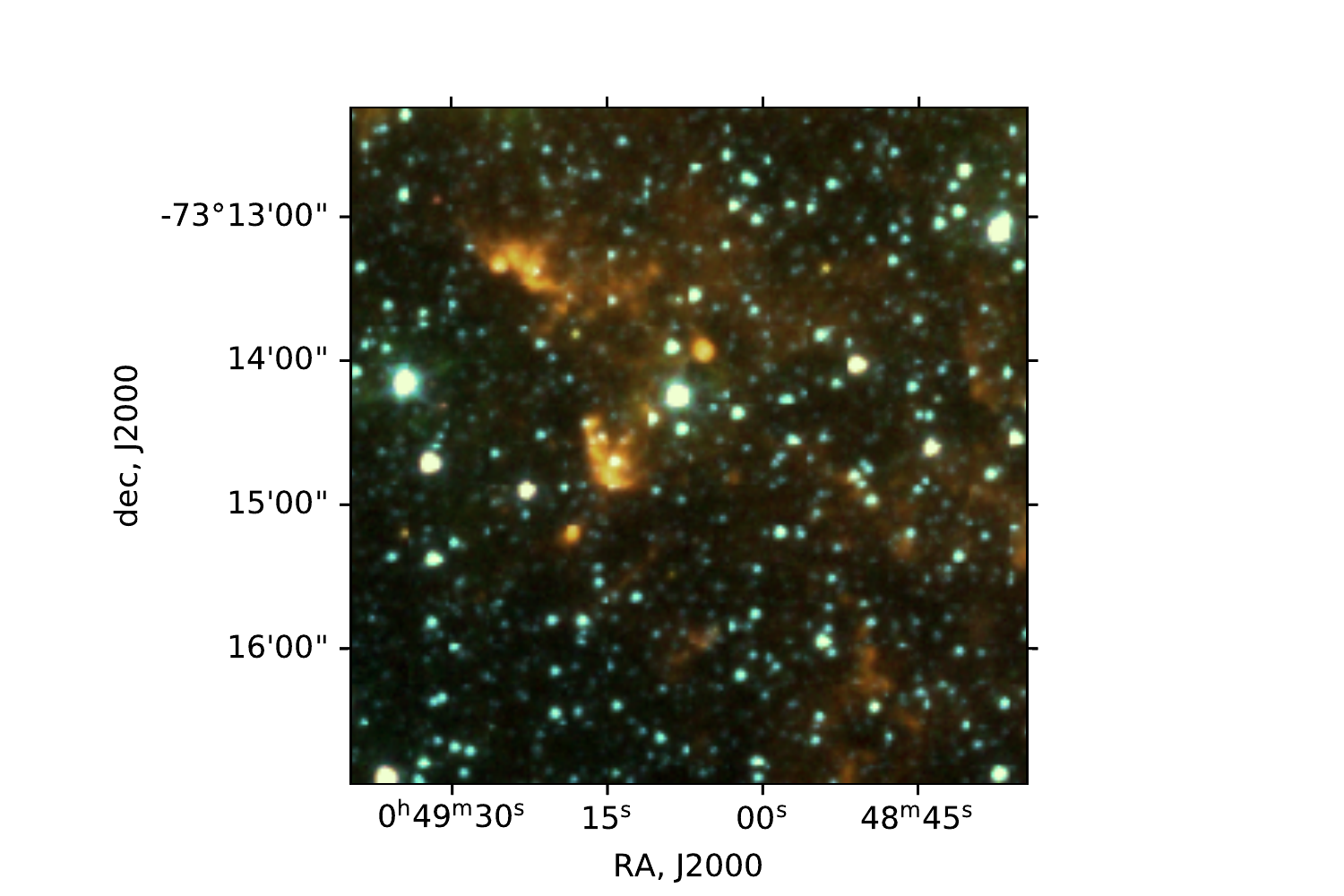}
    \caption{
    The 3.6 (blue), 5.8 (green) and 8.0\,$\mu$m  (red) three color composite image of SNR J0049.1$-$7314 field. Unrelated bow shock and nebulosity are detected in red colour.}
    \label{fig-00491color}
\end{figure}

A crescent-shaped emission arc next to a point source is detected on the east side of the white circle in the H$\alpha$ and 8\,$\mu$m images (Fig.\ref{fig-00491} h).
This is a bow shock with the crescent and point source also found in the radio image.
The crescent is not spatially resolved at longer wavelengths but still detected across all infrared images.
Fig.\,\ref{fig-00491color} shows a three-colour image of the region, composed of 3.6, 5.8 and 8.0\,$\mu$m.
This image clearly indicates the bow shock, as well as some nebulosity extending towards the north-east.

The source of this bow shock is most likely the outflow
from the young stellar object (YSO) candidate S3MC J004905.49$-$731356.31 
or possibly another YSO candidate S3MC J004914.84$-$731448.37 
\citep{Bolatto:2007hha}, both of which sit inside the bow shock.
Alternatively, another possible source is a red supergiant candidate SSTISAGEMA J004908.72$-$731355.0 
  \citep{Seale:2014kx}.
The bow shock itself is recorded as YSO candidate S3MC J004914.84$-$731448.37 by \citet{Bolatto:2007hha}.

The YSO candidate, S3MC J004905.49$-$731356.31  \citep{Bolatto:2007hha} appears to be a bright red point source in Fig.\,\ref{fig-00491color} and is detected across all {\it Spitzer} and {\it Herschel} bands (Fig.\ref{fig-00491}).
This source is located in a favourable position for causing both the bow shock and the nebulosity located to the north-east of that area (Fig.\,\ref{fig-00491color}). 
High-mass YSOs can cause bow shocks associated with their jets, however, in the LMC at 50\,kpc, the scale of the jets are recorded as less than 1\,arcsec \citep{McLeod:2018kd}.
In contrast, the detected bow shock in the SMC is more than 60\,arcsec away from the source, i.e. a much larger distance.
It could be associated with a high-mass star, but it needs to be confirmed if such a large jet is possible.
An alternative possibility is the red supergiant candidate SSTISAGEMA J004908.72$-$731355.0   \citep{Seale:2014kx}, which is seen as a faint green point source to the east of S3MC J004905.49$-$731356.31  in Fig.\,\ref{fig-00491color}.
Similar to  S3MC J004905.49$-$731356.31, its location favours it as a source of the bow shock given that mass loss from red-supergiants can trigger bow shocks, as observed in $\alpha$ Ori \citep{Ueta:2008p19184, Decin:bf}.
The bow shock in $\alpha$ Ori was about 7\,arcmin away from the star at a distance of 197\,pc \citep{Decin:bf}, corresponding to only about 1.4\,arcsec at SMC distance.
Therefore the scale of this SMC bow shock seems to be too large to result from mass lost from a red-supergiant.
The final possibility is the YSO candidate, S3MC J004914.84$-$731448.37, which is immediately behind the bow shock. It appears however that the star is too close to the bow shock, and the opening angle of the bow does not favour this star as a source. 
At this point the source of this bow shock remains unclear.

\subsubsection{SNR J0052.6$-$7238, B0050-728, DEM S68SE: detection}

\begin{figure*}
  \begin{minipage}[c]{0.70\textwidth}
    \includegraphics[width=\textwidth]{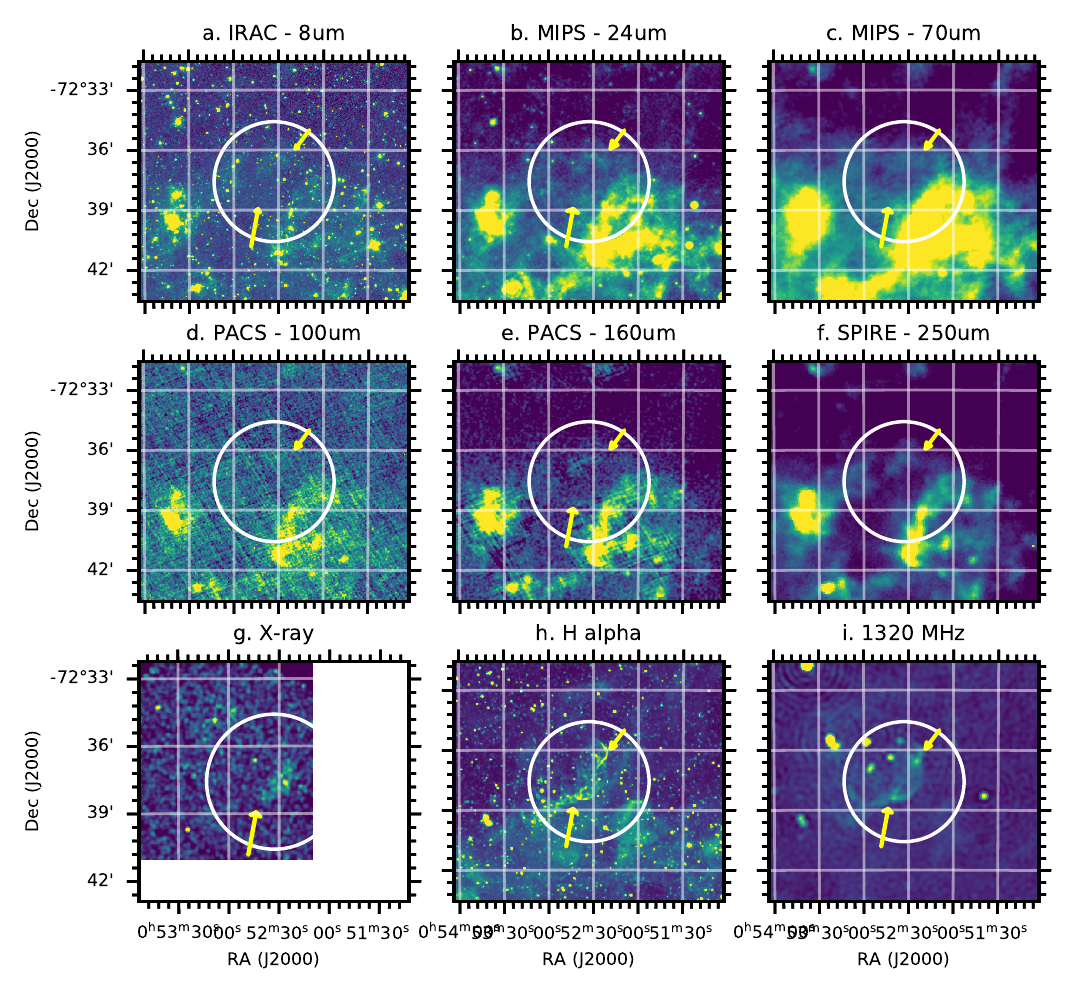}
  \end{minipage}\hfill
  \begin{minipage}[c]{0.30\textwidth}
    \caption{SNR J0052.6$-$7238, B0050-728, DEM S68SE: detection. 
    Part of shells or filaments  have been detected in the north, and potentially in the south at 8\,$\mu$m, 24\,$\mu$m and 70\,$\mu$m. 
    Panel (g) is  a {\it Chandra} image in the 0.3--2.1 keV band.
       } \label{fig-00526}
  \end{minipage}
\end{figure*}

SNR J0052.6$-$7238 the radio and H$\alpha$ images show several crescent-shaped emission features.
An intersecting double crescent feature can be clearly seen in the H$\alpha$ and 24\,$\mu$m images. The two crescents in the north are indicated with an arrow in Fig.\,\ref{fig-00526} b.
Interestingly, in H$\alpha$ the north-west crescent is brighter than the one just south of it, whereas at 24\,$\mu$m, the north-west crescent is much fainter than the one south of it.
Diffuse ISM emission is detected around the crescents at 8, and 70--250\,$\mu$m, but the crescent is not distinct at 8\,$\mu$m.

Additionally, a multitude of crescents lay out in the south of the SNR, indicated by an arrow in the H$\alpha$ and radio images of Fig.\,\ref{fig-00526}.
These crescents are probably detected as a diffuse cloud at 24\,$\mu$m.

\begin{figure}
	\includegraphics[width=1.3\columnwidth]{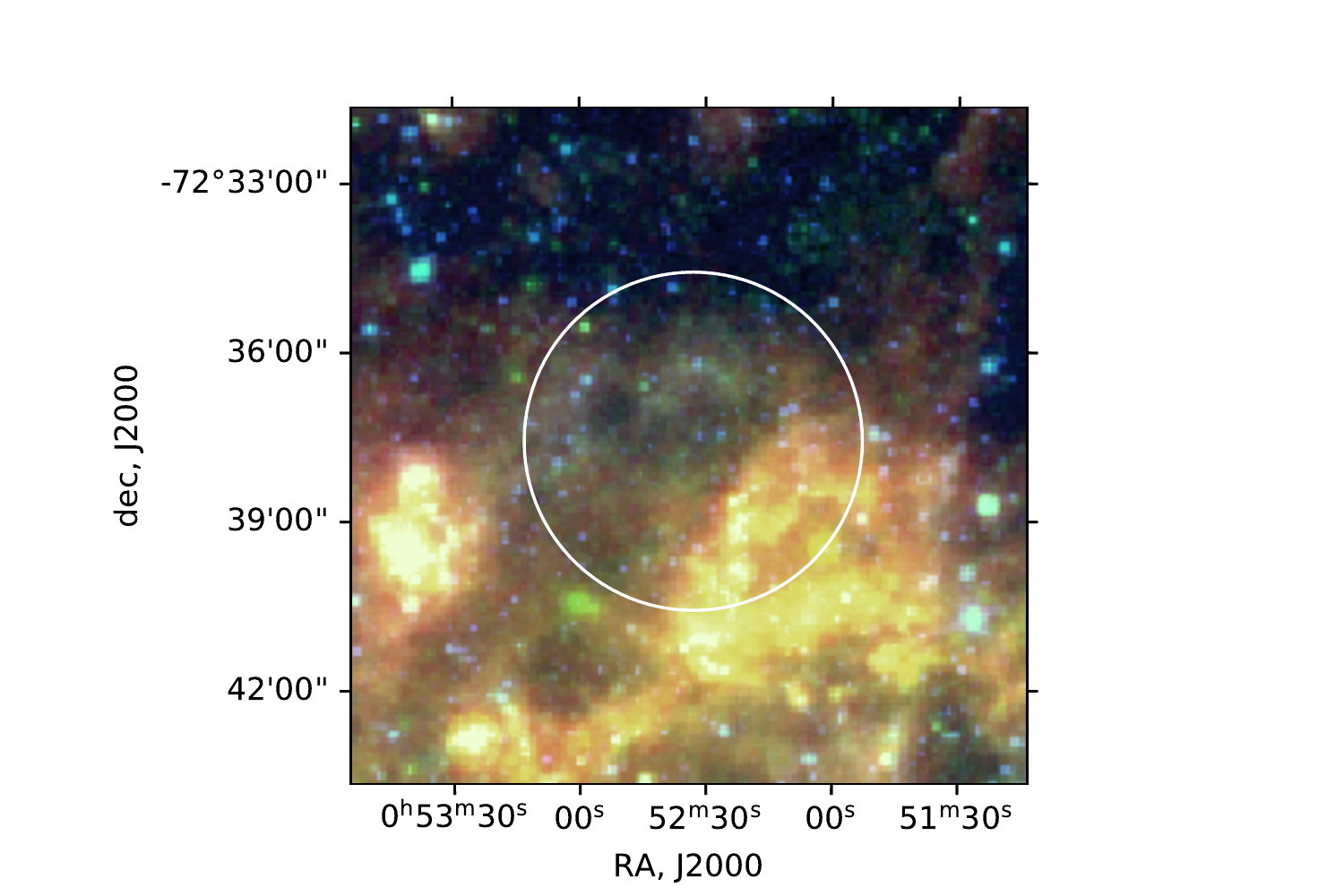}
    \caption{SNR J0052.6$-$7238, B0050-728, DEM S68SE. The 8 (blue), 24 (green) and 70 (red) \,$\mu$m three colour composite image - probably detected.}
    \label{fig-00475-3color}
\end{figure}



\subsubsection{SNR J0059.4$-$7210, B0057-72.2, DEM S103, IKT 18: possible detection} \label{subsec-0059.4}


This SNR is closely located in N66 H{\small II} region.
Although it is challenging to separate SNR emission from the intervening H{\small II} region at optical wavelengths, the SNR is clearly isolated in X-ray and radio images \citep{Ye:1991hl, Maggi:2019wq}.
Since the H{\small II} region is overwhelmingly bright in H$\alpha$ and infrared, it is not easy to separate the SNR using these wavelengths.
It is possible, however, to remove thermal emission estimated from the H$\alpha$ flux from the radio-continuum images \citep{Reid.2006}. 
The remaining non-thermal emission allows the identification of the SNR embedded in the dense N66 region.

\begin{figure*}
  \begin{minipage}[c]{0.70\textwidth}
    \includegraphics[width=\textwidth]{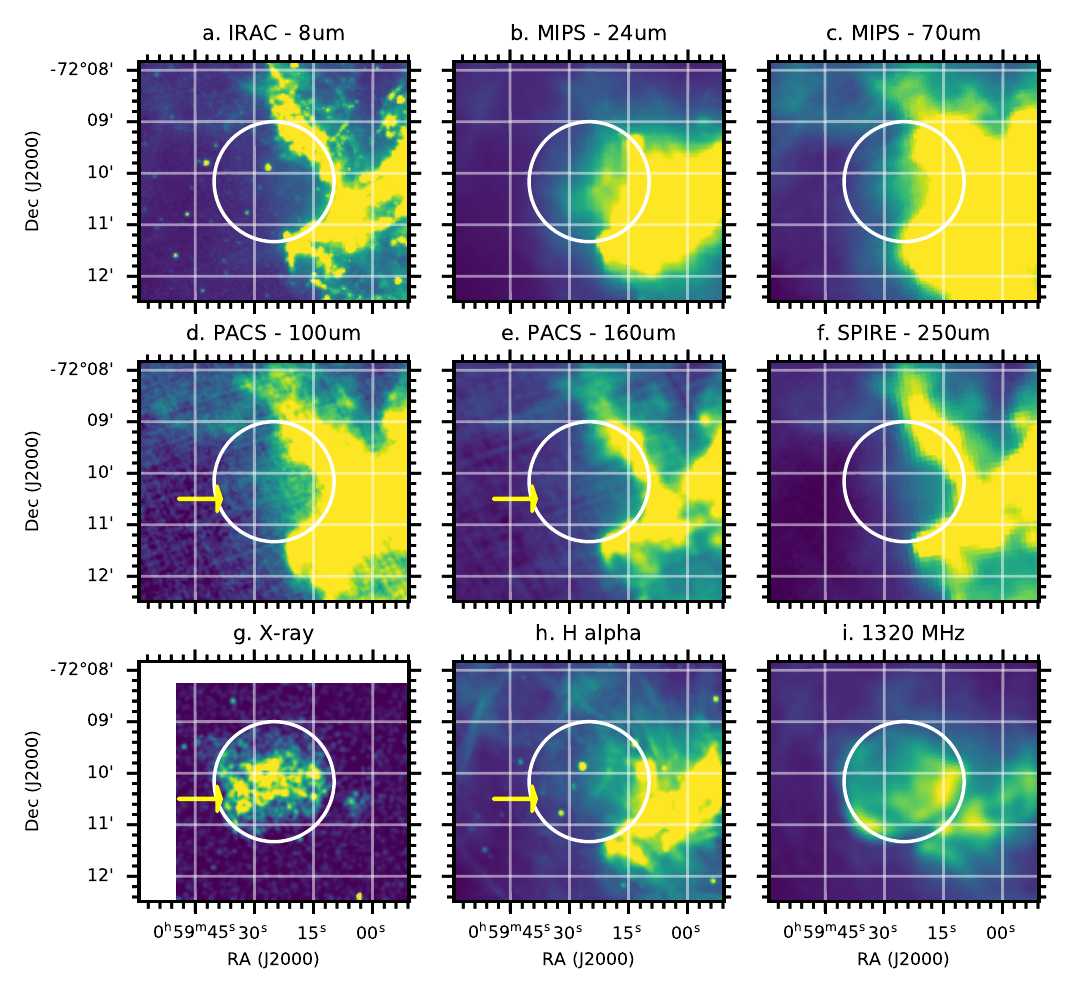}
  \end{minipage}\hfill
  \begin{minipage}[c]{0.30\textwidth}
    \caption{SNR J0059.4$-$7210, B0057-72.2, DEM S103, IKT 18: possible detection.  
    Panel (g) is a {\it Chandra} image in the 0.3--2.1 keV band.
           } \label{fig-00594}
  \end{minipage}
\end{figure*}

\begin{figure}
	\includegraphics[width=1.3\columnwidth]{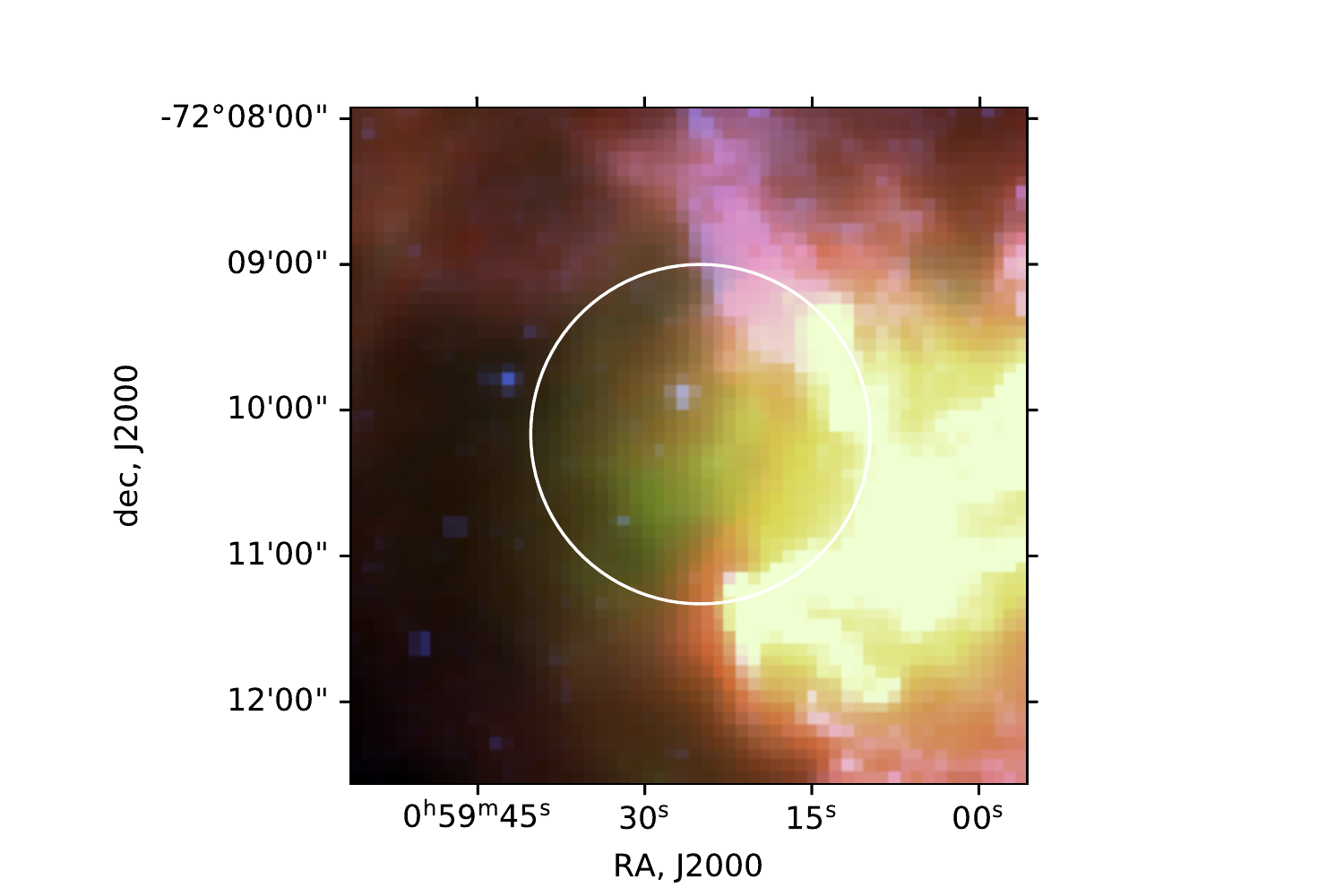}
    \caption{SNR J0059.4$-$7210, B0057-72.2, DEM S103, IKT 18. The 8\,$\mu$m (blue), 24\,$\mu$m (green) and 70\,$\mu$m (red)  three colour composite image with a possible detection of filaments, seen as green  near-circular' filaments within the white circle.}
    \label{fig-00594-3color}
\end{figure}

An arc, detected inside the white circle in the H$\alpha$ image (Fig.\,\ref{fig-00594}), could be associated with SNR J0059.4$-$7210 \citep{Danforth:ff, Rubio:uk}. 
The dynamical  age is estimated to be about $2\times10^5$\,years \citep{Danforth:ff, Nota:dk}, using expansion velocities measured in the UV spectrum.
Alternatively, this arc could potentially be a bubble formed by a mass-loss wind from Wolf-Rayet star, HD\,5980 \citep{HeydariMalayeri:ka}, which is seen as a bright point source in the H$\alpha$ image (Fig.\,\ref{fig-00594} h).
This arc is enlarged in Fig\,\ref{fig-00594-3color}, to emphasise the emission in 24\,$\mu$m which has been assigned a green colour. 
It can be seen that the 24\,$\mu$m  emission forms a near-circle, following inside the white circular line until it extends away to south east.
Because this arc is visible, we categorise the SNR as a possible infrared detection, however, the arc could be associated with a Wolf-Rayet star rather than the SNR.

The IR image gives the impression that as the SNR expands, it is compressing the surrounding ISM gas, creating a void where only infrared emission between 24 and 100$\mu$m appears survive.
That compression has resulted in a `gas deficiency' in the north-east of N66, whereas in the western side of N66, the H{\small II} region is gradually transiting towards a PDR (photon dominated region) \citep{Rubio:uk}.
Detailed analysis of this region is presented in Sect\,\ref{DEMS103-analysis}.

\begin{figure*}
  \begin{minipage}[c]{0.70\textwidth}
   \includegraphics[width=\textwidth]{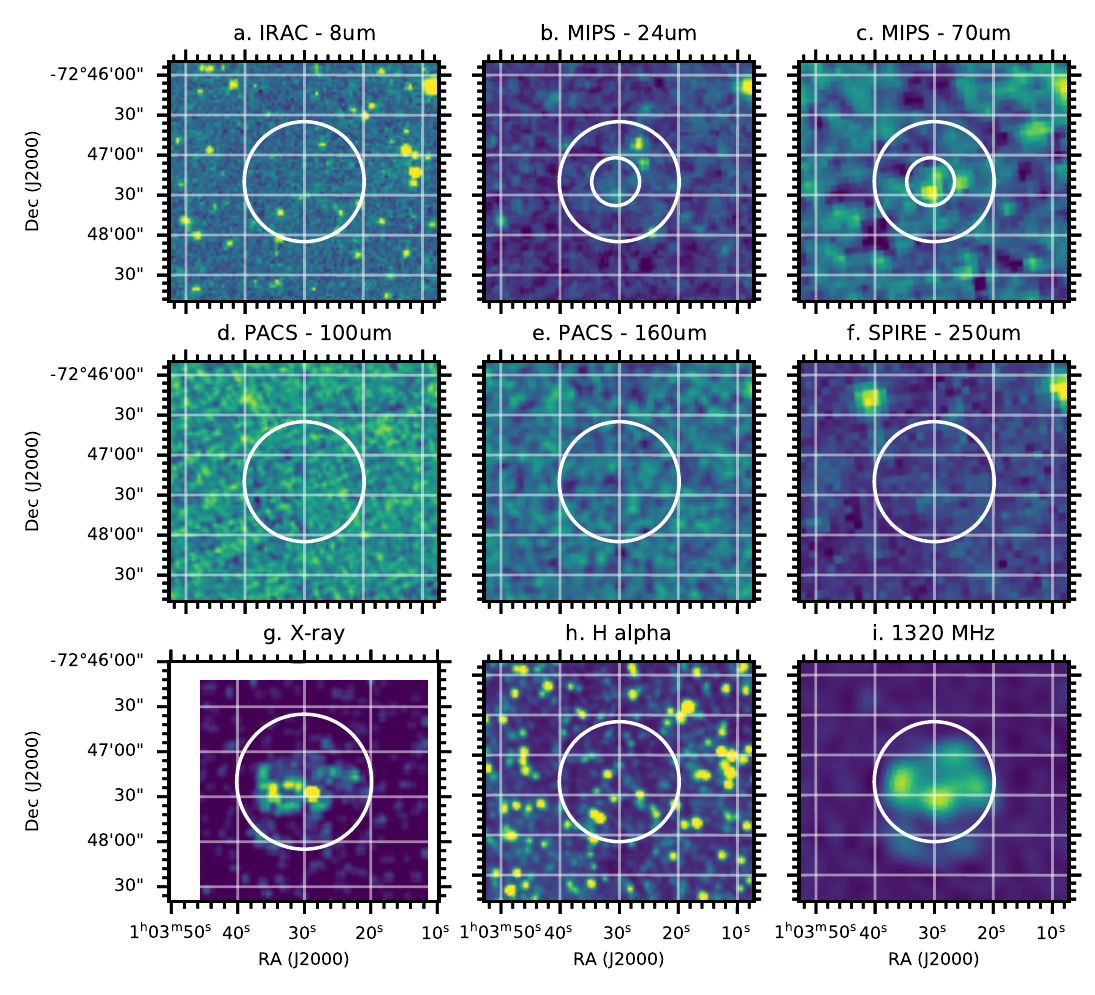}
  \end{minipage}\hfill
  \begin{minipage}[c]{0.30\textwidth}
    \caption{SNR J0103.5$-$7247, HFPK 334: possible detection.  
     Panel (g) shows  a {\it Chandra} image in the 0.44--1.17 keV band
            \label{fig-01035}}
  \end{minipage}
\end{figure*}

\subsubsection{SNR J0103.5$-$7247, HFPK 334: possible detection}

SNR J0103.5$-$7247 contains a bright point source seen in X-ray and radio images, though it is largely disputed whether this source is associated with the SNR or not  \citep{Crawford:2014ew}.
If the point source is associated with the SNR, it could be a compact source or pulsar wind nebula. 
 This point source is detected at 8, 24, 70 and 250\,$\mu$m (Fig.\,\ref{fig-01035}), though it is very faint at 8 and 250\,$\mu$m.
 This SNR is classified as a possible detection.



\subsubsection{ SNR J0103.6$-$7201: possible detection}

SNR J0103.6$-$7201  is a relatively new SNR reported by {\it XMM} \citep{Maggi:2019wq} and H$\alpha$ \citep{Gvaramadze:2019de}.
There is a point source detected in the 8\,$\mu$m image, and it may be a  counterpart to the Be X-ray binary   \citep{Gvaramadze:2019de}. 
This Be X-ray binary has been associated with the neutron star of SXP\,1323 that could have formed at the SN explosion about 40,000 years ago.

\begin{figure*}
  \begin{minipage}[c]{0.70\textwidth}
    \includegraphics[width=\textwidth]{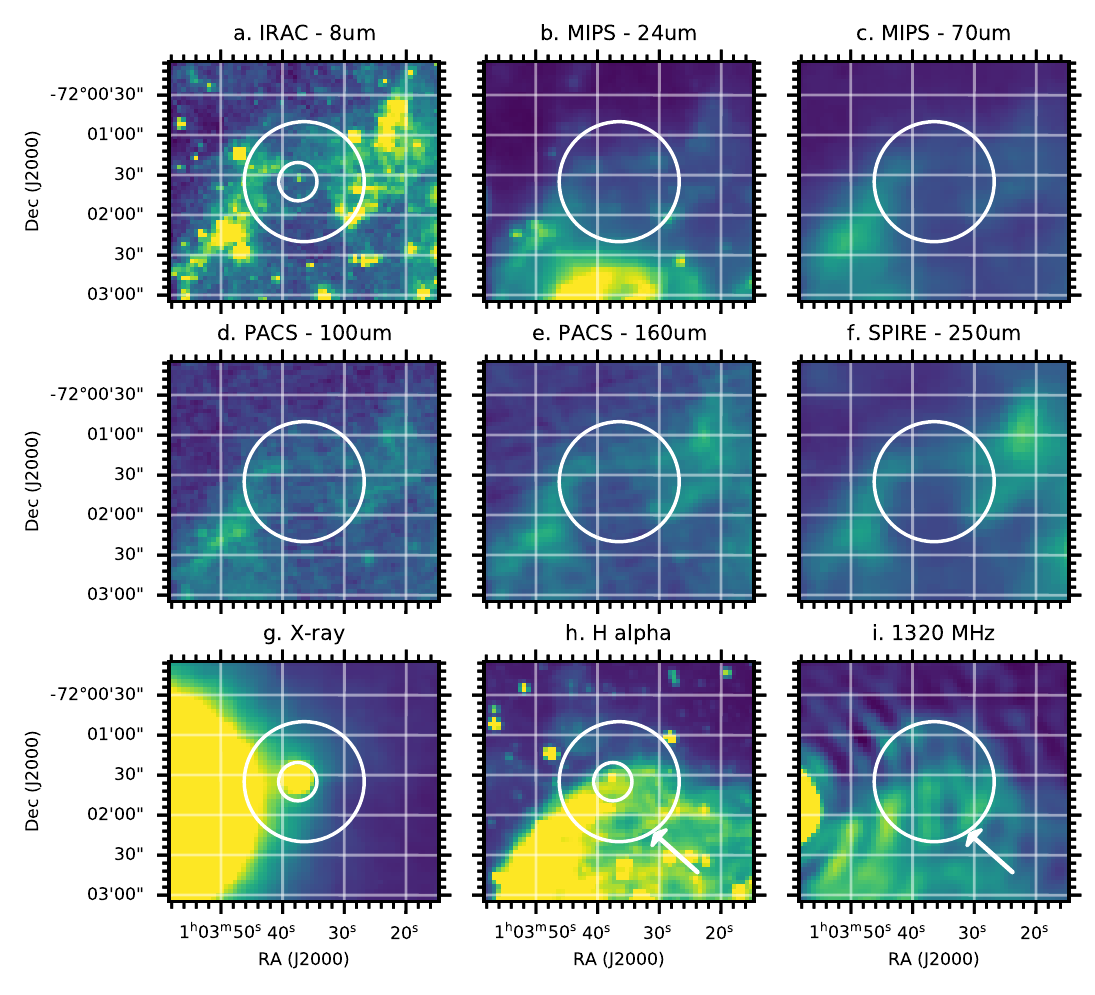}
  \end{minipage}\hfill
  \begin{minipage}[c]{0.30\textwidth}
\caption{ SNR J0103.6$-$7201: possible detection of a point source, which may be a neutron star associated with this SNR.
Note that arcs in the radio image (i) are an airy disc from the bright source to the east of the SNR.
           } \label{fig-01036}
  \end{minipage}
\end{figure*}

\citet{Gvaramadze:2019de} and  \citet{Maggi:2019wq} reported a very thin shell around SXP 1323, next to the H{\small II} region N\,76, which is very bright in the H$\alpha$ image \citep{Naze:2003cx}. 
The south-west part of this shell is pointed out with an arrow in the H$\alpha$ image in Fig.\ref{fig-01036}.
In the infrared, the emission is dominated by unrelated ISM emission, with some contribution from N\,76 at IRAC 8\,$\mu$m and 24\,$\mu$m. 
No evidence of an SNR shell is found in the infrared.

\subsubsection{SNR J0104.0$-$7202, B0102-72.3, DEM S124, IKT 22: detection }

SNR J0104.0$-$7202, or commonly known as 1E\,0102.2$-$7291, is one of the best studied SNRs in the SMC. 
Its dust emission was detected  by {\it Spitzer} \citep{Stanimirovic:2005p12420}, initially as a part of the SMC survey \citep{Bolatto:2007hha}, later followed by dedicated spectral mapping observations using {\it Spitzer} IRS \citep{Sandstrom:2009jxa, Rho:2009ky}.
From X-ray spectra, it is known to be oxygen-rich, suggesting that this SNR is a high-mass, core-collapse SN in origin \citep{Dopita:1981di}.
A candidate of the compact object has been reported by \citet{Vogt:2018jj}, though it has not been established whether this is really a compact object or not
\citep{Hebbar:gq, Long:fv}.
The estimated age of this SNR is from $\sim$1000 up to 3500\,yrs \citep{Tuohy:1983p3306, Alan:2019en}, and the progenitor  is potentially of high-mass stellar origin  \citep[40\,\Msun; ][]{Alan:2019en}.
The infrared emission is associated with the X-ray and radio ejecta with the brightest positions or clumps in 24\,$\mu$m closely resembling the brightest positions in [O\,{\small III}]. 
The estimated ejecta dust mass from  {\it Spitzer} IRS  is about $3\times10^{-3}$\,\Msun\,\citep{Sandstrom:2009jxa} to 0.01\,\Msun\,\citep{Rho:2009ky}.

\begin{figure*}
  \begin{minipage}[c]{0.70\textwidth}
    \includegraphics[width=\textwidth]{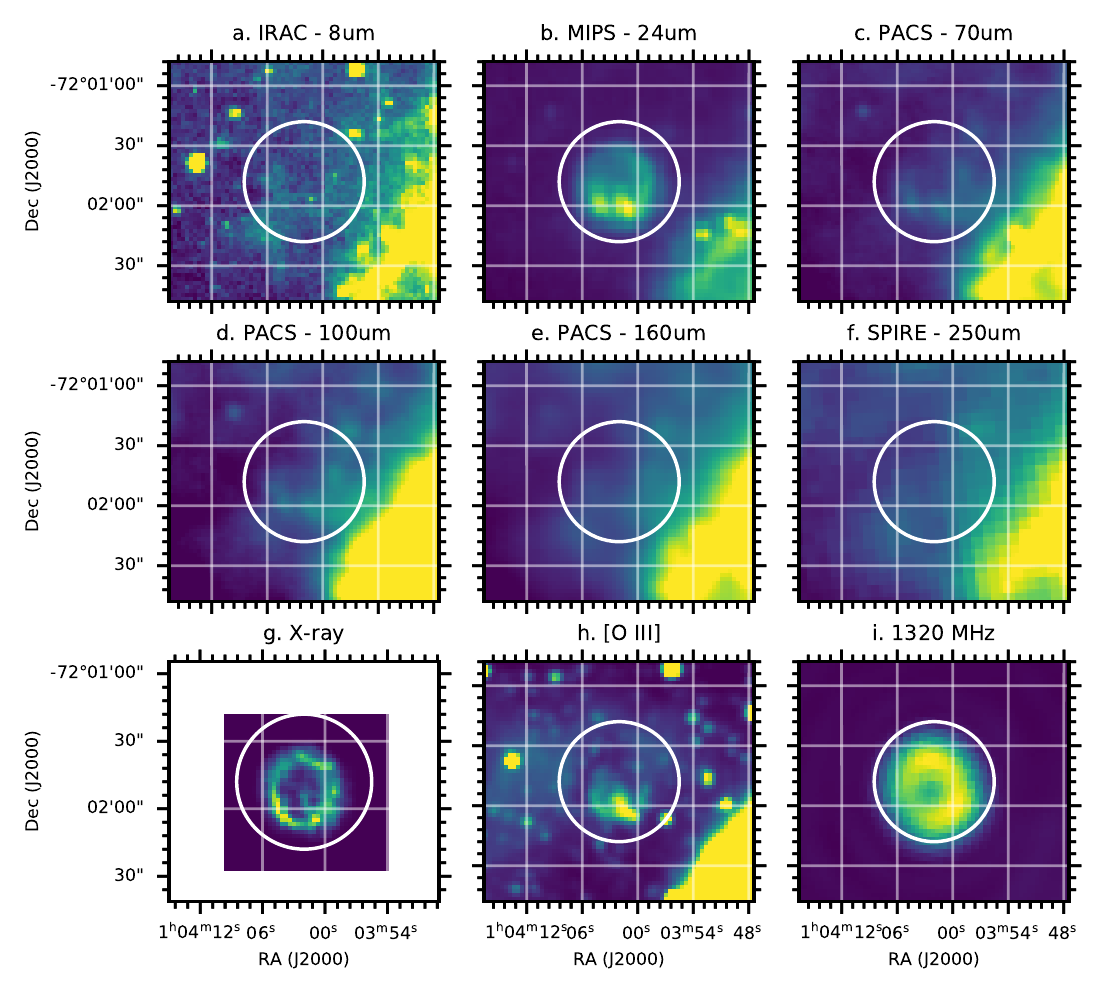}
  \end{minipage}\hfill
  \begin{minipage}[c]{0.30\textwidth}
    \caption{SNR J0104.0$-$7202, B0102-72.3, DEM S124, IKT 22, or commonly known as 1E\,0102.2$-$7291: detection. 
    Panel (g) is a {\it Chandra} image in the 0.3--10.0 keV band,
    and panel (h) is   an [O\,{\small III}] image, as this oxygen-rich SNR stands more at  [O\,{\small III}]  than at H$\alpha$.
           } \label{fig-01040}
  \end{minipage}
\end{figure*}

As reported by \citet{Stanimirovic:2005p12420} and \citet{Sandstrom:2009jxa}, the SNR is clearly visible at 24\,$\mu$m.
Herschel SNR dedicated observations (Proposal ID: 2011hers.prop.1802S) took images at 70 and 100\,$\mu$m, resulting in a clear detection at these wavelengths (Fig.\,\ref{fig-01040}). 
At 8\,$\mu$m, two very faint points are detected, the southerly one corresponds to 24\,$\mu$m bright clump, and the other in the east has a counterpart with an X-ray clump.
This SNR is located very close to the H{\small II} region LHA 115-N76C, which can be seen southwest of SNR J0104.0$-$7202 (Fig.\,\ref{fig-01040}).
This H{\small II} region is brighter than the SNR at infrared wavelengths. 
At 8\,$\mu$m, the SNR is faint but probably buried within unrelated ISM clouds, presumably with polycyclic aromatic hydrocarbons (PAHs) from the H{\small II} region and surroundings.
The cold dust emission from this H{\small II} region and surroundings dominates at 250\,$\mu$m, and contributes some emission at 160\,$\mu$m.
In Fig.\,\ref{fig-01040}, panel h shows [O\,{\small III}] image, instead of H$\alpha$, as this O-rich SNR stands out better at [O\,{\small III}] than H$\alpha$.

%

%
%
%


\begin{figure*}
  \begin{minipage}[c]{0.70\textwidth}
    \includegraphics[width=\textwidth]{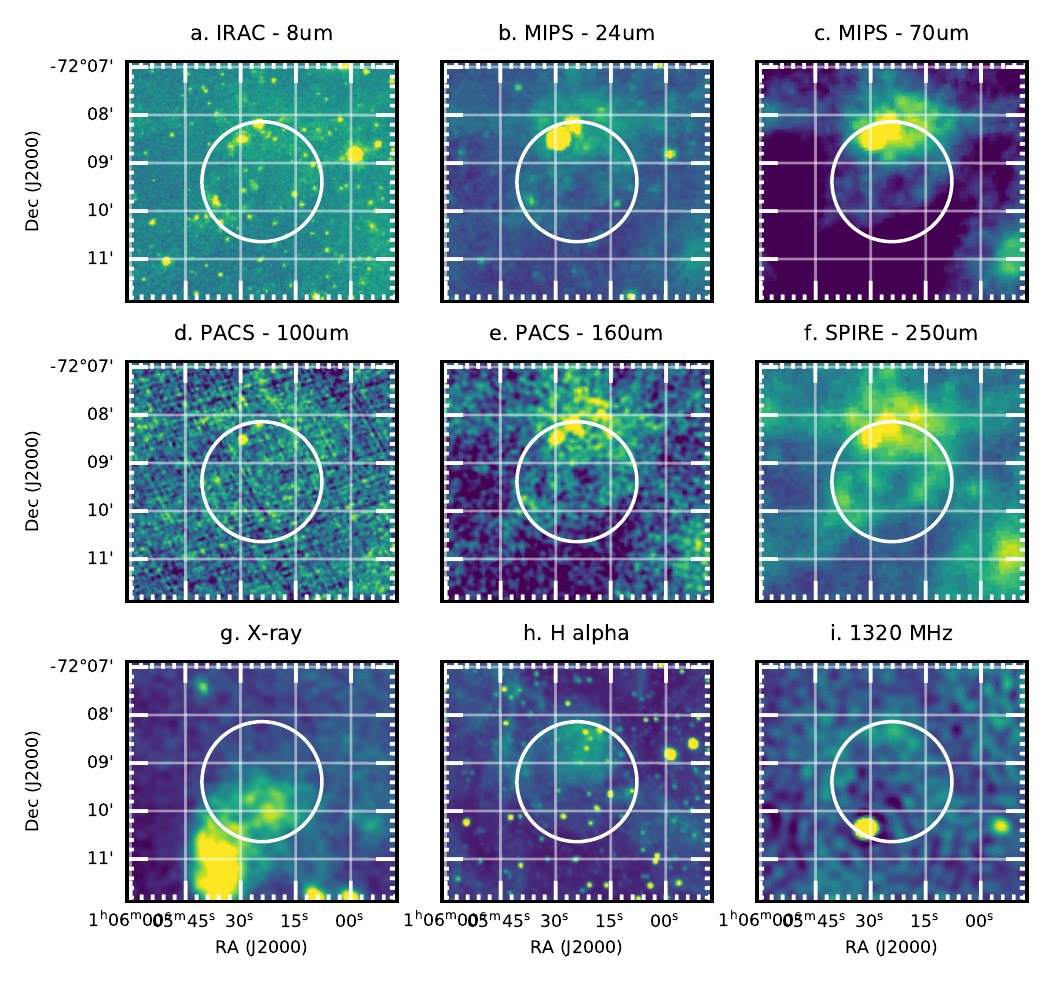}
  \end{minipage}\hfill
  \begin{minipage}[c]{0.30\textwidth}
\caption{ SNR J0105.6$-$7209, B0104-72.2, DEM S130: possible detection
           } \label{fig-01056}
  \end{minipage}
\end{figure*}

\subsubsection{SNR J0105.6$-$7204, DEM S128: possible detection }

There is a confusion of source names and coordinates for SNR J0105.6$-$7204 and SNR J0105.6$-$7209.
\citet{Davies:vz} listed SNR J0105.6$-$7204  (DEM S128) and SNR J0105.6$-$7209 (DEM S130)  in their H$\alpha$+[N\,{\small II}] photographic plates, however, the coordinates in the table and the location on the photographic plates have been mixed up: coordinates read from photographic plates for DEM S128 correspond to the coordinates for DEM S130 listed in their Table\,4.
\citet{2005MNRAS.364..217F} and \citet{Filipovic:2008fv} listed J010524$-$720923 as a SNR DEM S128, 
and J010539$-$720341 as a SNR candidate DEM S\,130.
\citet{Maggi:2019wq} concluded J010539$-$720341 is probably not an SNR.
Therefore, we investigate only SNR J0105.6$-$7204 as the SNR DEM\,S128.
The coordinates and the size of this SNR are taken from {\it XMM} analysis \citep{Haberl:to}.

 In the optical and radio, the north shell extends beyond the X-ray shell (Fig.\,\ref{fig-01056}), as pointed out by \citet{Maggi:2019wq}. 
That shell is also noticeable in infrared images, at least at 8 and 24\,$\mu$m, and potentially at 160 and 250\,$\mu$m (Fig.\,\ref{fig-01056}).
This SNR has been suggested to be type Ia origin  \citep{Roper:2015be}.
As it is bright at 250\,$\mu$m, the north shell might be tracing the place where the SNR interacts the most with nearby ISM, as found in Galactic type Ia, Tycho
\citep{Gomez:2012jka}






\subsubsection{SNR J0106.2$-$7205, B0104-72.3, IKT 25: detection}

There is a debate as to whether SNR J0106.2$-$7205 is of core-collapse or type Ia SN origin.
\citet{Lee:2011dr} and \citet{Roper:2015be} suggested it is a type Ia candidate, due to the high Fe abundance based on {\it XMM} and {\it Chandra} spectra.
\citet{Lopez:2014fo} and  \citet{Maggi:2019wq} found that the previous abundance analysis contains an error, and suggests core-collapse origin.
If it is a core-collapse SNR, the estimated age is 18,000 yrs old, with a progenitor mass of 18\,\Msun\, \citep{Katsuda:2018gs}

\begin{figure*}
  \begin{minipage}[c]{0.70\textwidth}
    \includegraphics[width=\textwidth]{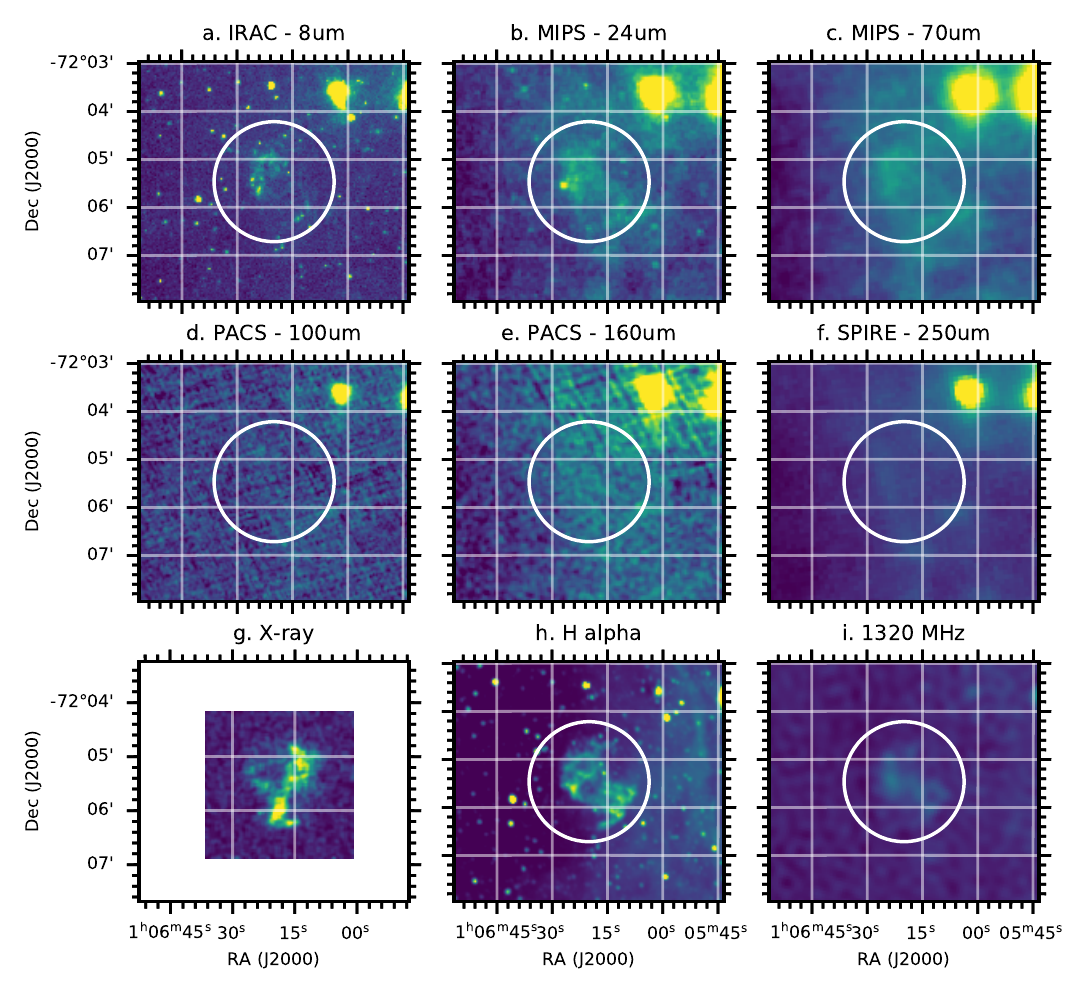}
  \end{minipage}\hfill
  \begin{minipage}[c]{0.30\textwidth}
\caption{ SNR J0106.2$-$7205, B0104-72.3, IKT 25: detection. 
Panel (g) is  a {\it Chandra} image in the 0.3--10.0 keV band
           } \label{fig-01062}
  \end{minipage}
\end{figure*}

\citet{Lee:2011dr} reported a detection using both IRAC 8\,$\mu$m and {\it AKARI} \citep{Koo:ty}.
We detect this SNR across {\it Spitzer} and {\it Herschel} images, though it is very faint at 100\,$\mu$m (Fig.\,\ref{fig-01062}).
There is a difference in morphology between X-ray and H$\alpha$+radio+IR, with the shapes and locations of IR resembling those of H$\alpha$ more closely.

\subsubsection{ SNR J0127.7$-$7332, SXP 1062: detection}

\begin{figure*}
  \begin{minipage}[c]{0.70\textwidth}
    \includegraphics[width=\textwidth]{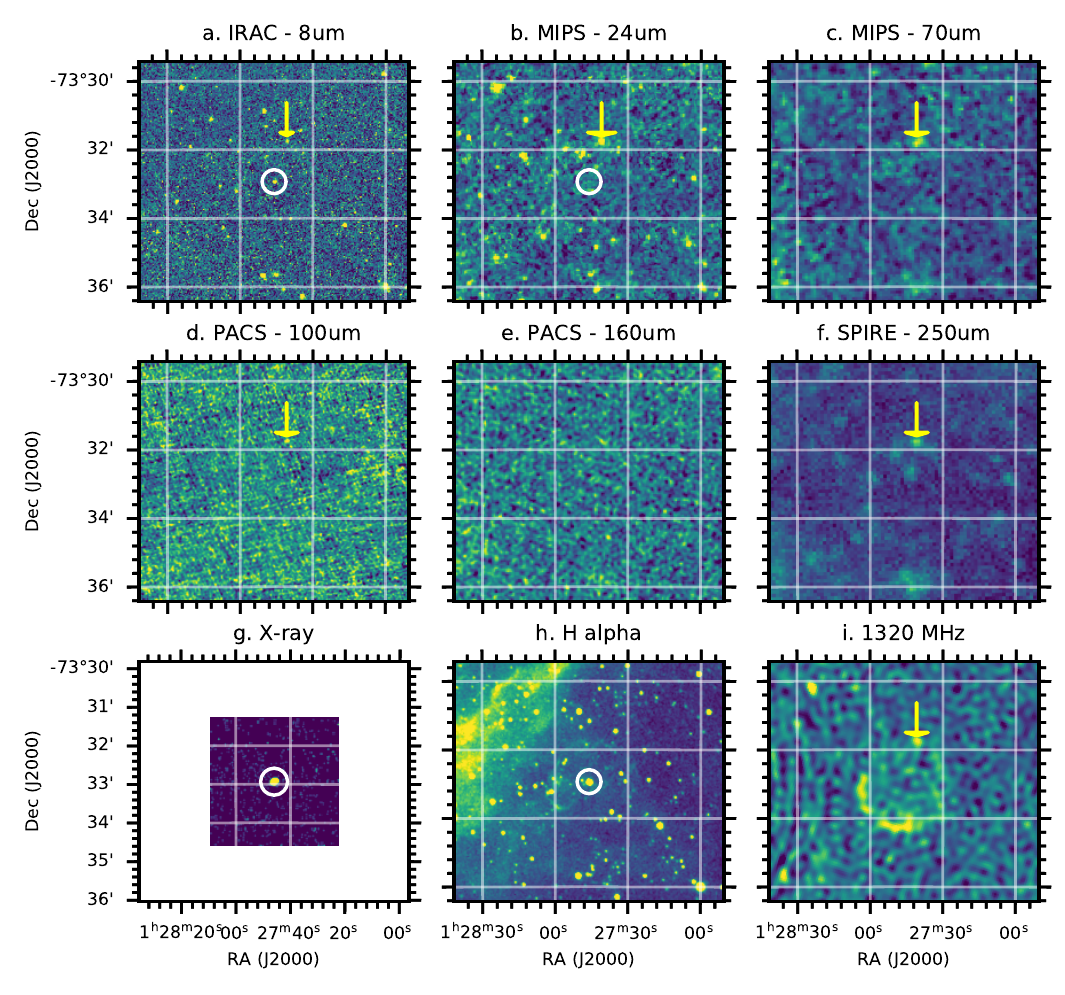}
  \end{minipage}\hfill
  \begin{minipage}[c]{0.30\textwidth}
\caption{ SNR J0127.7$-$7332 : detection of the pulsar SXP 1062, is indicated by a circle. There is an additional detection of small `blob' in the north possibly associated with the shell. 
Panel (g) is  a {\it Chandra} image in the 0.3--10.0 keV band
           } \label{fig-01277}
  \end{minipage}
\end{figure*}

SNR J0127.7$-$7332 shows shell-type structure, with the brightest rim at the south in the radio, while the northern shell is brighter in the optical image 
 \citep[Fig.\,\ref{fig-01277};][]{Maggi:2019wq}.
The central source is associated with a Be/X-ray binary, containing a pulsar SXP 1062 \citep{Maggi:2019wq} and is detected here at 8\,$\mu$m. 
Additionally, a small blob, which might be a part of north shell, is detected across infrared images  (Fig.\,\ref{fig-01277}).

\section{Analysis of dust emission} \label{dust_analysis}




 From the 24 SMC SNRs, we make a positive section from only 5. Two detections are associated with pulsars or pulsar wind nebulae, and only 3 detections are associated with expanding forward-shocked regions of the SNRs. 
 Another 5 have possible detections. The remaining 14 SNRs turned out to be non-detections both in {\it Spitzer} and {\it Herschel} bands. 
 The number of the detections of forward-shocked regions in the infrared is relatively small.
In order to understand this low detection rate, we make a simple calculation of the expected brightness of dust emission from SNRs under forward shocks, and compare it with the measured brightness.

\subsection{Predicted emission from collisionally heated dust grains} \label{section-model}

As a SNR expands, the gas plunges into the ISM, sweeping up the ambient ISM gas and dust and causing a shock at the expanding boundary.
In shocked regions, dust grains are heated up by the collision of fast moving electrons, and destroyed by the collision of high velocity  ions and atoms, as well as dust grains \citep{Jones:1994p8385}.
These forward shocks are considered to be one of the main dust destruction processes in the ISM, and determine the lifetime of the ISM dust grains 
 \citep[e.g.][]{Jones:1996bia, Micelotta:2018gl}.
There are several processes involved in dust destruction.
 Fast moving particles sputter atoms from the surface of the dust grains, and in high velocity shocks, this sputtering process is very efficient 
 \citep{Dwek:1992da}.
 Grain-grain collisions can destroy dust grains at much lower speeds, though more recent simulations 
 demonstrate that coupling sputtering and grain-grain collisions may cause much more damage than a simply adding these two processes 
 \citep{Kirchschlager:2019cq}.

In shocked regions, fast moving electrons play a role in heating dust via collisions.
 \cite{Dwek:1992da} described  the collisional heating  rate of dust grains,  $H(a)$ as
 \begin{equation} \label{eq1}
  H(a)= 2 \pi a^2 n_e ( \frac{8kT_e}{\pi m_e})^{1/2} kT_e \xi
 \end{equation}
where  $T_e$ is the electron temperature, $m_e$ is electron mass, $n_e$ is the electron density, and $a$ is the grain size.
The equation is based on a combination of the cross section of a grain, average Boltzmann velocity, and the kinetic energy of an electron.
Whether an electron goes through a dust grain or not depends on the grain size and the electron temperature, which $\xi$ represents.
 \citet{Dwek:bt} and \citet{Dwek:2008p28793} demonstrated that at a grain size of 0.01 and 0.1\,$\mu$m, the majority of electrons can be stopped., i.e. $\xi$=1.
If grain size is smaller than that range, it is likely to get slightly smaller than 1, however, it has only minor impact, and at a moment, for simplicity, we adopt $\xi$=1 across.
This heating rate is in equilibrium with the radiation from dust grains ($L_\lambda=4 \pi a^2 B_\lambda Q_\lambda$), where $Q_\lambda$ is the dust emissivity at wavelength $\lambda$.
For simplicity, $Q_\lambda$ is replaced with the Planck averaged dust emissivity 
 \begin{equation} \label{eq2}
\langle Q(T, a)\rangle=\frac{\int { B_\lambda(T) Q_{{\rm abs}, \lambda}}d\lambda}{\int {B_\lambda(T) d\lambda}}.
 \end{equation}
Although  \citet{Dwek:1992da} adopted 
 \begin{equation} \label{eq3}
\langle Q(T) \rangle /a \propto T_d^2
 \end{equation}
following \citet{Draine:1984p25590} and \citet{2011piim.book.....D}, this is a valid approximation up to $\sim$60\,K.
Unfortunately, small grains may exceed this temperature range, so  we therefore calculated $\langle Q(T_d, a) \rangle$, using the $Q_\lambda$ for silicates \citep{Ossenkopf:1992p27453}.
The equilibrium  temperature of the dust grains is then obtained by assimilating $H(a)$, whereby the total luminosity becomes \citep{Chawner2020}
 \begin{equation} \label{eq4}
L(a)=4 \pi a^2 \sigma T_d^4 \langle Q(a,T_d)\rangle. 
 \end{equation} 

We calculate the spectrum of the dust emission, using the grey body, i.e., $L_\lambda=4 \pi a^2 B_\lambda(T_d) Q_\lambda$, and $Q_\lambda$ as calculated for silicates \citep{Ossenkopf:1992p27453}.
We tested a mixture of silicates and amorhous carbon \citep{Zubko:1996p29442}, adopting dust-to-gas mass ratios of the Milky Way for silicates ($9\times10^{-3}$) and amorphous carbon ($1.4\times10^{-3}$) \citep{2010pcim.book.....T}, and the fact that elemental abundance ratios for the Milky Way and Magellanic Clouds are similar \citep{Gordon:2014bya}.
However, the mid-infrared emission is dominated by silicate emission, and for simplicity, we only used silicates for the rest of our discussions.

The total emission from all grains is calculated by integrating $L_\lambda$ across grain sizes $a$.
 We start with the smallest grain size ($a_{\rm min}$) of 0.005\,$\mu$m and the largest grain size ($a_{\rm max}$) is 0.25\,$\mu$m, following the SMC ISM value  \citep{Weingartner:2001p3411}.
 The smallest grain size is much smaller than the value of 0.023\,$\mu$m adopted by \citet{Dwek:2008p28793} for SN 1987A's circumstellar ring, but the largest grain size is the same as that given by \citet{Dwek:2008p28793}.
The power law index of $-3.5$ is used for the number distribution of dust grains \citep{Mathis:1977hp}.

We estimated the electron density $n_e$ and electron temperature $T_e$ from X-ray observations of SNRs.
\citet{Maggi:2019wq} listed  $T_e$ and emission measure $EM$ of SMC SNRs, and $n_e$ is estimated from $n_e= 1/f \sqrt{EM/V}$, where $f$ is the filling factor, and $V$ is the volume of the SNR. 
The filling factor is assumed to be 1 \citep{Maggi:2019wq}, initially, and $V$ is calculated  from  the diameters in Table\,\ref{detectionlevel}.
\citet{Maggi:2019wq} also used $EM$ to estimate hydrogen density $n_{\rm H}$ as $n_{\rm H}= 1/f \sqrt{EM/1.2 V}$, where a factor of 1.2 represents the difference between hydrogen and electron numbers for  a fully ionised gas.
The diameters in this table have been measured in radio even though radio and X-ray diameters might be different.
\citet{Maggi:2019wq} showed that the majority of SMC SNRs have a hydrogen density $n_{\rm H}$ of about 0.03--0.06 cm$^{-3}$ (Table\,\ref{table-detection}).

Finally, we estimate the dust mass swept up by SNRs, assuming all swept up dust grains survive the shocks.
That is calculated from the volume $V$ of the SNR, ISM hydrogen density $n_{\rm H}$ or $n_{\rm H, ISM}$ and the gas-to-dust mass ratio in the ISM.
\citet{temim:2015bs} estimated the local ISM hydrogen density, using the H{\small I} emission line, and estimated the line of sight depth of the SMC, the values of which we used for $n_{\rm H, ISM}$. 
All SMC SNRs have an $n_{\rm H, ISM}$ range of 0.5--1.9\,cm$^{-3}$.
As found in Table\,\ref{table-detection}, the values of $n_{\rm H, ISM}$ are about 10--20 times higher than those of $n_{\rm H}$, estimated from SNR EM by 
\citet{Maggi:2019wq}. Subsequently, this also affects the estimated swept up dust mass by the same factor.  
We will discuss this discrepancy later in Sect.\,\ref{section-dust-analysis}.  
Finally the gas-to-dust mass is assumed to be 500 \citep{Gordon:2014bya}. 
This is the smallest value within the range (500--1300), expected from the elemental abundance \citep{Gordon:2014bya}.

\subsection{Dust emission from a SNR} \label{section-dust-analysis}

We take SNR J0106.2$-$7205 (ITK\,25) as an example of the predicted emission from dust grains, swept-up by an SNR.

\begin{figure}
    \includegraphics[width=\columnwidth]{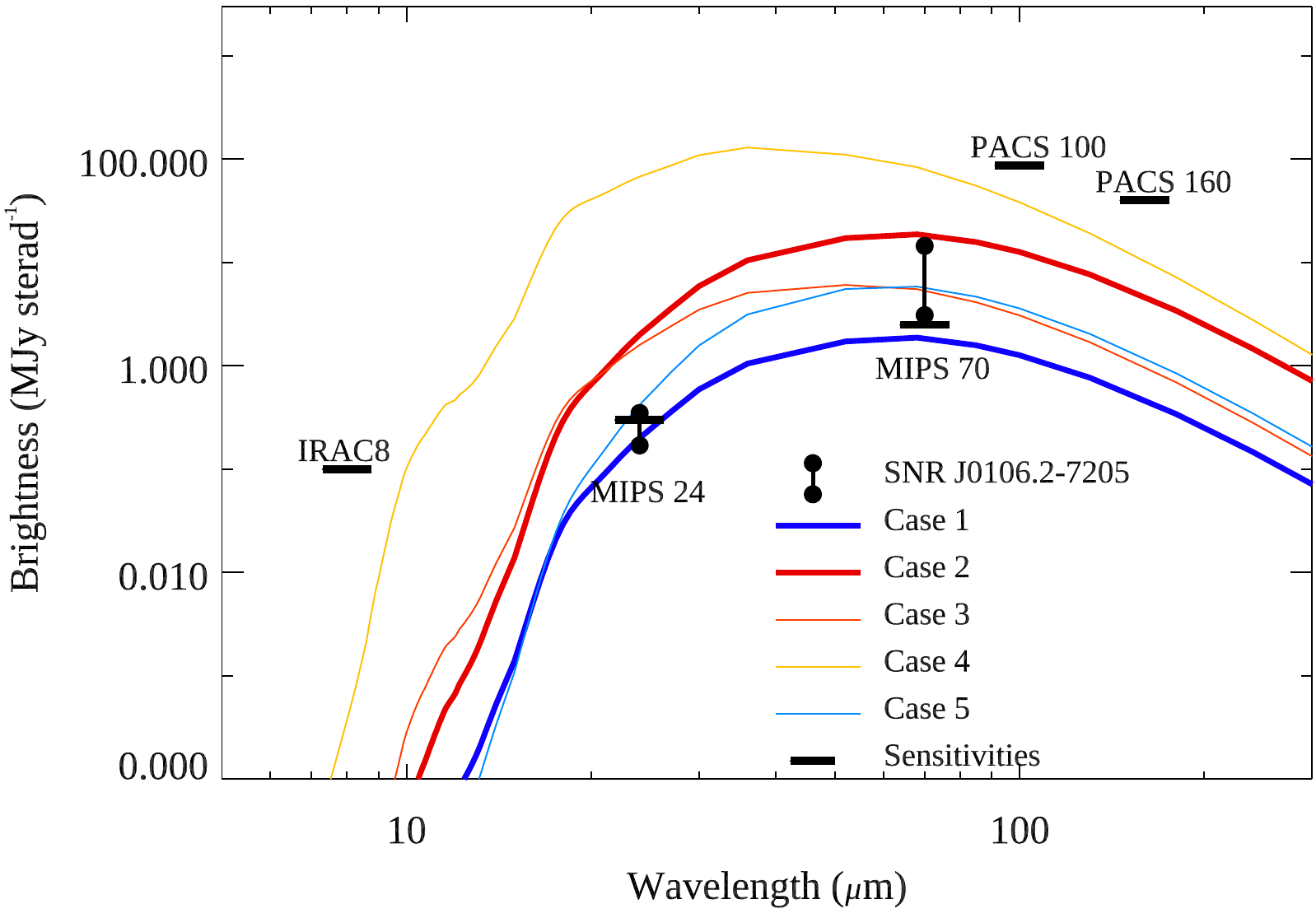}
\caption{ 
The expected dust emission from an SNR, an example using SNR J0106.2$-$7205 (ITK\,25), comparing approximate brightness at MIPS 24 and 70\,$\mu$m fluxes (black circles).
Case 1  and case 2 show expected dust emission, using $n_{\rm H}$ and $n_{\rm H, ISM}$, respectively, 
assuming that all ISM swept up dust fills a disk within the SNR radius.
Case 3 demonstrates the emission, if the radius of the SNR is half of the one used for cases 1 and 2.
Case 4 using the filling factor of 0.1, instead of 1 for the SNR, increasing the value of $n_{\rm H}$ to match $n_{\rm H, ISM}$.
That would also increase $n_e$, hence, more heating source.
Case 5 (blue thick line) shows the case when only  large dust grains (0.18--0.25\,$\mu$m) are included.
Case 5 approximates closely to the MIPS 24 and 70\,$\mu$m brightness (black circles).
    } \label{SED}
\end{figure}

Case 1 (blue line) in Figure\,\ref{SED} shows the predicted dust emission from SNR J0106.2$-$7205 (ITK\,25), assuming that all swept up dust can survive SNR-ISM shocks.
%
The swept-up dust mass is estimated, using the hydrogen density $n_{\rm H}$ from the X-ray emission measure \citep{Maggi:2019wq}.
Case 2 (red line) shows the predicted dust emission, using the hydrogen density $n_{\rm H, ISM}$ from the H{\small I} emission line \citep{temim:2015bs}.
 Fig.\,\ref{SED}  includes approximate average brightnesses of SNR J0106.2$-$7205 at MIPS 24 and MIPS 70\,$\mu$m bands, indicated by circles. 
The large uncertainties in the surface brightnesses are caused by uncertainties in estimating `background' levels, as they are inconsistent across the span of the SNR.
The total surface brightness of the background and SNR in MIPS 24\,$\mu$m  is above the sensitivity limit however MIPS 24 brightness range goes below the sensitivity limit (5\,$\sigma$) after `background' subtraction.
The predicted emission of case 1 is much closer to the approximate surface brightness of SNR J0106.2$-$7205 (ITK\,25).
This one assumes that all swept-up dust grains survive shocks, i.e. no dust destruction.
However, there is a caveat to this interpretation.

As mentioned in Sect.\,\ref{section-model}, there is a factor of 10--20 discrepancy in two different methods of estimating hydrogen densities.
The one from X-ray EM is $n_{\rm H}$=0.04\,cm$^{-3}$, while  H{\small I} along the line of sight gives $n_{\rm H, ISM}$=0.9cm$^{-3}$.
The obvious cause of this difference is the assumed size of the SNR, hence, its volume used for converting EM to $n_{\rm H}$.
As seen in Fig.\,\ref{fig-01062}, this SNR has an irregular shape, and the size we applied is the largest radius, assuming symmetric shape, so that reducing the radius is reasonable.
Increasing the radius by a factor 2 can increase $n_{\rm H}$ to 0.3\,cm$^{-3}$.
Naturally, by reducing the estimated radius, and hence the volume, the total dust mass also reduces.
On the other hand, by reducing the volume, the electron density, $n_e$ goes up by a factor of $\sqrt V$, increasing the total heating.
Therefore, reducing the size by a factor of 2 causes the emission to look like case 3 in Fig.\,\ref{SED}, where dust is warmer than case 1 and overall emission is increased.
The temperature of the smallest grain is 60\,K for case 1, and 71\,K for case 3.
In this case, the 24\,$\mu$m surface brightness is over-predicted by a factor of 5--10.


 We can make a further experiment by matching the $n_{\rm H}$ and $n_{\rm H, ISM}$ by decreasing the filling factor $f$ from 1 to 0.1 for $n_{\rm H}$, as $n_{\rm H}= 1/f \sqrt{EM/1.2 V}$. 
 The radius is kept at the initial size.
 In this case, $n_e$ goes up by a factor of 10, i.e. an increase in the heating source, and the total dust mass increases by a factor of 10 over and above case 1 (but the dust mass remains the same as case 2), resulting in a much hotter and brighter dust emission (case 4 in Fig.\,\ref{SED}). 
The smallest grain has an equilibrium temperature of 86\,K in case 4, much hotter than 60\,K for case 1.
Case 4 is very unlikely, because it over-predicts 24\,$\mu$m flux by a factor of 200--400.

One possible way to accept the low filling factor described in case 4, but to match the observed brightness would be to remove the small dust grains from the equation.
Because smaller grains tend to reach higher equilibrium temperature, removing small grains can result in reducing emission at shorter wavelengths, thereby also reducing overall emission.
Case 5 demonstrates dust emission only limited from grains between 0.18--0.25\,$\mu$m, instead of default 0.005--0.25\,$\mu$m as in case 4.
The total dust mass is reduced to only 18\,\% of case 4, i.e. equivalent to a scenario where smaller dust grains have been destroyed by SN shocks, and only larger dust grains survived.
Since large grains tend to have lower dust temperatures than small grains, the model can reproduce the measured surface brightness reasonably well.

Note that very small grains might not reach equilibrium temperature, but instead they can reach higher temperatures.
Once small grains absorb the energy from photons, their temperature spikes up, instantly.
However, this makes the overall emission much hotter and much brighter, so that the discrepancies between the measured brightness and predicted brightness in case 2 or 3 would worsen.

There is about 10\,\% numerical error in the model brightness, so that the difference between $n_{\rm H}$  and $n_{\rm H, ISM}$ does not linearly reflected in $F_{24}$/$F'_{24}$ or $F_{70}$/$F'_{70}$ in Table \,\ref{table-detection}.

\section{Dust temperature and density analysis of SNR J0059.4$-$7210 (DEM\,S103)}\label{DEMS103-analysis}

%

\begin{figure*}
   \includegraphics[width=1.00\textwidth]{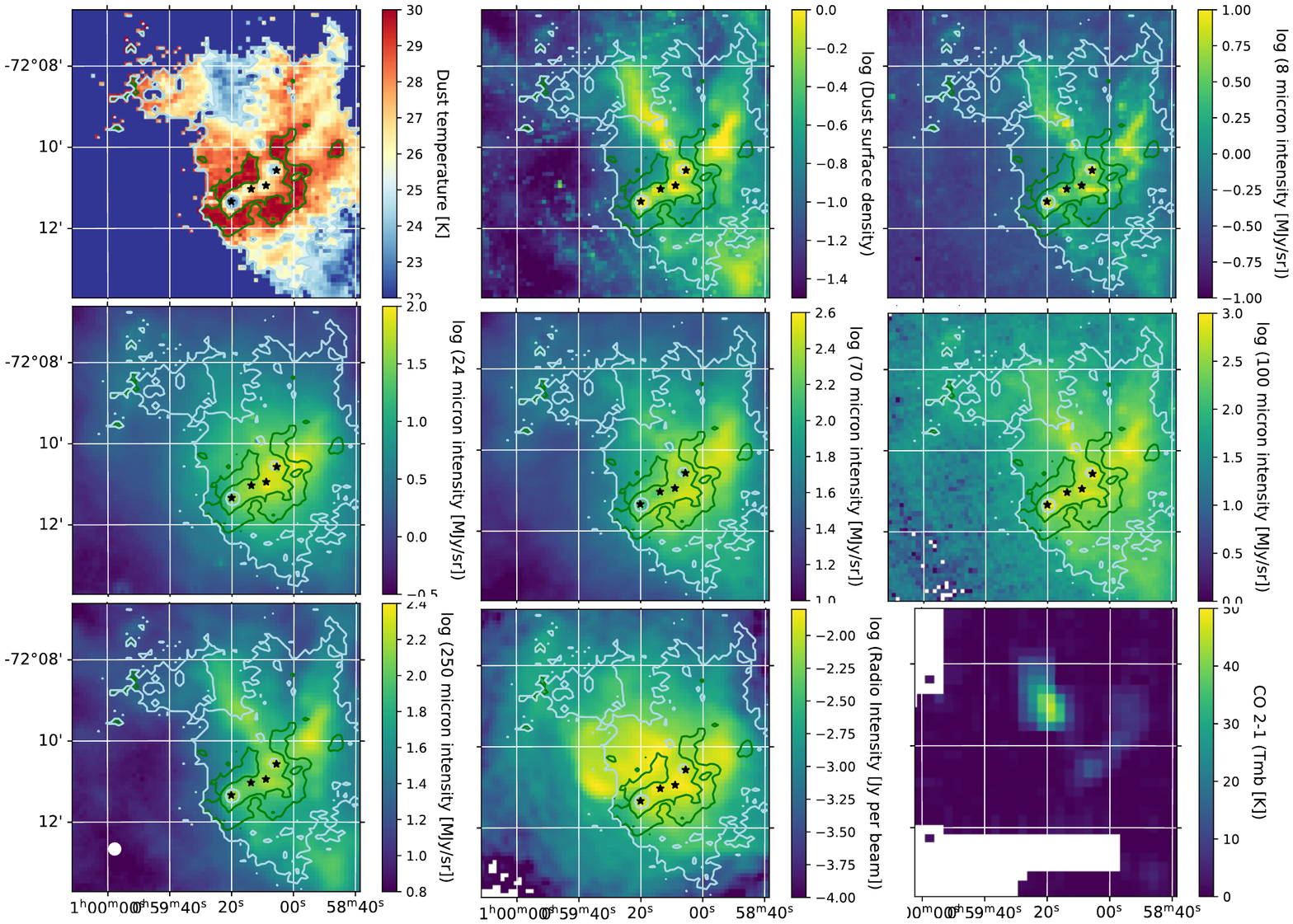}
\caption{ 
Dust temperature $T_d$ , and surface density $\Sigma_{d}$ maps, together with {\it Spitzer}, {\it Herschel}, radio and CO 2--1 maps of SNR J0059.4$-$7210 (DSM\,S103).
The line contours trace the dust temperature of  29\,K (green) and 15\,K (light blue).
The white circle in 250\,$\mu$m map shows the SPIRE 250\,$\mu$m beam size.
{\it Spitzer} and {\it Herschel} images are smoothed to the SPIRE 250\,$\mu$m angular resolution. 
Black star marks show the locations of clusters which exhibit colder dust temperatures.}
  \label{DEMS103_dust} 
\end{figure*}

As the SNR J0059.4$-$7210 (DSM\,S103) looks as if it collides with the star-forming region NGC\,346, we investigate its effect on dust, using {\it Spitzer} and {\it Herschel} images.

\subsection{Analysis}

We used  {\it Spitzer} MIPS 24 and 70\,$\mu$m,  {\it Herschel}  PACS 100 and 160\,$\mu$m and SPIRE 250\,$\mu$m images.

Only for this analysis, we used residual images from SAGE-SMC  for MIPS 24 and 70\,$\mu$m bands.
The residual images are the images after point sources were subtracted. 
These removed point sources include bright clusters in the star-forming regions 0f NGC\,346.
All images were smoothed to have a 6 arcsec pixel pitch, matching to the coarsest angular resolution of the {\it Herschel} SPIRE 250\,$\mu$m band.
{\it Spitzer} and {\it Herschel} uncertainty maps were used as a $\sigma$ for fitting a grey body.
As found in the images, the PACS\,100\,$\mu$m image has much larger uncertainties, due to its lower spatial sampling rate.
The  {\it Herschel} uncertainty maps reflect this issue, and the weight of 100\,$\mu$m intensities is much lower than the rest of the {\it Spitzer} and {\it Herschel} images.
We also removed S/N$<$10 regions in the final fitted results.

The {\it Spitzer} and {\it Herschel} images were fitted pixel to pixel, using the grey body of the dust emission.
The surface brightness of the dust $S_{\nu}$ is associated with the temperature as $S_{\nu}=\kappa_{\nu} \Sigma_{d} B_{\nu}(T_d)$, where $B_{\nu}(T_d)$ is the Planck function.
We used an equation from \citet{Gordon:2014bya}, $S_{\nu}=(2.0891\times10^{-4}) \kappa_{\nu} \Sigma_{d} B_{\nu}(T_d)$, where the dust surface mass density $\Sigma_{d}$ is in \Msun pc$^{-2}$, the dust emissivity $\kappa_{\nu}$ in  cm$^2$\,g$^{-1}$ and $S_{\nu}$ and $B_{\nu}(T_d)$ in MJy\,sr$^{-1}$.
The  $\kappa_{\nu}$, was calculated by dust emissivity law $\kappa_{\lambda}$=$\kappa_{\lambda_0}(\lambda/{\lambda_0})^{-\beta}$ from \citet{Hildebrand:1983tm}.
We adopted the power law index of $\beta=1.9$ 
from Planck Galactic plane studies \citep{PlanckCollaboration:2014cx}, which is consistent with the estimated  $\beta$ at SNR-ISM interacting regions in the Galactic plane \citep{Chawner:2019dn}.
Grain density $\rho$ of 0.27\,m$^2$\,kg$^{-1}$ at 450\,$\mu$m is adopted from \citet{Gomez:2010p29834}.
These properties are overall consistent with the estimated the ranges of dust properties, reported in the SMC ISM \citep{Gordon:2014bya}.

CO $J$=2--1 observations of SNR DEM\,S103 were obtained, using the Atacama Pathfinder Experiment  (APEX) with Heterodyne SIS detector, nFLASH.
The data were acquired intermittently from 9th  to 14th of December 2020, in OTF mapping mode, and its $T_{mb}$ map is plotted in right bottom panel of Fig.\,\ref{DEMS103_dust}.

\subsection{Results}

Fig.\,\ref{DEMS103_dust} shows the resultant dust temperature $T_d$ and the dust surface mass density $\Sigma_{d}$ in SNR DEM\,S103 and the neighbouring star-forming region NGC\,346.
The top panel shows $T_d$ (left) and $\Sigma_{d}$ (right).
The contour of $T_d$ is overlaid in {\it Spitzer} 8, 24, {\it Herschel} 70, 100, 250 and radio images in the top right, the middle, the bottom left and central panels.

 In the south-east region of, NGC\,346, the $T_d$ is lower ($\sim$22\,K) at the locations of stellar clusters, marked by stars in Fig.\,\ref{DEMS103_dust}. 
These stellar clusters are surrounded by  
 higher (about 30\,K) dust temperature regions.
 We carefully examined these low temperature regions in the {\it Spitzer} and {\it Herschel} images as they contain stellar clusters which could have potentially been removed as point sources in the {\it Spitzer} residual images, thereby artificially lowering the inferred dust temperature. 
However, the low temperature is mainly determined through the  {\it Herschel} images, and these images show little contribution from the stars, hence, we can be confident that the cold dust at these stellar cluster locations is real.
 Surprisingly, these low $T_d$ regions have bright radio emission. 
 In contrast, the surrounding higher $T_d$ region has fainter radio emission. 
Between DEM\,S103 and NGC\,346, there is a  higher $T_d$ region, and this region corresponds to the gap in radio emission between DEM\,S103 and NGC\,346.

The stellar cluster regions, marked with stars in Fig.\,\ref{DEMS103_dust}, have colder $T_d$, with the highest $\Sigma_{d}$.
 Although  the dust mass has $T_d$ dependence, because the fitting procedure involves $B_{\nu}(T_d)$, the difference in dust mass is much larger than what can be explained by $B_{\nu}(T_d)$ at 22\,K and 30\,K: the difference is 0.49 and 0.33 dex at 160 and 250\,$\mu$m, so that bright infrared emission from these cluster regions is explained by more dust mass towards these regions.
 There is negligible contribution ($<0.1\,\%$) of synchrotron radiation at 24--250\,$\mu$m wavelengths, estimated from radio 960\,MHz and 1320\,MHz images \citep{Joseph2019}.  
 Very low levels of synchrotron and free-free agree with the analysis of SMC molecular clouds by \citet{Takekoshi:2017cr}.
Further,  the 250\,$\mu$m emission, which picks up colder dust, is brighter within this region.
Hence, indeed, this area has a higher dust mass, as confirmed in the dust mass map.
Some of the dust emission corresponds to filaments being extinguished against the H{\small II} region, seen by  the optical {\it HST} image
 \citep{Nota:dk}.
 Although not clearly detected in our nFLASH CO image, CO emission was detected in this region in previous mapping 
\citep{Rubio:uk, Hony:hl}.

  In the north-west of NGC\,346, the dust temperature is low ($\sim$22--23\,K).
  The low temperature region is not localised at the stellar clusters only, but throughout the entire region, covering multiple stellar clusters and their surrounding regions.
That is in contrast to the southern part of NGC\,346, where the low dust temperature is localised at stellar clusters only.
Also in the north-western region, there is little radio continuum emission detected, but CO is brightest.

 On the west, outside of the green contour line, infrared emission is bright, but it is caused by combination of both moderately high temperature (26\,K) and high dust mass.


Note that if the S/N is below 10 at 100\,$\mu$m intensity, the temperature has been masked. In the figure, this masked region was assigned a 20\,K colour, but this is not a true reflection of the actual temperature. 
Low temperature regions tend to have less far-infrared emission, hence, fainter and lower S/N in {\it Herschel} 100\,$\mu$m images.
Therefore, we are unable to measure the dust temperature below 20\,K.  \citet{Gordon:2014bya} analysed the overall dust in the SMC ISM, and estimated the mean SMC dust temperature of below 20\,K, hence it is obvious that we cannot analyse such low temperature regions with insufficient S/N.



\subsection{Discussion of the dust temperature map for DEMS103 and NGC 346}
The combination of high-surface density and low dust temperatures at the location of the stellar clusters bears a striking resemblance to the G\,316.75$-$00.00 ridge in the Milky Way, as seen by \citet{Watkins:2019gc}. 
In the case of G316.75, a high-column density and low-dust temperature filament, containing four embedded O-type stars, is located at the centre of a bipolar H{\small II} region, powered by those stars. 
The photoionising UV radiation from these embedded high-mass stars is inefficient at heating and disrupting gas along the axis of the filament, but can much more easily escape in the perpendicular direction, ionising the surrounding lower-density gas that is exposed to the radiation.

A similar scenario could explain many of the features in the west of NGC\,346, such as the presence of the cold, high-density dust ridge surrounded by the ring of warmer and more diffuse dust. 
The warm dust is associated with the H{\small II} region, and has probably been heated by massive stars contained within the clusters embedded within the ridge. 
This ridge could well be the natal ridge from which the stellar clusters formed, and may itself be the result of compression by the SNR on the north-east side (Sect.\,\ref{subsec-0059.4}). 
Because the ionised gas mass (and therefore heated dust grains) within it probably accounts for only a small fraction of the total mass, the cold dust dominates the temperature map.
The H{\small II} region on the south-western side of the ridge explains the free-free radiation seen within the radio and H$\alpha$ data, while on the north-western side of the ridge, the SNR is responsible for the synchrotron radiation. The H{\small II} region and the SNR may partially overlap, and the lower gas density on the south-western compared to the south-eastern side may explain the relatively low level of H$\alpha$ emission there.



\section{Discussions}

\subsection{SNR detection rate}


After examining all 24 SMC SNRs, we report 5 detections and 5 possible detections of associated dust.
The MIPS 24\,$\mu$m is the most effective way to identify the SNRs, produced by various combinations of the dust continuum and lines \citep{PinheiroGoncalves:2011ff}. 
The next most useful band is IRAC 8\,$\mu$m.
This band is composed of contributions from  synchrotron radiation,  [Ar{\small II}] and [Ar{\small III}] lines, with possible H$_2$ lines, as well as dust emission  \citep{Ennis:gz, Reach:2006cl}.
MIPS\,70\,$\mu$m is not effective, because of its lower sensitivity and lower angular resolution compared to the 24- and 8-$\mu$m bands.
In the Galactic survey, the PACS 70\,$\mu$m was effective at identifying SNRs together with the other {\it Herschel} bands \citep{Chawner:2019dn, 2020MNRAS.493.2706C}, 
however, this band is not available for most of the SMC SNRs.

Amongst five detections, two (SNR J0040.9$-$7337 and SNR J0127.7$-$7332) are associated with pulsars/point sources, and only three have detections of a nebulous component  associated with SNRs.
That leads to a low 13\,\% detection rate for extended emission in SMC SNRs.
Even including possible detections, in total only 6 SNRs have shown extended emission in the infrared.

Far-infrared surveys of SNRs in the Milky Way show that younger (less than 5\,000 years) SNRs tend to be more likely to be detected \citep{2020MNRAS.493.2706C}.
Amongst 5 detected and 5 possible detected sources, only one (SNR J0104.0$-$7202) has a relatively well determined age (2050 years), so it is very difficult to verify this.
With limited number, it is difficult to evaluate if the detection rate has the progenitor mass dependence (Table\,\ref{table-detection}).
Either lower (8--12.5\,\Msun) and relatively higher ($> 21$\,\Msun) progenitor origins may  have a higher detection rate, with little detection in between.
However, this is a general tendency of the progenitor mass distribution in the SMC SNRs from \citet{Anonymous:c6obe9b8} rather than dust specific results.

\citet{Dokara:2021vw} reported a spatial anti-correlation between mid-infrared emission and radio emission, however, with a limited sample we did not see such a tendency.

\subsection{ Dust destruction or something else?} 


It has been pointed out that the collisionally heated dust model over-predicts dust emission from SNRs \citep{Seok:2013fz, Seok:2015jo, Chawner2020, Priestley:2021tt}.
Indeed, our analysis in Sect.\,\ref{dust_analysis}  may offer this as a possibility: 
if the $n_{\rm H, ISM}$, hydrogen density estimated from the H{\small I} line is used, the model indeed predicts a much higher brightness than found from those that were measured (case 4).
In order to reduce mid-infrared emission, the simplest way is to remove the small grains, leaving only large ($\sim$0.18\,$\mu$m) grains, thus reducing the dust mass by a factor of $\sim$20\,\%.
That might replicate the situation that small dust grains have been destroyed by SN shocks, while large dust grains survive.
Since the mass of dust grains is dominated by large grains, only a small reduction in dust mass can substantially reduce the IR emission.
In contrast, when case 1, $n_{\rm H}$, hydrogen from X-ray EM, is used, the mode predicts a reasonable brightness.
Uncertainty of the hydrogen density makes it even more challenging to determine if dust grains have been destroyed or not.

An over-prediction of dust emission by the collisionally heated model does exist in all SMC SNRs in general, not limited to  SNR J0106.2$-$7205 (ITK\,25) demonstrated in Sect.\,\ref{dust_analysis}.
Table\,\ref{table-detection} shows the model predicts  $F'_{24}$ of 0.4--7.2\,MJy\,sterad$^{-1}$, based on  $n_{\rm H, ISM}$ (equivalent to case 2 for SNR J0106.2$-$7205), meaning that all SNRs could have been detected at 24\,$\mu$m, as the sensitivity limit is 0.3 MJy\,sterad$^{-1}$ (5\,$\sigma$),  if there is no ISM confusion.
In reality, there are only two detections with two more possible detections out of 8 sources.
The detection rates seem to be  lower, compared to the prediction from the collisionally heated dust model.
One possible explanation is that small dust grains have been destroyed, so the  24\,$\mu$m emission is much lower than predicted.
However, we need to be very cautious about this interpretation, as $n_{\rm H, ISM}$  tends to be a factor of 10 or more higher than $n_{\rm H}$.
It could be possible that the adopted thickness of the SMC for $n_{\rm H, ISM}$ given by \citet{temim:2015bs} may be a few factors too small, while the adopted radius is a few factors too large. 
In such a scenario a combination of the two adjustments,  even with some ISM confusion, could potential resolve the issue of low detection rate.
Therefore, at the present time and from our data, there is no conclusive impartial evidence that dust destruction is commonly occurring in SNR shocks.


One concern is that X-ray and dust emission do not always correspond, spatially.
Such an example is found in SNR\,J0052.6$-$7238 (Fig.\,\ref{fig-00526}), which shows dust emission in the north, corresponding to H$\alpha$, whereas X-ray emission originates in the south with low detections of dust emission.
In this case it is probable that the main SNR ejecta and the densest regions of ISM matter do not spatially correspond. 
Hot plasma emitting from X-ray emission would be missing the dust grains needed to produce collisional heated dust infrared emissions. 
If hot plasma is not responsible for dust emission in SNRs,  the past dust destruction rate through  SNR might have been overestimated.
%

\citet{Slavin:2015in} demonstrated that the dust destruction rate reduces substantially, once the shocks change from collisional shocks to radiative shocks.
Although there is a dispute over the type of SN involved, the age of SNR\,J0106.2$-$7205 is estimated to be 17,600\,years
\citep{Lopez:2014fo}.
At this age it is in general considered to be in the Sedov phase, typically between a few hundred years old to $\sim10^5$ years old, and the shock is still expected to be collisional.
\citet{Slavin:2015in} argues that at about $6\times10^4$\,years old, the radiative cooling is becoming more significant, but SNR J0106.2$-$7205 has not yet reached that age.
The formal distinction of radiative or non-radiative shocks should be made according to shock velocity \citep{McKee:1980jz}.
Unfortunately, in this case there is no shock velocity estimate present. We therefore assume that this SNR still experiences collisional (and non-radiative) shocks.

 In SNR\,J0106.2$-$7205 the X-ray and IR emitting regions do not always correspond, while IR emitting regions tend to correspond with H$\alpha$.
That link may be caused by the effect of the density.
From eqs.\,(\ref{eq1}) and (\ref{eq3}), the dust luminosity approximately follows  $L = H \propto n_e T_e^{3/2}$. 
The infrared emission has a $T_e$ dependence, and $T_e$ is proportional to $v_s^2$, where $v_s$ is the shock velocity. 
On the other hand, $EM \sim n_e^2 \times V$.
Assuming that the electron density $n_e$ in the shocks is linearly related to the upstream ISM gas density $n_{{\rm ISM}}$,
the ISM gas density, hence, $n_e$ may vary by an order of magnitudes in the ISM, while the shock velocity $v_s$ variation within a SNR can be only by a factor of a few, so that ISM gas density could be the major driving factor of infrared brightness. 
A higher ISM density with a reasonably high velocity shock would therefere cause higher IR emission. 
X-ray brightness has even stronger dependence on $n_e$.
Assuming that the SPIRE 250-$\mu$m image represents ISM density along our line of the sight,  the ambient ISM density might be higher to the north.
Reasonably high ISM density might drive a detectable level of dust emission in the north of the SNR.
Also H$\alpha$ emission is seen in the north of the SNR, and shocks in this region may be radiative shocks, i.e. shocks with H-recombination lines, weak forbidden lines of low ionised metals and non-thermal X-ray emission
\citep{yie, Chevalier80, heng2010}.
This demonstrates that the hydrogen density is crucial for evaluating infrared brightness, hence, whether dust grains being destroyed or not.


In conclusion,  within the given uncertainties, we find no conclusive evidence that SNRs have destroyed local swept-up ISM dust.
It has previously  been suggested that collisional heating of dust grains may over-predict infrared emission in some SNRs \citep[e.g.][]{Williams:2006cn, Seok:2015jo}, and indeed, this scenario may be accepted if we assume that small dust grains have been destroyed as in the case of SNR J0052.6$-$7238.
However, we advise caution here when interpreting emission levels.
First, electron densities measured from X-ray emission do not necessarily spatially correspond to infrared dust emission.
Second, there are uncertainties involved in the size of the SNRs and hydrogen density, even though SMC SNRs have very good distance estimates to date. 
Considering the uncertainties involved in estimating hydrogen and electron densities, IR emission can be explained without the need to invoke dust destruction.



\section*{Acknowledgements}

We thank Dr. Piere Maggi for very useful input about X-ray emission.
{\it Herschel} is an ESA space observatory with science instruments provided by European-led Principal Investigator consortia and with important participation from NASA. 
This work is based in part on observations made with the {\it Spitzer Space Telescope}, which was operated by the Jet Propulsion Laboratory, California Institute of Technology under a contract with NASA. Support for this work was provided by NASA.
This publication is based on data acquired with the Atacama Pathfinder Experiment (APEX). APEX is a collaboration between the Max-Planck-Institut fur Radioastronomie, the European Southern Observatory, and the Onsala Space Observatory. The ESO program no is 106.218.
MM acknowledges support from STFC Ernest Rutherford fellowship (ST/L003597/1), MJB, AB, and RW acknowledge support from European Research Council (ERC) Advanced Grant SNDUST 694520, and HLG  acknowledges support from the European Research Council (ERC) in the form of Consolidator Grant COSMICDUST (ERC-2014-CoG-647939).
FDP acknowledges support from STFC Consolidated grant.

\section*{Data Availability}
{\it Spitzer} and {\it Herschel} survey data are available at {\it Spitzer} Heritage Archive (SHA) 
(https://sha.ipac.caltech.edu/applications/Spitzer/SHA/)
and Herschel Science Archive
(http://archives.esac.esa.int/hsa/whsa/)



\bibliographystyle{mnras}
\bibliography{smc_ref} 




\appendix

\clearpage  

\section{Description of infrared emission from SNRs -- non detections and SNR candidates}

\subsection{Non detections} \label{sect-non-detection}


\subsubsection{SNR J0047.2$-$7308, B0045-73.4, IKT 2:  unlikely detection}

\begin{figure*}
  \begin{minipage}[c]{0.70\textwidth}
    \includegraphics[width=\textwidth]{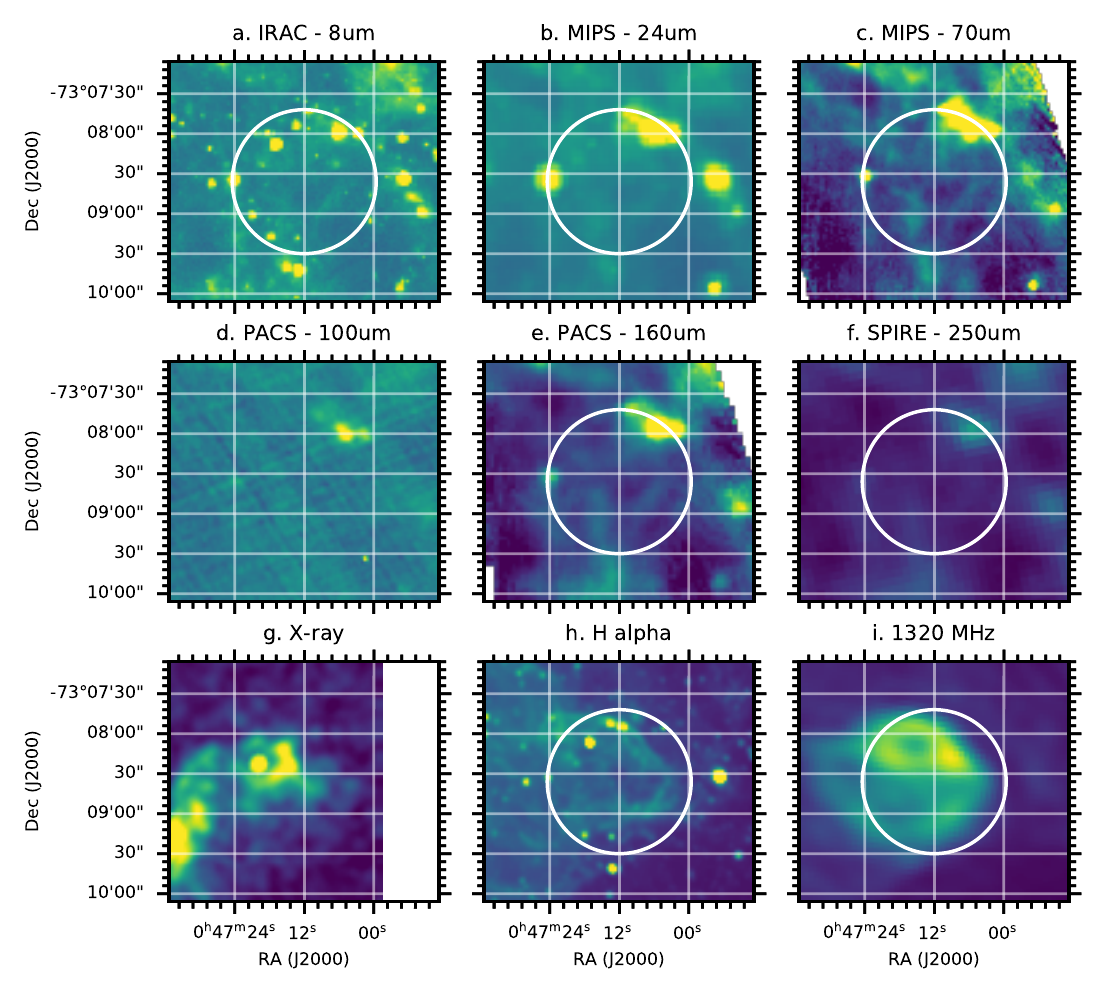}
  \end{minipage}\hfill
  \begin{minipage}[c]{0.30\textwidth}
    \caption{SNR J0047.2$-$7308, B0045-73.4, IKT 2: unlikely detection. The 70 and 160\,$\mu$m images are from PACS pointing observations, and the X-ray image is a  {\it Chandra} image in the 0.3--10.0 eV band.
    It is a bright shell type SNR in radio, but it is faint in  H$\alpha$. Part of the shell might be detected at 8, 70, 160 and possibly 24\,$\mu$m images near the centre, but it is not totally clear whether they are associated with the SNR or unrelated ISM emission. The bright infrared clouds in north-west of the shell are probably not part of the SNR. 
     } \label{fig-00472-photo}
  \end{minipage}
\end{figure*}

\begin{figure}
	\includegraphics[width=1.3\columnwidth]{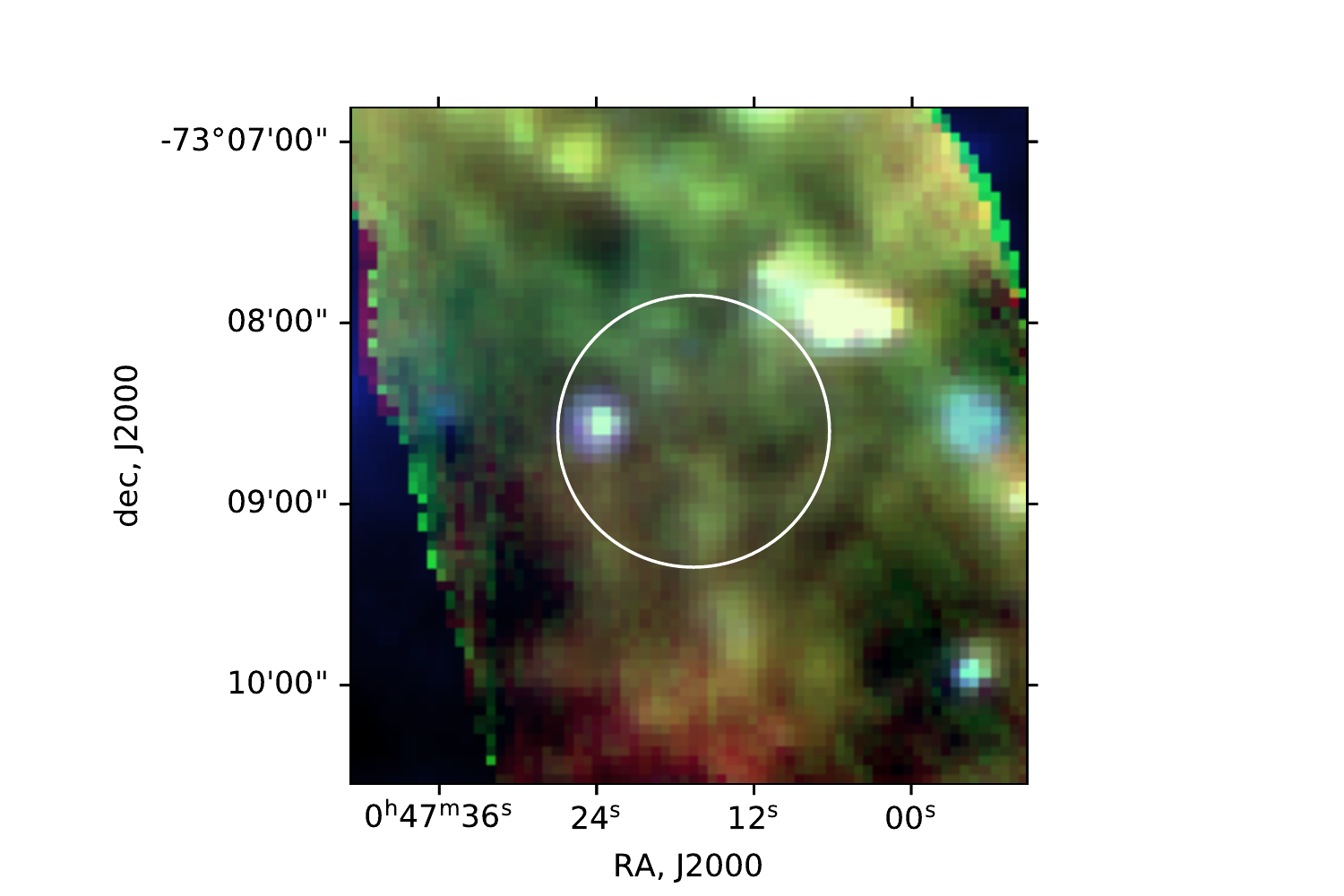}
    \caption{SNR J0047.2$-$7308, B0045-73.4, IKT 2. - 24, 70 and 160\,$\mu$m 3-color image. The cirrus within the circle appears to be a continuation of general ISM emission with homogeneous temperatures across the field, rather than dust emission clearly associated with the SNR.
}
      \label{fig-00472-3color}
\end{figure}

This SNR is radio bright (panel i of Fig\,\ref{fig-00472-photo}), and there are associated circular regions found in H$\alpha$ to the north-west and south-west.
There is a filamentary structure in the south-west of the SNR which
can be seen in H$\alpha$. Similar structure is found all of the infrared images, 
however, it is most likely part of a larger ISM structure, as the
location of infrared filament is slightly off from H$\alpha$ emitting region.
Also, this filament does not stand out in the 3-colour image shown in
Fig.\ref{fig-00472-3color} -- if the filament was part of the SNR, it
would tend to stand out in a blue colour against ISM.

There are bright infrared clouds in north-west of the shell, just inside the white circle  in Fig\,\ref{fig-00472-photo}, 
however, these clouds are probably not part of the SNR, but associated with molecular clouds or an H{\small II} region, as there is no associated emission found in either X-ray or radio images.

\subsubsection{SNR J0047.5$-$7306, B0045$-$73.3: non detection}

This SNR was reported in {\it ROSAT} and {\it XMM-Newton} X-ray images, 
with \citet{2005MNRAS.364..217F} identifying the corresponding radio emission at J004728$-$730601. 
 \citet{Maggi:2019wq}  showed that there are two SNRs in very close proximity (SNR J0047.5$-$7306 and SNR J0047.2$-$7308), separated by slightly different X-ray abundance patterns.
 
\begin{figure*}
  \begin{minipage}[c]{0.70\textwidth}
    \includegraphics[width=\textwidth]{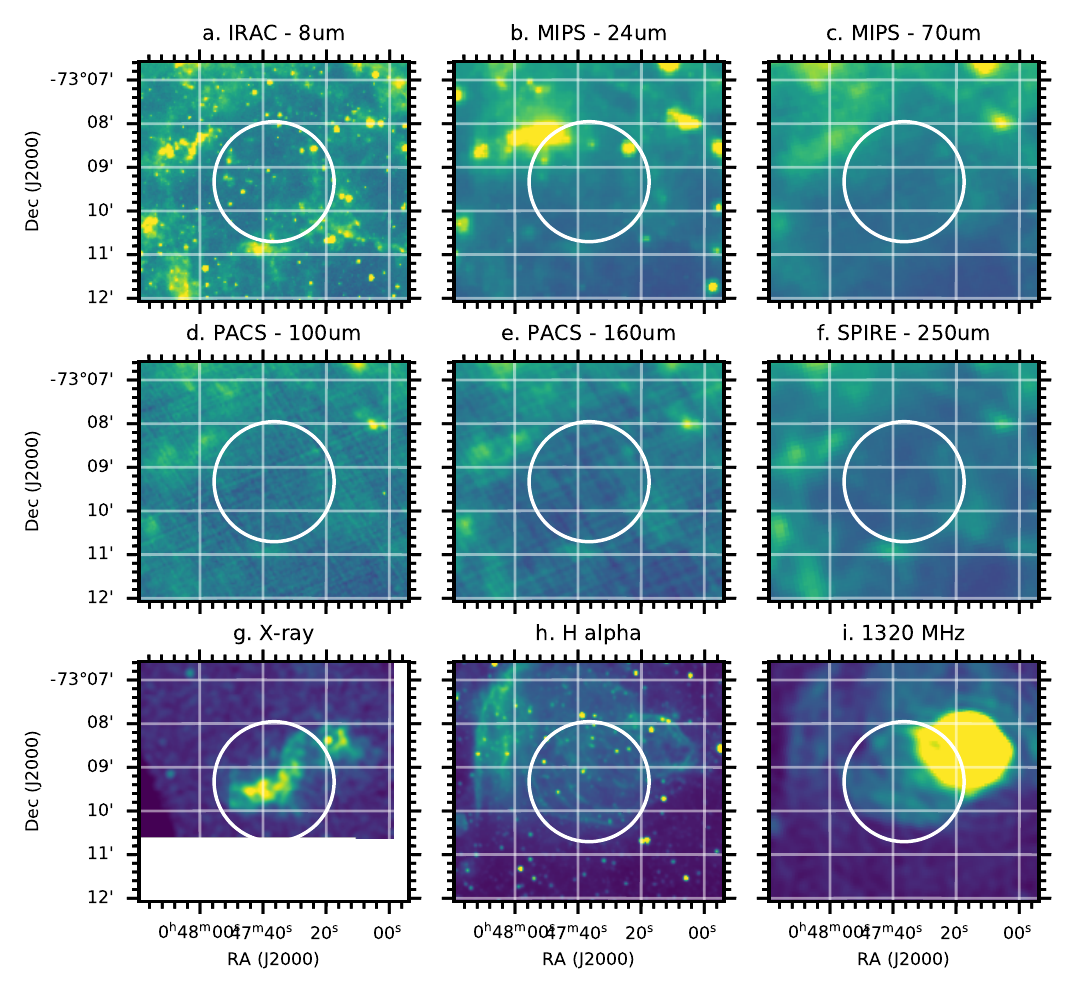}
  \end{minipage}\hfill
  \begin{minipage}[c]{0.30\textwidth}
    \caption{SNR J0047.5$-$7306, B0045-733 - non-detection. This SNR is located just south of SNR J0047.2$-$7308.
     } \label{fig-00475}
  \end{minipage}
\end{figure*}

In Fig.\,\ref{fig-00475}, the H$\alpha$ image (panel h) shows that
there are at least two separate nebulae which somewhat overlap. The feature also found in the radio image (panel i).
The circle indicates the coordinates and size of the SNR from \citet{2005MNRAS.364..217F}.
Filamentary structure is found across the circle in H$\alpha$, and
some the structure may extend outside the circle.

Although there is some infrared emission within the white circle of
Fig.\,\ref{fig-00475}, none of the structure resembles H$\alpha$ ones, leading to its listing as a non-detection. The IR emission is most likely of ISM origin.

\subsubsection{SNR J0048.4$-$7319, B0046-73.5, DEM S42E, IKT 4: unlikely detection}

\begin{figure*}
  \begin{minipage}[c]{0.70\textwidth}
    \includegraphics[width=\textwidth]{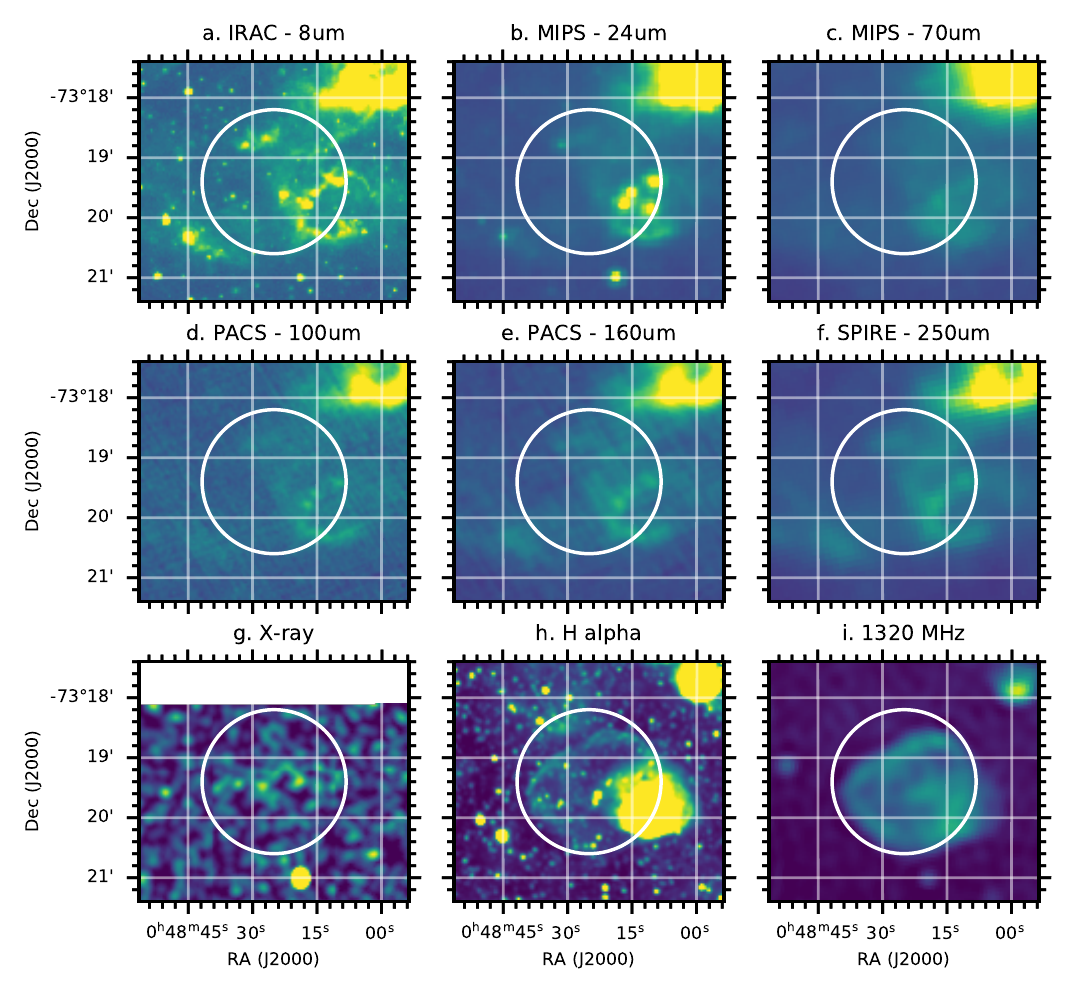}
  \end{minipage}\hfill
  \begin{minipage}[c]{0.30\textwidth}
    \caption{SNR J0048.4$-$7319, B0046-73.5, DEM S42E, IKT 4:  unlikely detection. 
   Panel (g) is  a {\it Chandra} image in the 0.3--10.0  keV band.
  A structure is found in all the IR images  from 8\,$\mu$m to 250\,$\mu$m  in the south west, but any association with the SNR is not clear.
   The nebular structure found to the north in the H$\alpha$, 8 and 24\,$\mu$m images just inside the white circle is probably not related with SNR.
       } \label{fig-00484}
  \end{minipage}
\end{figure*}

In radio, this SNR shows a shell structure, and the corresponding shell is also found in the H$\alpha$ image  (Fig.\,\ref{fig-00484}).
In H$\alpha$, a dense and bright emission nebula is seen to the west. 
 Although this bright nebula dominates the west of the SNR, a
 combination of MCELS [S{\small II}] and H$\alpha$ images shows that
 the strong sulphur throughout the SNR is not present in the nebula,
 hence,  we deem this bright  H$\alpha$ source  unlikely to be part of the SNR.
The rim of this emission is found in all of the infrared images, some of which have a necklace-like appearance.
One of the blobs is the young stellar object candidate S3MC\,J004815.34-731935.61 \citep{Bolatto:2007hha}.
It may be possible that the infrared and H$\alpha$ emission are associated with the young stellar object and H{\small II} region. 

There is a discrete patch of emission found in north-east of the circle, not only at 8\,$\mu$m but faintly across the other mid-infrared images.
This emission is located between SNR J0048.4$-$7319 and a cluster of stars Bruck\,47 \citep{Bica:1995ic}, both of which are clearly seen in the H$\alpha$ image. 
It is unclear whether the detected filaments  with two blobs inside at 8\,$\mu$m belong to SNR, the cluster, or unrelated ISM, even though they are quite strong in sulpuur.

There is no clear SNR associated emission detected in the infrared, so this SNR is classified as `unlikely detection'.

\subsubsection{ SNR J0049.0$-$7306: unlikely detection}

\begin{figure*}
  \begin{minipage}[c]{0.70\textwidth}
    \includegraphics[width=\textwidth]{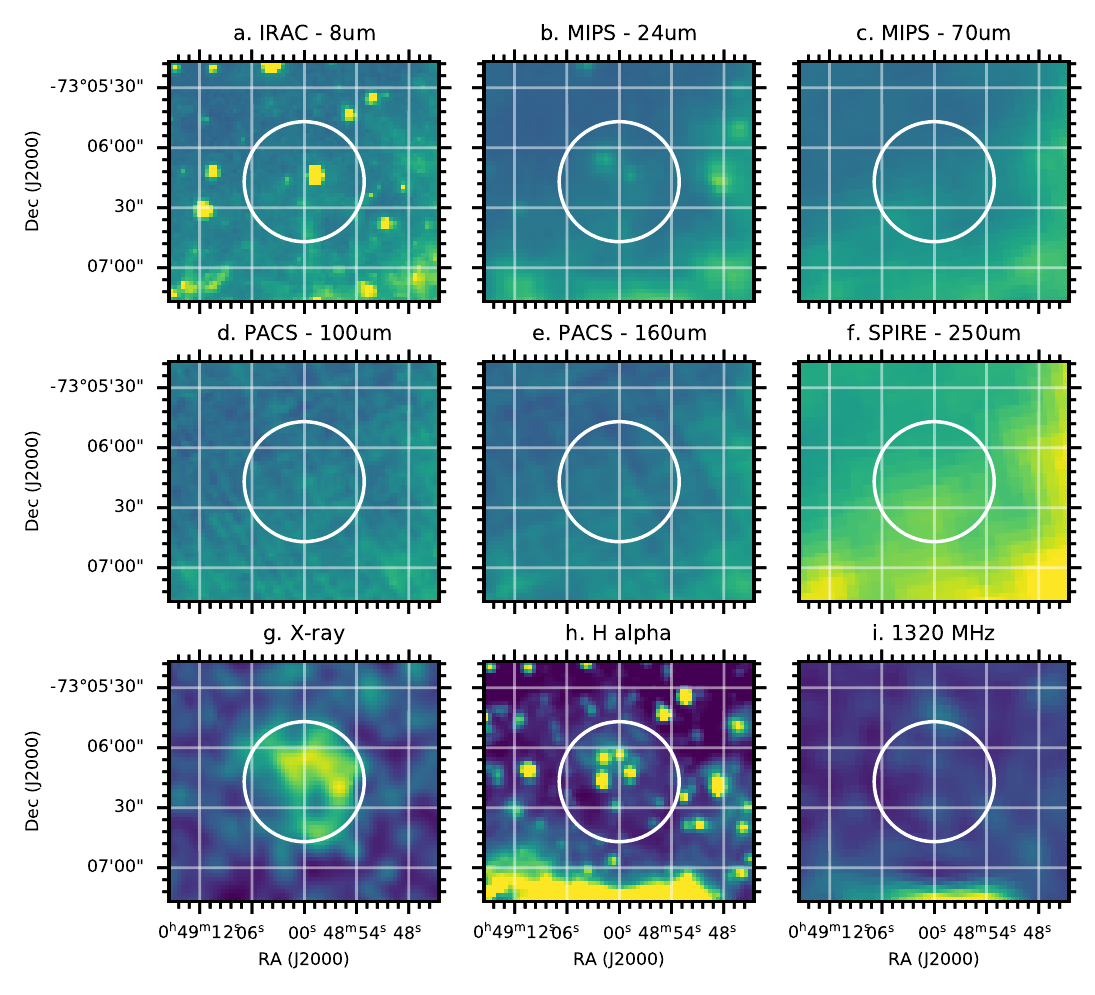}
  \end{minipage}\hfill
  \begin{minipage}[c]{0.30\textwidth}
    \caption{SNR J0049.0$-$7306: unlikely detection
       } \label{fig-00490}
  \end{minipage}
\end{figure*}

\citet{Hendrick:2005bk} reported an analysis of {\it Chandra} spectra, and concluded this to be a core-collapse SNR with O and Ne-rich gas.
 The Fe/Ne abundance ratio might suggest a progenitor mass in the range of 20--40\,\Msun \citep{Takeuchi:2016hs}. 
\citet{Haberl:to} listed it as an SNR candidate with faint emission in
{\it XMM-Newton} images, but has not been detected in radio.
In optical (Fig.\,\ref{fig-00490} h), there are four point sources,
surrounded by some nebulocity. It is difficult to evalute if this
nebulocity is associated with SNR. 
There is some diffuse emission in the infrared images at 8 and
24\,$\mu$m but it is not clear if emission is related to the SNR, thus, the SNR is unlikely detected at infrared.

\subsubsection{SNR J0051.1$-$7321, DEM S53, B0049-73.6, IKT 6: unlikely detection}

\begin{figure*}
  \begin{minipage}[c]{0.70\textwidth}
    \includegraphics[width=\textwidth]{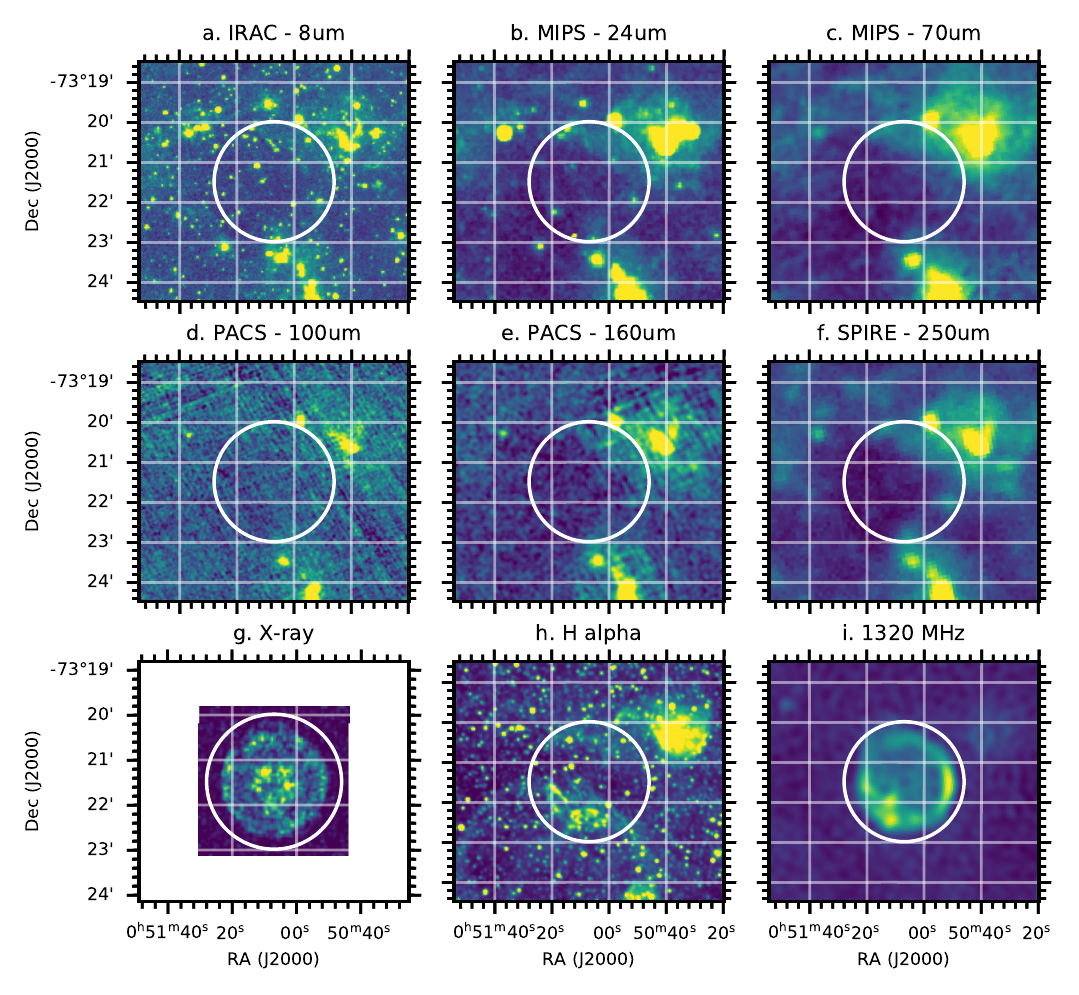}
  \end{minipage}\hfill
  \begin{minipage}[c]{0.30\textwidth}
    \caption{SNR J0051.1$-$7321, DEM S53, B0049-73.6, IKT 6: non detection. 
    Panel (g) is  a {\it Chandra} image in the 0.3--0.6 keV band.
       } \label{fig-00511}
  \end{minipage}
\end{figure*}

This SNR has an estimated age of 14,000 years with a progenitor mass of 13--15\,\Msun\, \citep{Hendrick:2005bk, Katsuda:2018gs}.
So far, 170\,\Msun\, has been swept up already \citep{Hendrick:2005bk}.
Although this SNR has a clear circular shell with an X-ray detection
of ejecta inside (Fig.\,\ref{fig-00511}), there is little emission corresponding to this feature in the infrared images.
Although the 8\,$\mu$m image shows some faint shell emission on the
south-east of the white circle, that part of the shell is relatively
faint in the radio image.
Since it would be unlikely that only this part of the SNR shell would be detected, we conclude that this must be recorded as an `unliklely detection'.

\subsubsection{SNR J00565$-$7208: non detection}

\begin{figure*}
  \begin{minipage}[c]{0.70\textwidth}
    \includegraphics[width=\textwidth]{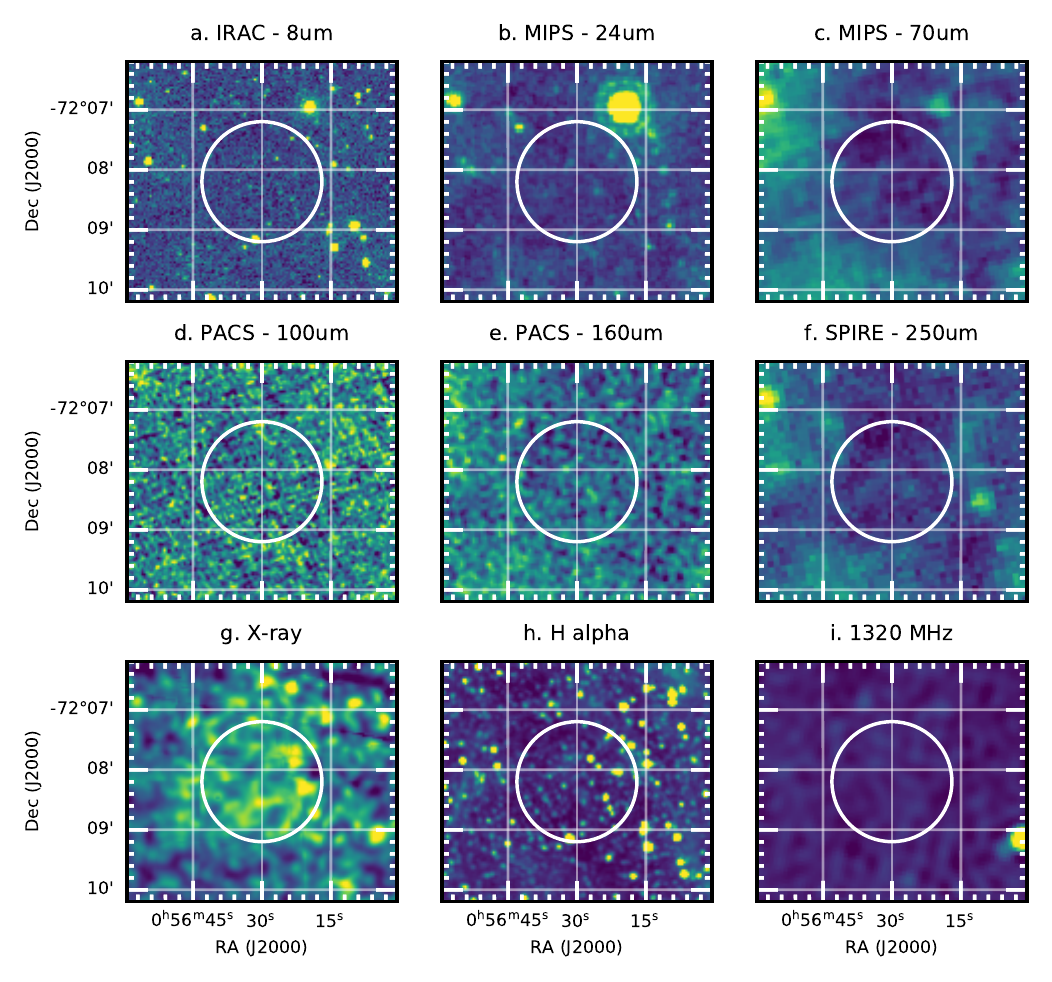}
  \end{minipage}\hfill
  \begin{minipage}[c]{0.30\textwidth}
    \caption{SNR J00565$-$7208: non detection
           } \label{fig-00565}
  \end{minipage}
\end{figure*}

From {\it XMM-Newton} observations, \citet{Haberl:to} listed SNR J00565$-$7208 as an SNR candidate.
This SNR has a low-surface brightness in X-ray \citep{Haberl:to} images.
Fig.\,\ref{fig-00565} shows that this SNR candidate is very faint in radio and H$\alpha$. The lack of a clear shape in these  reference images makes it difficult to identify the SNR candidate in infrared images.

\subsubsection{SNR J0057.7$-$7213: non detection}

\begin{figure*}
  \begin{minipage}[c]{0.70\textwidth}
    \includegraphics[width=\textwidth]{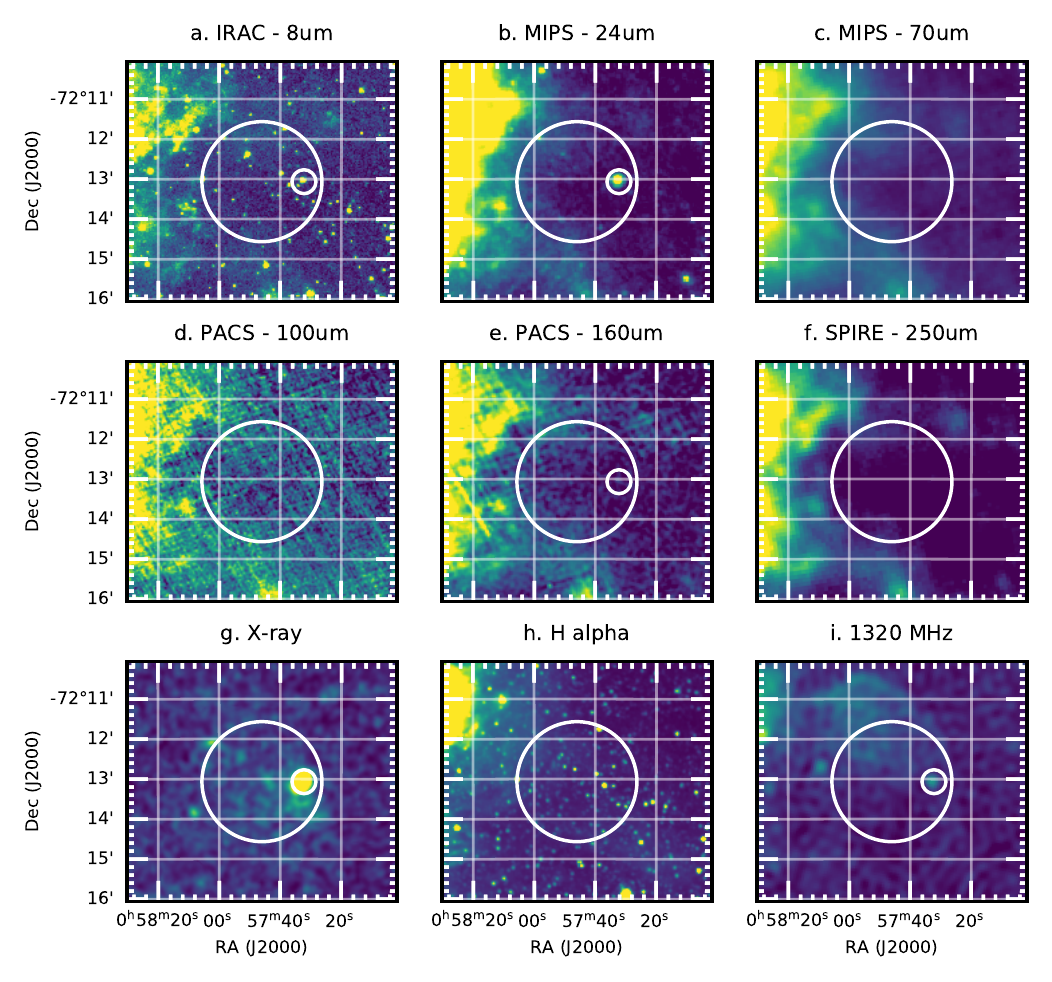}
  \end{minipage}\hfill
  \begin{minipage}[c]{0.30\textwidth}
    \caption{SNR J0057.7$-$7213: non detection. A bright point source, detected in X-ray, radio and some infrared images is unrelated to the SNR.
           } \label{fig-00577}
  \end{minipage}
\end{figure*}

\begin{figure*}
  \begin{minipage}[c]{0.70\textwidth}
    \includegraphics[width=\textwidth]{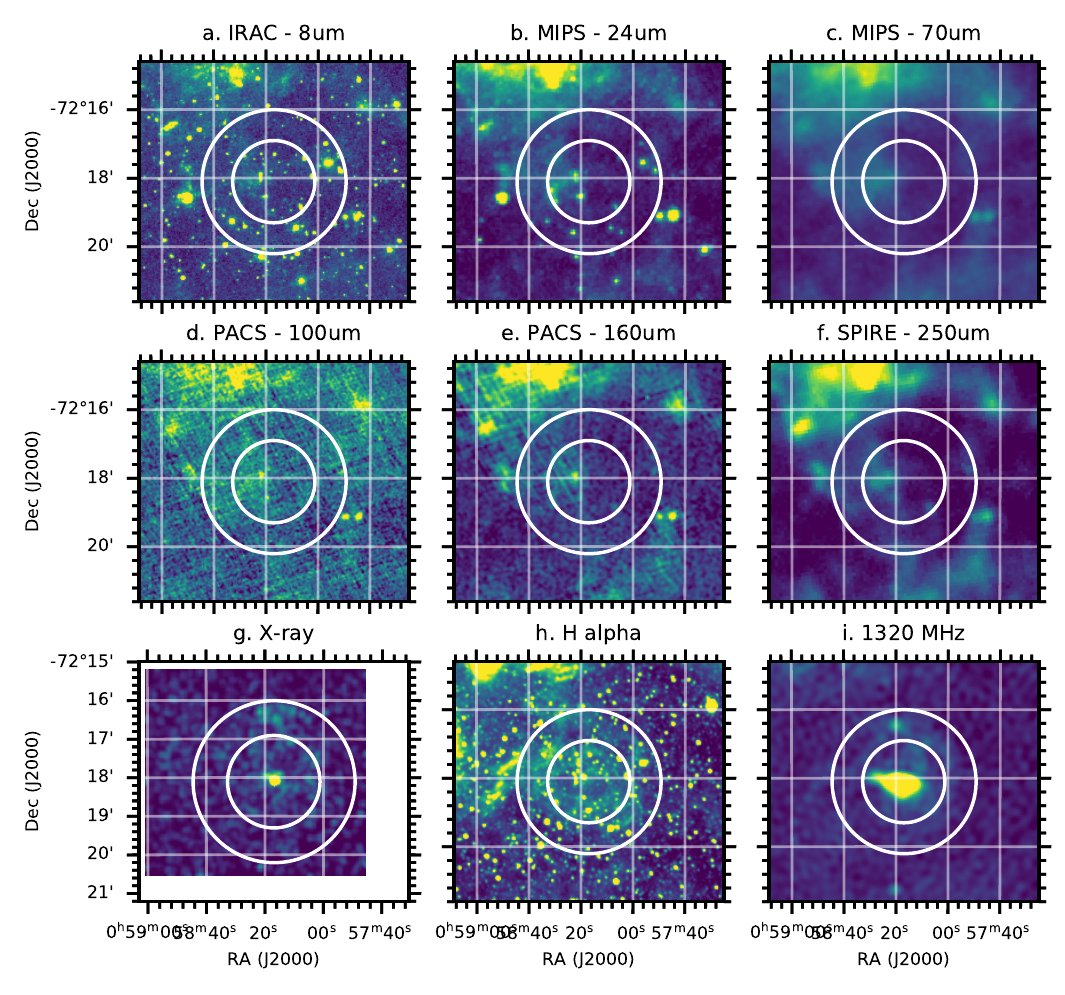}
  \end{minipage}\hfill
  \begin{minipage}[c]{0.30\textwidth}
    \caption{SNR J0058.3$-$7218,  B0056-72.5, IKT 16: unlikely detection. 
    Panel (g) is  a {\it Chandra} image in the 0.58--1.17 keV band.
           } \label{fig-00583}
  \end{minipage}
\end{figure*}

SNR J0057.7$-$7213 is listed as an SNR candidate by \citet{Haberl:to}, based on the {\it XMM-Newton} survey of SMC SNRs.
Radio emission found next to the X-ray emission (Fig.\ref{fig-00577}) belongs to SNR J0058.3$-$7218 \citep{2005MNRAS.364..217F}, which is displayed in Fig.\ref{fig-00583}.
It is unclear if these are two individual SNRs or if they are linked regions of one single SNR \citep{Haberl:to}.
There is no infrared detection, made worse by the bright ISM emission that is overwhelming the region.

There is a bright point source detected in X-ray and radio, as well as at 8, 24, and 160\,$\mu$m.
This is 2MASS J00573272$-$7213022, also known as [NHS 2003] 4 from \citet{Naze:2003dw}.
It was initially proposed to be an X-ray binary, due to 40\,\% variabilities in X-ray fluxes 
\citep{Anonymous:Ks2vG43g}, but  it was later classified as a candidate non-nuclear extragalactic source 
\citep{Lin:2012kx}.
This bright X-ray and infrared source is unlikely to be related to the SNR.

\subsubsection{SNR J0058.3$-$7218, B0056-72.5, IKT 16: unlikely detection}

  SNR J0058.3$-$7218 is the only pulsar wind nebula known in the SMC \citep{Alsaberi:2019gl}, with an estimated SNR age of 14,700 yrs \citep{Owen:2011hp}.
There is a bright point source found in X-ray, which has been identified as emission from the neutron star  \citep{Owen:2011hp, Maitra:2015ig}.
It has also been suggested as a possible AGN \citep{Sturm:2013vy}, however, \citep{Alsaberi:2019gl} determined it to be a pulsar. 

There are two point sources found in H$\alpha$ (Fig.\ref{fig-00583}) near the centre: XMMU\,J005822.1$-$721759 and XMMU\,J005820.7$-$721754.
Both of them were listed as AGN candidates before \citep{Sturm:2013vy} but XMMU\,J005822.1$-$721759 was later revised as a pulsar.
The source in the east, XMMU\,J005820.7$-$721754, is bright in the infrared and it is elongated, but it is probably not associated with the SNR, if it is an AGN candidate.

The H$\alpha$ image shows bright nebular inside the 1.2 arcmin radius circle  (smaller circle in Fig.\ref{fig-00583}).
Although there is some emission on the east side of the inner circle seen in all the infrared images with a point source within, this emission is probably not associated with the SNR, as its morphology does not resemble that of H$\alpha$.




\subsubsection{SNR J0100.3$-$7134, B0058-71.8, DEM S108: non-detection}

\begin{figure*}
  \begin{minipage}[c]{0.70\textwidth}
    \includegraphics[width=\textwidth]{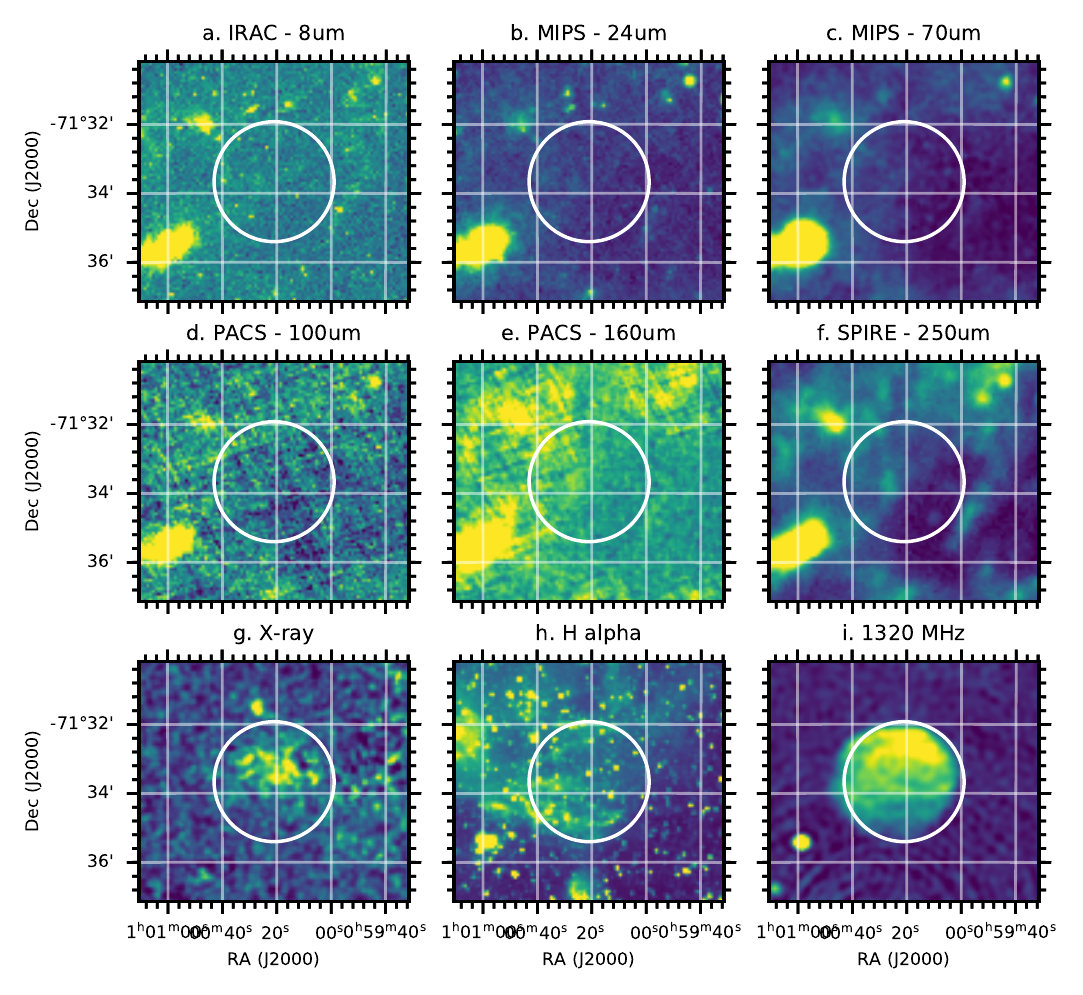}
  \end{minipage}\hfill
  \begin{minipage}[c]{0.30\textwidth}
    \caption{SNR J0100.3$-$7134, B0058-71.8, DEM S108: non detection. 
           } \label{fig-01003}
  \end{minipage}
\end{figure*}

The radio emission of SNR J0100.3$-$7134 is bright in the north, while in H$\alpha$,  emission in the southern part, tracing a shell structure, is brighter.
Although there is some infrared emission within and surrounding this
SNR, the morphology does not resemble radio or H$\alpha$ (Fig.\ref{fig-01003}), so this SNR is listed as a non detection.

\begin{figure*}
  \begin{minipage}[c]{0.70\textwidth}
    \includegraphics[width=\textwidth]{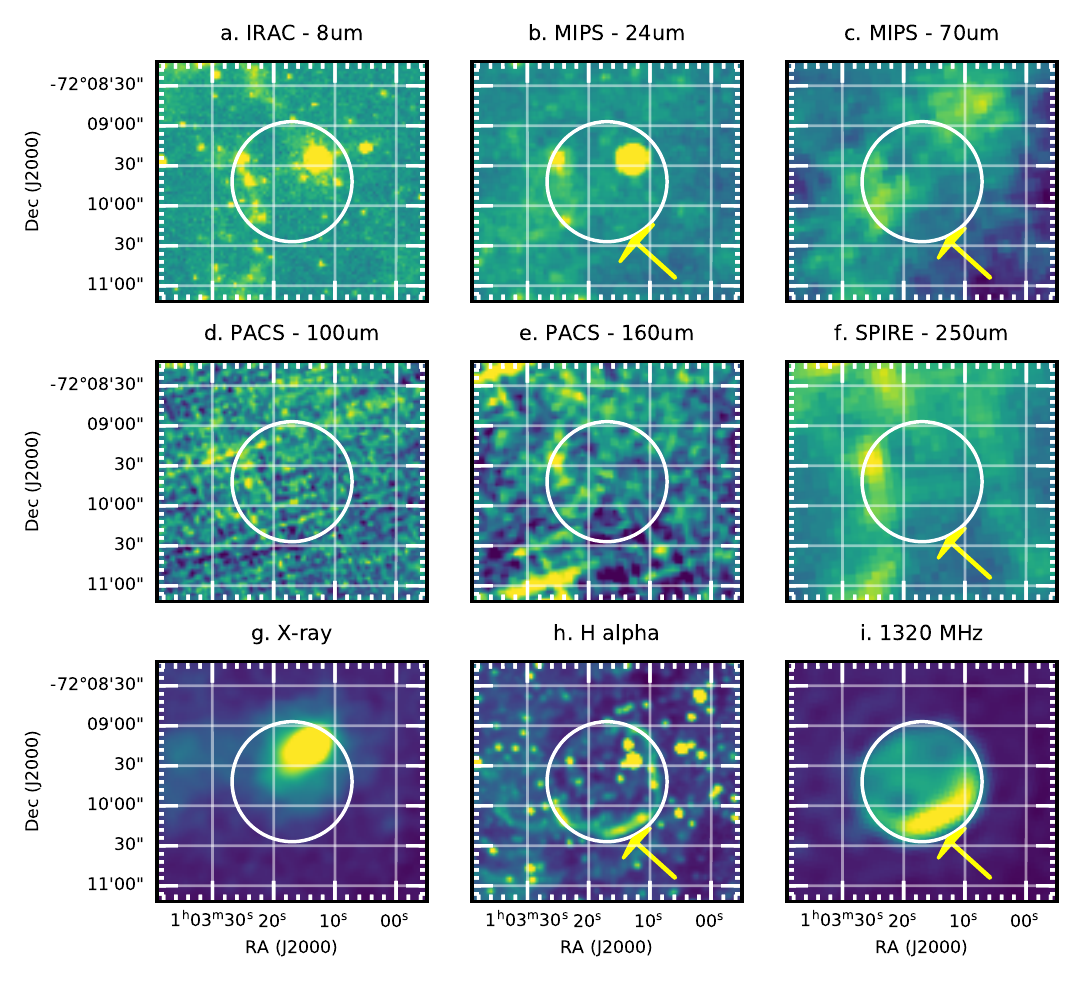}
  \end{minipage}\hfill
  \begin{minipage}[c]{0.30\textwidth}
    \caption{SNR J0103.2$-$7209,  B0101-72.6, IKT 21: unlikely detection.  
           } \label{fig-01032}
  \end{minipage}
\end{figure*}

\subsubsection{SNR J0103.2$-$7209,  B0101-72.6, IKT 21: unlikely detection}

SNR J0103.2$-$7209 has a well-defined shell structure in radio and H$\alpha$ (Fig.\,\ref{fig-01032}). 
Observations in [S{\small II}] were used to determine a diameter of about 90\,arcsec \citep{Mathewson:1984ke, Maggi:2019wq}.
In X-ray, the reported size is much larger at 270 arcsec \citep{Maggi:2019wq}.

H$\alpha$ and radio images both show the brightest part of the rim is in the south-west (Fig.\,\ref{fig-01032}). This rim may correlate with emission seen in 24\,$\mu$m and 250\,$\mu$m, however, due to the limited sensitivity, it is not clearly defined. This SNR is therefore classified as an unlikely detection in infrared.


\subsubsection{SNR J0105.1$-$7223, DEM S128, ITK 23: non detection}

\begin{figure*}
  \begin{minipage}[c]{0.70\textwidth}
    \includegraphics[width=\textwidth]{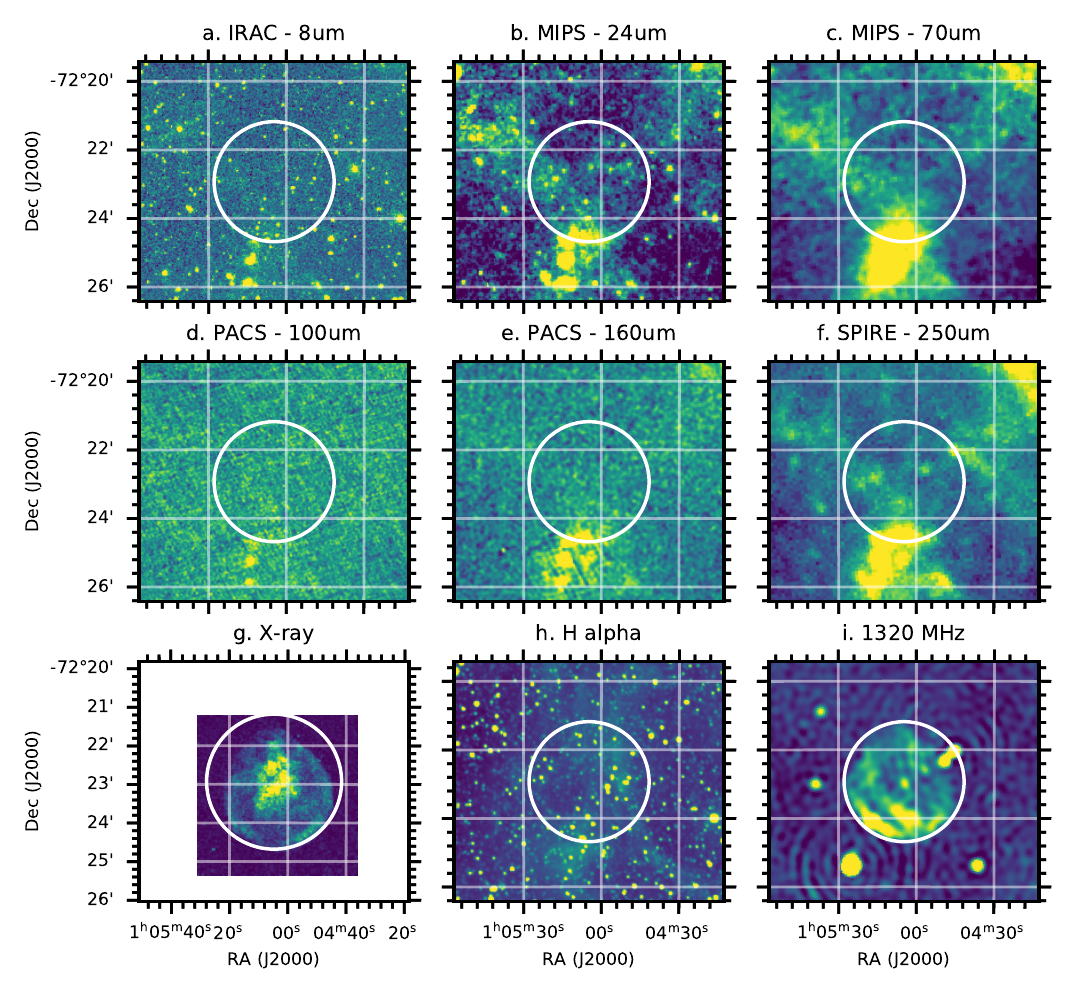}
  \end{minipage}\hfill
  \begin{minipage}[c]{0.30\textwidth}
    \caption{SNR J0105.1$-$7223, B0103-72.6, DEM S125, IKT 23: non-detection. 
    Panel (g) is  a {\it Chandra} image in the 0.3--10.0 keV band.
           } \label{fig-01051}
  \end{minipage}
\end{figure*}

SNR J0105.1$-$7223 is the second X-ray brightest SNR in the SMC, and has a core-collapse origin \citep{vanderHeyden:2004ee, Maggi:2019wq}.
Any optical and infrared emission from the SNR, if it exists, is buried within unrelated ISM,
which includes the bright region in the south  (Fig.\,\ref{fig-01051}).
This SNR is categorised as a non detection in the infrared.

\subsubsection{ SNR candidate J01065$-$7242: non detection}

\begin{figure*}
  \begin{minipage}[c]{0.70\textwidth}
    \includegraphics[width=\textwidth]{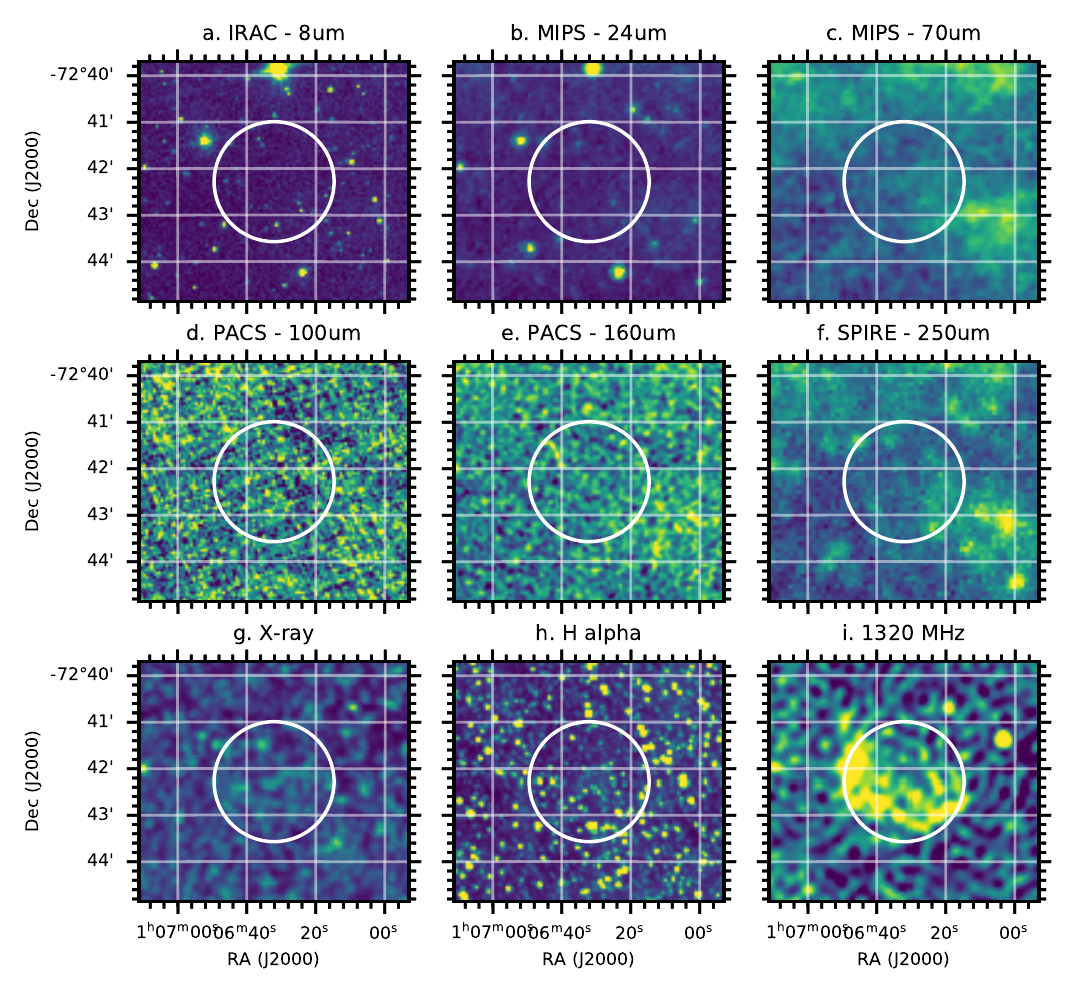}
  \end{minipage}\hfill
  \begin{minipage}[c]{0.30\textwidth}
\caption{ SNR candidate J01065$-$7242: non detection
           } \label{fig-1065}
  \end{minipage}
\end{figure*}

This object was relatively recently nominated as a candidate SNR, after detecting radio continuum emission from the ASKAP survey. However, it is yet to be confirmed with an X-ray spectrum \citep{Maggi:2019wq}.
Indeed, although the radio image shows a clear spherical  shell
(Fig\,\ref{fig-1065} i), it is not clearly found in X-ray or
H$\alpha$ and there is no sign of infrared emission associated with
this SNR.


\subsubsection{ SNR candidate J01097$-$7318: non detection}

\begin{figure*}
  \begin{minipage}[c]{0.70\textwidth}
    \includegraphics[width=\textwidth]{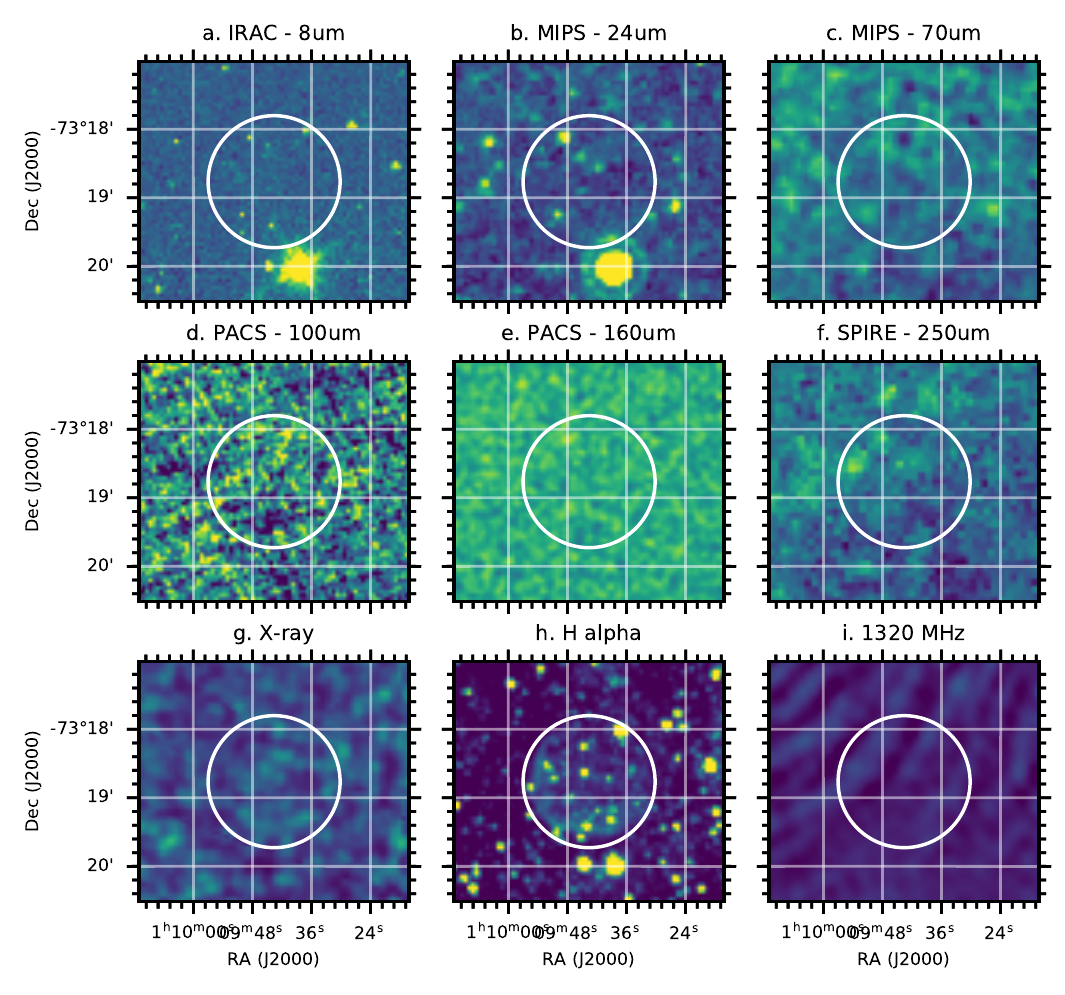}
  \end{minipage}\hfill
  \begin{minipage}[c]{0.30\textwidth}
\caption{ SNR candidate J01097$-$7318: non-detection
           } \label{fig-01097}
  \end{minipage}
\end{figure*}

SNR candidate J01097$-$7318 was recognised by a shell-type optical structure, and its high [S\,{\small II}]/H$\alpha$ ratio in the MCELS survey \citep{Maggi:2019wq}.
It has a clearly defined sphere in the optical (Fig.\,\ref{fig-01097} h). 
\citet{Maggi:2019wq} stated some issue with the X-ray image, which is also included in this paper  (Fig.\,\ref{fig-01097} g). 
Unfortunately, we don't see any corresponding infrared emission in Fig.\,\ref{fig-01097}.

\subsection{SNR candidates/Non-SNRs?}\label{sect-non-SNR}

In this section, we present SNR J0047.8$-$7317 which had been thought to be SNR, but recent study by \citet{Maggi:2019wq} classified as non SNRs.
We display infrared images with possible detection.


\begin{figure*}
  \begin{minipage}[c]{0.70\textwidth}
    \includegraphics[width=\textwidth]{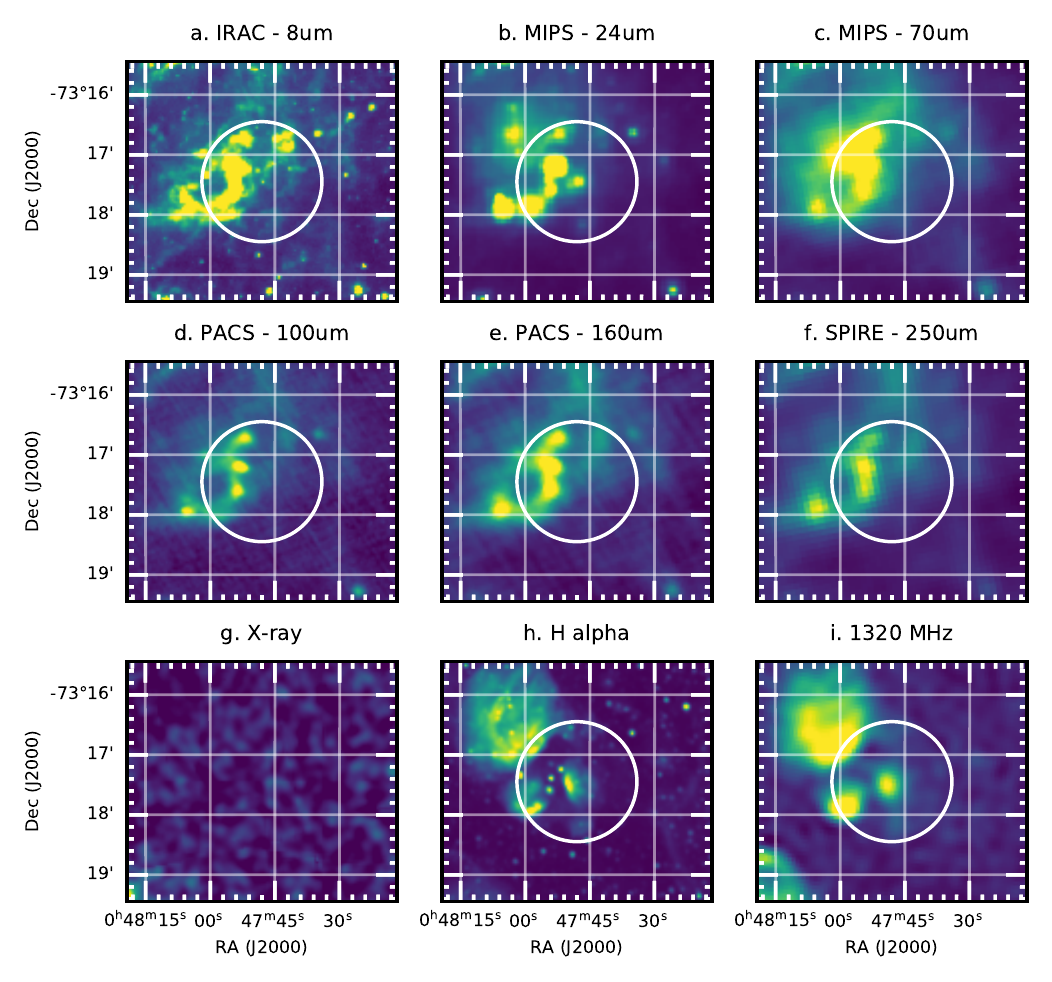}
  \end{minipage}\hfill
  \begin{minipage}[c]{0.30\textwidth}
    \caption{Historically, SNR J0047.8$-$7317 but recent work found this SNR to be a part of H{\small II} region N19.  A bright point source is detected at 24\,$\mu$m, which has corresponding in H$\alpha$ and radio emission. Only with a single band detection, it is difficult to evaluate its nature: possible detection.
       } \label{fig-00478}
  \end{minipage}
\end{figure*}



\subsubsection{SNR J0047.8$-$7317, NS21: possible detection}

This source was initially listed as a SNR candidate from radio emission \citep{Filipovic:2002kx, 2005MNRAS.364..217F},  with X-ray detections  later \citep{Haberl:to}.
However, \citet{Pellegrini:2012ii} identified this object (LHA 115-N21) as an H{\small II} region, from optical [S{\small II}] and [O{\small III]} ratio.
Recent radio images by ASKAP did not resolve a shell, hence, considered to be a part of H{\small II} region N19 \citep{Maggi:2019wq}.

A point source, corresponding to the radio source, is detected at the centre of this nebula at 24\,$\mu$m only (Fig.\,\ref{fig-00478}).
Because it is only one band detection, its identification is undetermined.

An H{\small II} region/molecular cloud NGC 267 is located north-east of this SNR, as traced by  H$\alpha$ emission,
and molecular cloud [RLB 93] SMC-B2\,1 (RA 00 47 47.4, Dec $-$73 16 48) and several young stellar objects are found in between the location of the SNR and NGC\,267. These young stellar objects are bright in 8--250\,$\mu$m.


\clearpage  

\subsubsection{SNR J0051.9$-$7310, IKT 7: unlikely detection}

SNR J0051.9$-$7310 is a candidate of SNR, reported by {\it Einstein} X-ray detection \citep{Inoue:uy}, however, its identification is still disputed.
It could be associated with X-ray Binary, SXP 172 \citep{Anonymous:JdFjFa_s}, but could be still core-collapse SNR \citep{Anonymous:c6obe9b8}
It has listed as 97'' diameter in X-ray \citet{2010MNRAS.407.1301B} based on on-line source MCSNR, which is no longer available, however, \citet{Anonymous:c6obe9b8} state it as a point source. There is a contradicting information about this potential SNR.

\begin{figure*}
  \begin{minipage}[c]{0.70\textwidth}
    \includegraphics[width=\textwidth]{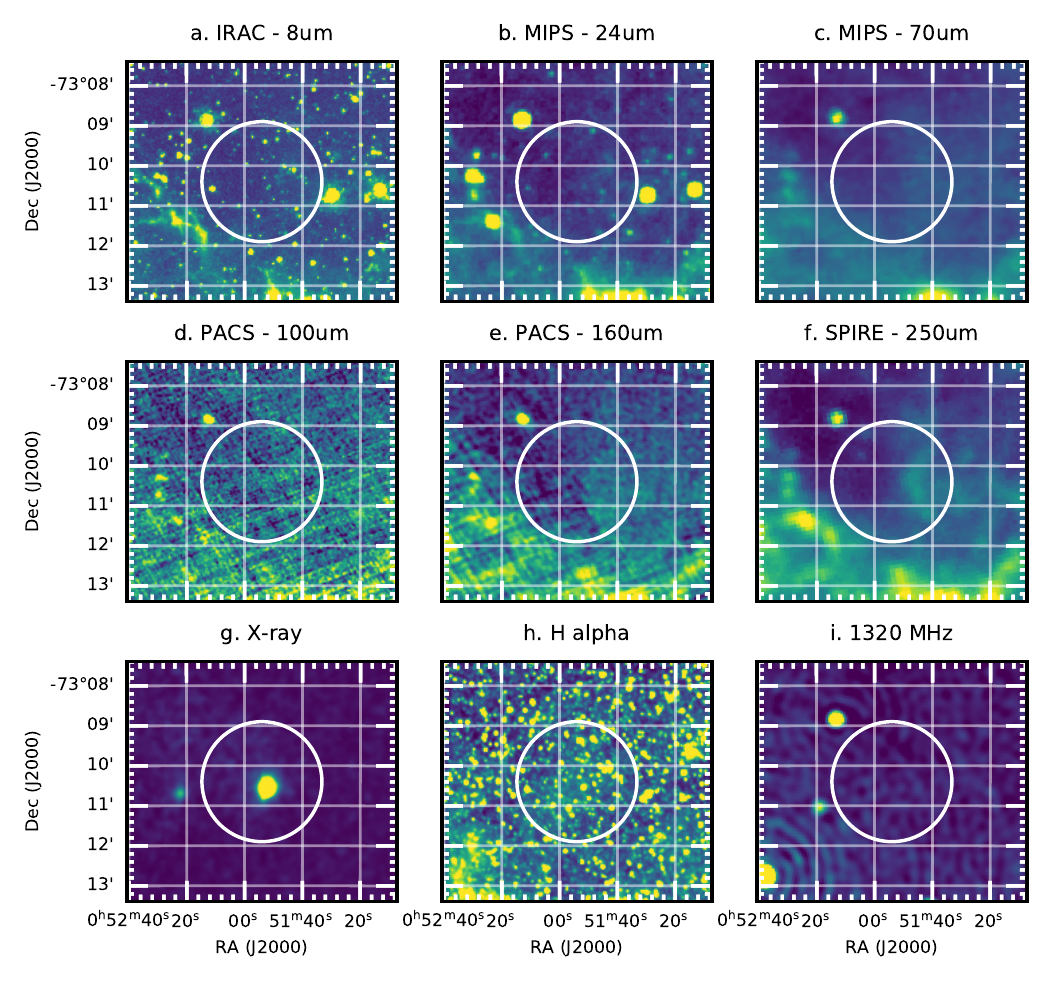}
  \end{minipage}\hfill
  \begin{minipage}[c]{0.30\textwidth}
    \caption{SNR J0051.9$-$7310, IKT 7: unlike detection.
       } \label{fig-00519}
  \end{minipage}
\end{figure*}

There is no infrared emission associated to this point source Fig.\ref{fig-00519}. 
H$\alpha$ image contains so many point sources, so that it is extremely difficult to identify extended emission, if any from SNR.
There is also diffuse emission in infrared within the white circle, however, it is difficult to claim that this is associated with possible SNR, so that this SNR is classified as unlikely detection.

\subsubsection{ SNR J0114.0$-$7317, LHA 115-N 83C: non detection}

SNR J0114.0$-$7317 is listed as SNR candidate by  \citet{Anonymous:TowVvoI8}.
However, no X-ray emission was detected, and it is likely to be a compact H{\small II} region \citep{Maggi:2019wq}.

The infrared image is overwhelmed with nearby star forming region, and can not find any emission from this possible SNR, if any.

\begin{figure*}
  \begin{minipage}[c]{0.70\textwidth}
    \includegraphics[width=\textwidth]{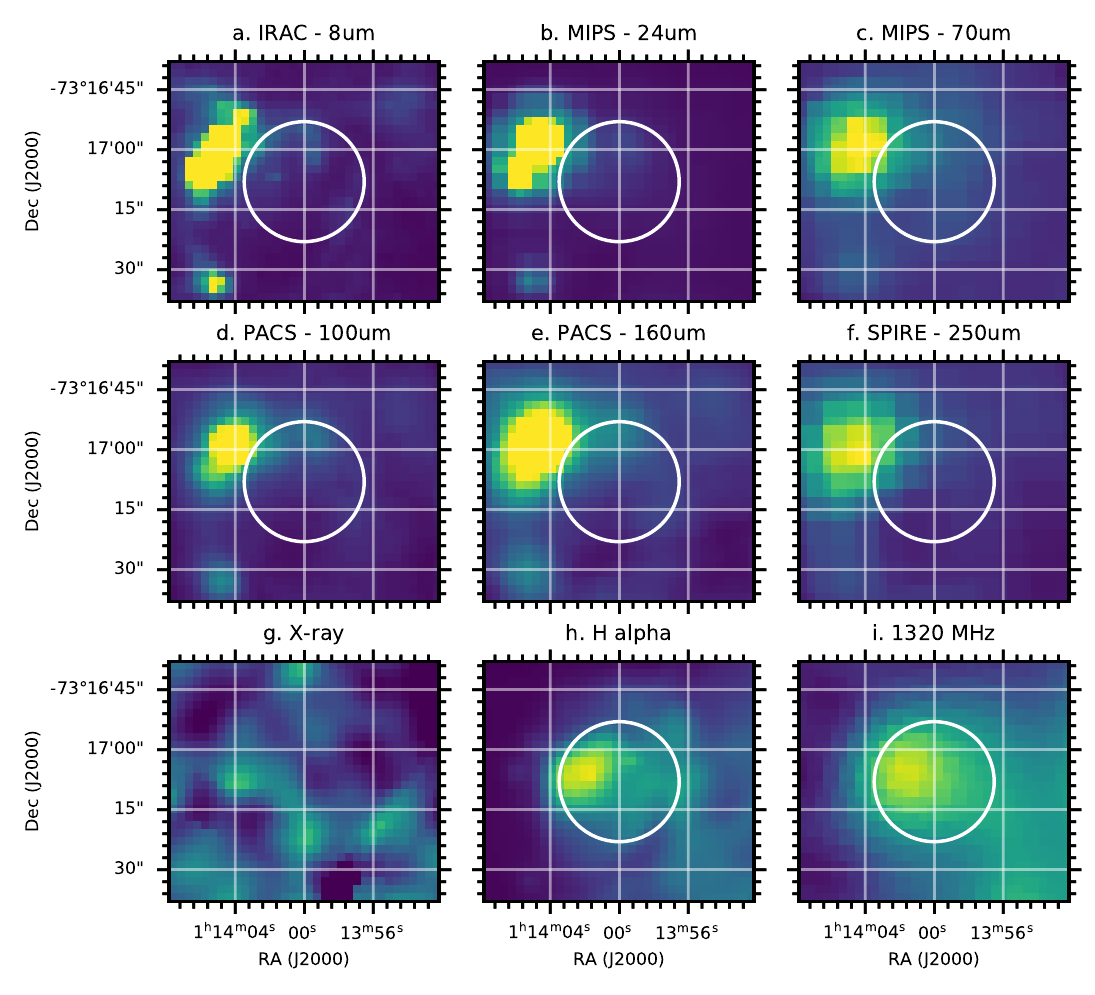}
  \end{minipage}\hfill
  \begin{minipage}[c]{0.30\textwidth}
\caption{ SNR J0114.0$-$7317, LHA 115-N 83C: no detection.
           } \label{fig-01140}
  \end{minipage}
\end{figure*}



\bsp	
\label{lastpage}
\end{document}